\documentclass[aps]{revtex4}
\usepackage{amsmath}
\usepackage{amssymb}
\usepackage{hyperref}
\hypersetup{colorlinks,linkcolor=red,citecolor=cyan,anchorcolor=black} 
\begin{document}
\allowdisplaybreaks
 
\title{Cosmological perturbations in conformal gravity II}

\author{Asanka Amarasinghe, Matthew G. Phelps and Philip D. Mannheim}
\affiliation{Department of Physics, University of Connecticut, Storrs, CT 06269, USA \\
asanka.amarasinghe@uconn.edu, matthew.phelps@uconn.edu,\\ philip.mannheim@uconn.edu\\ }

\date{December 31, 2018}

\begin{abstract}

In this paper we continue a study of cosmological perturbations in the conformal gravity theory. In previous work we had obtained a restricted set of solutions to the cosmological fluctuation equations, solutions that were required to be both transverse and synchronous. Here we present the general solution. We show that in a conformal invariant gravitational theory fluctuations around any background that is conformal to flat (backgrounds that include the cosmologically interesting Robertson-Walker and de Sitter geometries) can be constructed from the (known) solutions to fluctuations around a flat background. For this construction to hold it is not necessary that the perturbative geometry associated with the fluctuations itself be conformal to flat. Using this construction we show that in a conformal Robertson-Walker cosmology early universe fluctuations grow as $t^4$. We present the scalar, vector, tensor decomposition of the fluctuations in the conformal theory, and compare and contrast our work with the analogous treatment of fluctuations in the standard Einstein gravity theory.

\end{abstract}

\maketitle

\section{Introduction}
\label{S1}

In a recent paper \cite{Mannheim2012a} we presented the first steps in an analysis of cosmological fluctuations in the fourth-order derivative conformal gravity theory. Conformal gravity has been advanced by one of us as a candidate alternative to standard Einstein gravity and reviews of its status at both the classical and quantum levels may be found in \cite{Mannheim2006,Mannheim2012b,Mannheim2017}, with the establishment of its unitarity and the positivity of its inner product at the quantum level being found in \cite{Bender2008a,Bender2008b,Mannheim2011a,Mannheim2018} and reviewed briefly in Appendix \ref{SG}. Various other studies of conformal gravity and of higher derivative gravity theories in general can be found in \cite{Hoyle1964,Stelle1977,Stelle1978,Adler1982,Lee1982,Zee1983,Riegert1984a,Riegert1984b,Teyssandier1989,'tHooft2010a,'tHooft2010b,'tHooft2011,'tHooft2015a,Maldacena2011}. In the study of \cite{Mannheim2012a} we found some specific perturbative solutions that are of cosmological interest, and in this paper we present the general and exact perturbative solutions to fluctuations around any background that is conformal to flat. Since both the Robertson-Walker and de Sitter geometries are conformal to flat, our results are immediately of relevance to cosmology. As we show both in \cite{Mannheim2012a} and here, since Robertson-Walker and de Sitter  background geometries are conformal to flat, treatment of fluctuations around them is greatly facilitated by working in a gravitational theory that possesses conformal symmetry. In fact by the judicious choice of gauge that we make in this paper (specifically a gauge condition that is itself conformally invariant), we are able to show that in the conformal theory fluctuations around any background that is conformal to flat can be constructed from fluctuations around a flat background, with this being the case even though the perturbative geometry associated with the fluctuations need not itself be conformal to flat.

As a possible candidate alternative to standard Einstein gravity, conformal gravity is attractive in that it is a pure metric theory of gravity that possesses all of the general coordinate invariance and equivalence principle structure of standard gravity while augmenting it with an additional symmetry, local conformal invariance, in which  the action is left invariant under local conformal transformations on the metric of the form $g_{\mu\nu}(x)\rightarrow e^{2\alpha(x)}g_{\mu\nu}(x)$ with arbitrary local phase $\alpha(x)$. Under such a symmetry a gravitational action that is to be a polynomial function of the Riemann tensor is uniquely prescribed, and with use of the Gauss-Bonnet theorem is given by (see e.g. \cite{Mannheim2006}) 
\begin{eqnarray}
I_{\rm W}=-\alpha_g\int d^4x\, (-g)^{1/2}C_{\lambda\mu\nu\kappa}
C^{\lambda\mu\nu\kappa}
\equiv -2\alpha_g\int d^4x\, (-g)^{1/2}\left[R_{\mu\kappa}R^{\mu\kappa}-\frac{1}{3} (R^{\alpha}_{\phantom{\alpha}\alpha})^2\right].
\label{AP1}
\end{eqnarray}
Here $\alpha_g$ is a dimensionless  gravitational coupling constant, and
\begin{eqnarray}
C_{\lambda\mu\nu\kappa}= R_{\lambda\mu\nu\kappa}
-\frac{1}{2}\left(g_{\lambda\nu}R_{\mu\kappa}-
g_{\lambda\kappa}R_{\mu\nu}-
g_{\mu\nu}R_{\lambda\kappa}+
g_{\mu\kappa}R_{\lambda\nu}\right)
+\frac{1}{6}R^{\alpha}_{\phantom{\alpha}\alpha}\left(
g_{\lambda\nu}g_{\mu\kappa}-
g_{\lambda\kappa}g_{\mu\nu}\right)
\label{AP2}
\end{eqnarray}
is the conformal Weyl tensor,  a tensor that vanishes in geometries that are conformal to flat,  and that for any metric $g_{\mu\nu}(x)$ transforms as  $C^{\lambda}_{\phantom{\lambda}\mu\nu\kappa} \rightarrow  C^{\lambda}_{\phantom{\lambda}\mu\nu\kappa}$ under $g_{\mu\nu}(x)\rightarrow e^{2\alpha(x)}g_{\mu\nu}(x)$, with all derivatives of $\alpha(x)$ dropping out. Since conformal invariance requires that there be no intrinsic mass scales at the level of the Lagrangian, in the conformal theory all mass scales must come from the vacuum via spontaneous symmetry breaking. With such mass generation particles can then localize and bind into inhomogeneities such as the stars and galaxies that are of interest to astrophysics. Since the Weyl tensor would not vanish in the presence of  inhomogeneities, the transition from a cosmological background geometry to the cosmological fluctuations associated with inhomogeneities is thus a transition from conformal to flat geometries to geometries that are not conformal to flat. Despite this, and as we show in this paper, precisely because the theory does have an underlying conformal symmetry, one can still use the conformal symmetry to control the fluctuations.

In order to explicitly implement this objective our paper is organized as follows. In Sec. \ref{S2} we introduce the general formalism associated with fluctuations in the conformal gravity theory, and discuss the implications of coordinate invariance and conformal invariance for them. While this formalism had already been presented in \cite{Mannheim2012a} and is included here in order to make our presentation be self-contained, what is new here is our identifying of the new and convenient gauge condition given in  (\ref{AP23}), a condition that is conformal invariant. Its utility is that once we have solved for fluctuations around a flat background we can then readily construct fluctuations around a background that is conformal to flat. We are of course not the first to discuss fluctuations in higher derivative theories (see the flat background fluctuation studies of e.g. \cite{Stelle1977}, \cite{Stelle1978}, \cite{Lee1982}, \cite{Riegert1984a}, \cite{Riegert1984b}, \cite{Teyssandier1989}), but we are the first to study fluctuations around conformal to flat backgrounds in conformal theories. In fact our results are not even restricted to de Sitter or Robertson-Walker cosmological backgrounds, two specific backgrounds that are conformal  to flat, with our key fluctuation equation, (\ref{AP61}), holding for fluctuations around any background that is conformal to flat. In Sec. \ref{S3} we obtain the equations of motion that describe conformal gravity fluctuations around a completely general and arbitrary background, and in Sec. \ref{S4} we study fluctuations around an arbitrary conformal to flat background, and derive our key fluctuation equation, (\ref{AP61}), an equation that is new to the literature. So as to be able to find some exact solutions, in our earlier paper \cite{Mannheim2012a} we had restricted to transverse gauge fluctuations that in addition were required to be synchronous. In the present paper we have no need for the synchronous requirement, and by working in the conformal gauge we are able to find all solutions to the conformal theory cosmological fluctuation equations without approximation. Interestingly, in the specific cosmological models that we treat in Appendix \ref{SA} and Appendix \ref{SB} we find that  the solutions that are also synchronous are non-leading at late time.

In cosmological fluctuation theory it has been found very convenient to use the SVT (scalar, vector, tensor) decomposition of the fluctuations as it naturally incorporates gauge invariance. In Secs. \ref{S5} and \ref{S6} we present the general SVT formalism in a presentation that is greatly facilitated by our use of the projection technique that we provide in Appendix \ref{SE}. Our approach to the SVT decomposition is new to the literature, and in it we obtain a conformal gravity fluctuation equation around a conformal to flat background, (\ref{AP75}), that is new to the literature. And in addition, using the projection technique alone we obtain a completely gauge invariant conformal gravity fluctuation equation, (\ref{E38}), for fluctuations around a flat background. This equation is also new to the literature, and involves no need to make any choice of gauge at all, with it being gauge invariant in its own right. And in appendix \ref{SF} we generalize this result by providing a conformal gravity fluctuation equation, (\ref{F3}), for fluctuations around a general conformal to flat background, using a procedure that again requires no choice of gauge. In this general case the fluctuation equation contains 151 terms, and they can all be combined into just one single term. This shows the power of conformal symmetry. The projection technique that we develop in this paper thus provides a way to implement gauge invariance that is distinct from the SVT approach. In Sec. \ref{S7} we compare and contrast the SVT decompositions of  the fluctuation equations in conformal gravity and Einstein gravity. We augment our paper with seven appendices, and in them we apply our fluctuation studies to some specific cosmologies. Also in Appendix \ref{SG} we briefly discuss the unitarity problem in the quantum version of the conformal gravity theory, and show that the quantum theory is free of any ghost states with negative norm. With conformal gravity also being renormalizable (unlike standard gravity), it is thus a consistent quantum theory of gravity, and one can consistently quantize the classical conformal gravity fluctuations that we study in this paper.

\section{Formalism and Coordinate and Conformal Invariance}
\label{S2}

\subsection{General Formalism}

With the Weyl action $I_{\rm W}$ given in  (\ref{AP1}) being a fourth-order derivative function of the metric, functional variation with respect to the metric $g_{\mu\nu}(x)$ generates fourth-order derivative gravitational equations of motion of the form \cite{Mannheim2006} 
\begin{eqnarray}
-\frac{2}{(-g)^{1/2}}\frac{\delta I_{\rm W}}{\delta g_{\mu\nu}}=4\alpha_g W^{\mu\nu}=4\alpha_g\left[2\nabla_{\kappa}\nabla_{\lambda}C^{\mu\lambda\nu\kappa}-
R_{\kappa\lambda}C^{\mu\lambda\nu\kappa}\right]=4\alpha_g\left[W^{\mu
\nu}_{(2)}-\frac{1}{3}W^{\mu\nu}_{(1)}\right]=T^{\mu\nu},
\label{AP3}
\end{eqnarray}
where the functions $W^{\mu \nu}_{(1)}$ and $W^{\mu \nu}_{(2)}$ are given by
\begin{eqnarray}
W^{\mu \nu}_{(1)}&=&
2g^{\mu\nu}\nabla_{\beta}\nabla^{\beta}R^{\alpha}_{\phantom{\alpha}\alpha}                                             
-2\nabla^{\nu}\nabla^{\mu}R^{\alpha}_{\phantom{\alpha}\alpha}                          
-2 R^{\alpha}_{\phantom{\alpha}\alpha}R^{\mu\nu}                              
+\frac{1}{2}g^{\mu\nu}(R^{\alpha}_{\phantom{\alpha}\alpha})^2,
\nonumber\\
W^{\mu \nu}_{(2)}&=&
\frac{1}{2}g^{\mu\nu}\nabla_{\beta}\nabla^{\beta}R^{\alpha}_{\phantom{\alpha}\alpha}
+\nabla_{\beta}\nabla^{\beta}R^{\mu\nu}                    
 -\nabla_{\beta}\nabla^{\nu}R^{\mu\beta}                       
-\nabla_{\beta}\nabla^{\mu}R^{\nu \beta}                          
 - 2R^{\mu\beta}R^{\nu}_{\phantom{\nu}\beta}                                    
+\frac{1}{2}g^{\mu\nu}R_{\alpha\beta}R^{\alpha\beta},
\label{AP4}
\end{eqnarray}                                 
and where $T^{\mu\nu}$ is the conformal invariant energy-momentum tensor associated with the matter source. Since $W^{\mu\nu}$ is obtained from an action that is both general coordinate invariant and conformal invariant, in consequence, and without needing to impose any equation of motion or stationarity condition, $W^{\mu\nu}$ is automatically covariantly conserved and traceless and obeys $\nabla_{\nu}W^{\mu\nu}=0$, $g_{\mu\nu}W^{\mu\nu}=0$ on every variational path used for the functional variation of $I_{\rm W}$.
   
Despite its somewhat formidable appearance, especially compared to the standard second-order derivative Einstein equations
\begin{eqnarray}
-\frac{1}{8\pi G}\left(R^{\mu\nu} -\frac{1}{2}g^{\mu\nu}R^{\alpha}_{\phantom{\alpha}\alpha}\right)=T^{\mu\nu},
\label{AP5}
\end{eqnarray}
(\ref{AP3}) immediately admits of two key vacuum solutions, namely solutions with vanishing Weyl tensor and solutions with vanishing Ricci tensor. Solutions with vanishing Weyl tensor include the cosmologically relevant de Sitter and RW geometries, since the line elements of both geometries can be written in the conformal to flat (and thus vanishing Weyl tensor) form
\begin{eqnarray}
ds^2=-\Omega^2(t,x,y,z)\eta_{\mu\nu}x^{\mu}x^{\nu}=\Omega^2(t,x,y,z)[dt^2-dx^2-dy^2-dz^2]
\label{AP6}
\end{eqnarray}
for appropriate choices of the conformal factor $\Omega(t,x,y,z)$. (Here $\eta_{\mu\nu}$ is the flat Minkowski metric with ${\rm diag}[\eta_{\mu\nu}]=(-1,1,1,1)$ in our notation, which follows \cite{Weinberg1972}). Solutions with vanishing Ricci tensor include all vacuum solutions to Einstein gravity such as the Schwarzschild solution exterior to a static, spherically symmetric source, with the Schwarzschild  solution geometry not being conformal to flat. 

\subsection{Conformal Invariance}

Because the Weyl action is locally conformal invariant, the function $W^{\mu\nu}(x)$ has the property that under 
\begin{eqnarray}
g_{\mu\nu}(x)\rightarrow \Omega^2(x) g_{\mu\nu}(x)=\bar{g}_{\mu\nu}(x),\qquad
g^{\mu\nu}(x)\rightarrow \Omega^{-2}(x) g^{\mu\nu}(x)=\bar{g}^{\mu\nu}(x),
\label{AP7}
\end{eqnarray}
$W^{\mu\nu}(x)$ and $W_{\mu\nu}(x)$ transform as 
\begin{eqnarray}
W^{\mu\nu}(x)\rightarrow \Omega^{-6}(x) W^{\mu\nu}(x)=\bar{W}^{\mu\nu}(x),\qquad
W_{\mu\nu}(x)\rightarrow \Omega^{-2}(x) W_{\mu\nu}(x)=\bar{W}_{\mu\nu}(x),
\label{AP8}
\end{eqnarray}
where the dependence of $\bar{W}_{\mu\nu}(x)$ on $\bar{g}_{\mu\nu}(x)$ is the same as that of 
$W_{\mu\nu}(x)$ on $g_{\mu\nu}(x)$. The great utility of (\ref{AP8}) is that it holds regardless of whether or not the metric $g_{\mu\nu}(x)$ is conformal to flat. Moreover,  if we decompose each of $g_{\mu\nu}(x)$ and $\bar{g}_{\mu\nu}(x)$ into a background metric and a fluctuation according to
\begin{eqnarray}
ds^2&=&-[g^{(0)}_{\mu\nu}+h_{\mu\nu}]dx^{\mu}dx^{\nu},\qquad g_{\mu\nu}(x)=g^{(0)}_{\mu\nu}(x)+h_{\mu\nu}(x),\qquad g^{\mu\nu}(x)=g_{(0)}^{\mu\nu}(x)-h^{\mu\nu}(x),
\nonumber\\
\bar{g}_{\mu\nu}(x)&=&\bar{g}^{(0)}_{\mu\nu}(x)+\bar{h}_{\mu\nu}(x), \qquad\bar{g}^{\mu\nu}(x)=\bar{g}_{(0)}^{\mu\nu}(x)-\bar{h}^{\mu\nu}(x),
\label{AP9}
\end{eqnarray}
then $W_{\mu\nu}(x)$ and $\bar{W}_{\mu\nu}(x)$ will decompose as 
\begin{eqnarray}
W_{\mu\nu}(g_{\mu\nu})= W^{(0)}_{\mu\nu}(g^{(0)}_{\mu\nu})+\delta W_{\mu\nu}(h_{\mu\nu}),\qquad
\bar{W}_{\mu\nu}(\bar{g}_{\mu\nu})=\bar{W}^{(0)}_{\mu\nu}(\bar{g}^{(0)}_{\mu\nu})+\delta \bar{W}_{\mu\nu}(\bar{h}_{\mu\nu}),
\label{AP10}
\end{eqnarray}
where $W_{\mu\nu}(h_{\mu\nu})$ is evaluated in a background geometry with metric $g^{(0)}_{\mu\nu}(x)$, while $\bar{W}_{\mu\nu}(\bar{h}_{\mu\nu})$ is evaluated in a background geometry with metric $\bar{g}^{(0)}_{\mu\nu}(x)$. In addition, since the theory is conformal invariant, the matter sector $T_{\mu\nu}$ must transform as $\Omega^{-2}(x) T_{\mu\nu}(x)=\bar{T}_{\mu\nu}(x)$,  and decompose as
\begin{eqnarray}
T_{\mu\nu}(g_{\mu\nu})= T^{(0)}_{\mu\nu}(g^{(0)}_{\mu\nu})+\delta T_{\mu\nu}(h_{\mu\nu}),\qquad
\bar{T}_{\mu\nu}(\bar{g}_{\mu\nu})=\bar{T}^{(0)}_{\mu\nu}(\bar{g}^{(0)}_{\mu\nu})+\delta \bar{T}_{\mu\nu}(\bar{h}_{\mu\nu}).
\label{AP11}
\end{eqnarray}
Thus if we know how to solve for fluctuations $h_{\mu\nu}(x)$ around a background $g^{(0)}_{\mu\nu}(x)$, i.e. if $g^{(0)}_{\mu\nu}(x)$ is such that we can actually find solutions to $\delta W_{\mu\nu}(h_{\mu\nu})=\delta T_{\mu\nu}(h_{\mu\nu})/4\alpha_g$, we can then obtain solutions to  $\delta \bar{W}_{\mu\nu}(\bar{h}_{\mu\nu})=\delta \bar{T}_{\mu\nu}(\bar{h}_{\mu\nu})/4\alpha_g$  for fluctuations $\bar{h}_{\mu\nu}(x)$ around a background metric $\bar{g}^{(0)}_{\mu\nu}(x)$ simply by setting 
\begin{eqnarray}
\bar{h}_{\mu\nu}(x)=\Omega^2(x)h_{\mu\nu}(x),\qquad \delta \bar{W}_{\mu\nu}(\bar{h}_{\mu\nu})=\Omega^{-2}(x)\delta W_{\mu\nu}(h_{\mu\nu}).
\label{AP12}
\end{eqnarray}
Since the structure of the fluctuations around a flat background has already been obtained in \cite{Mannheim2006}, via. (\ref{AP12}) we can construct the fluctuations around any background that is conformal to flat. Since all cosmologically relevant background geometries happen to be conformal to flat, this is extremely convenient, showing that despite its fourth-order derivative nature, there are simplifications in the conformal cosmological case that do not occur in the standard second-order theory. 

Using the conformal properties of the theory we can also isolate the contribution of the trace $h=g_{(0)}^{\mu\nu}h_{\mu\nu}$ to the fluctuation $\delta W_{\mu\nu}$. Specifically, we note that under a general conformal transformation a  general metric transforms as $g_{\mu\nu}\rightarrow \Omega^2(x)g_{\mu\nu}$ while a general $W_{\mu\nu}$ transforms as $W_{\mu\nu}\rightarrow \Omega^{-2}(x)W_{\mu\nu}$. Thus if we set $\Omega^2(x)=(1+h/4)$, then to lowest order in $h$ we can set  $\Omega^2(x)g_{\mu\nu}=g_{\mu\nu}+\delta g_{\mu\nu}=g_{\mu\nu}+hg_{\mu\nu}/4$, $\Omega^{-2}(x)W_{\mu\nu}=W_{\mu\nu}+\delta W_{\mu\nu}(h)=(1-h/4)W_{\mu\nu}=
W_{\mu\nu}-hW_{\mu\nu}/4$, to thus find that the contribution  of the trace is given by  $\delta W_{\mu\nu}(h)= -hW_{\mu\nu}/4$. Thus, as noted in \cite{Mannheim2012a}, if the background $W^{\mu\nu}$ is zero (background $C^{\mu\lambda\nu\kappa}$ either zero or more generally obeying $2\nabla_{\kappa}\nabla_{\lambda}C^{\mu\lambda\nu\kappa}-R_{\kappa\lambda}C^{\mu\lambda\nu\kappa}=0$) the trace  of the fluctuation decouples completely from the fluctuation in $W^{\mu\nu}$. As well as being a very convenient property of conformal gravity fluctuations, obtaining the general relation $\delta W_{\mu\nu}(h)= -hW_{\mu\nu}/4$ for any background provides a nice internal check on our calculations, just as is made manifest  in (\ref{AP50}) below.

To take advantage of the nature of the dependence of $\delta W_{\mu\nu}$ on $h$ we introduce a quantity $K_{\mu\nu}(x)$ defined as 
\begin{eqnarray}
K_{\mu\nu}(x)=h_{\mu\nu}(x)-\frac{1}{4}g^{(0)}_{\mu\nu}(x)g_{(0)}^{\alpha\beta}h_{\alpha\beta},
\label{AP13}
\end{eqnarray}
with $K_{\mu\nu}$ being traceless with respect to the background metric $g_{(0)}^{\mu\nu}$. If we now evaluate $\delta W_{\mu\nu}(h_{\mu\nu})=\delta W_{\mu\nu}(K_{\mu\nu}+hg^{(0)}_{\mu\nu}/4)$ for some general fluctuation $h_{\mu\nu}$ around some general $g_{(0)}^{\mu\nu}$ background we will obtain $\delta W_{\mu\nu}(h_{\mu\nu})=\delta W_{\mu\nu}(K_{\mu\nu})-hW_{\mu\nu}/4$, with the contribution of the trace indeed being isolated. Then if the background is conformal to flat, the dependence on $h$ will drop out identically and $\delta W_{\mu\nu}(h_{\mu\nu})$ will be given as $\delta W_{\mu\nu}(h_{\mu\nu})=\delta W_{\mu\nu}(K_{\mu\nu})$. Thus rather than be a function of the ten-component $h_{\mu\nu}$, $\delta W_{\mu\nu}$ must instead be a function of the nine-component $K_{\mu\nu}$ alone if the background is conformal to flat. Note that we are not asserting here that $h_{\mu\nu}$ has been made traceless by a conformal transformation (in fact it could not be since $g_{(0)}^{\mu\nu}h_{\mu\nu}$ is conformal invariant). Rather, we are asserting that for any background that is conformal to flat, the first-order fluctuation in $\delta W_{\mu\nu}$ can only depend on the traceless combination $K_{\mu\nu}=h_{\mu\nu}-g^{(0)}_{\mu\nu}h/4$ rather than on $h_{\mu\nu}$ itself, an extremely convenient simplification. In \cite{Mannheim2006} we had already found this to explicitly be the case for perturbations around flat spacetime, and in \cite{Mannheim2012a} had explicitly shown it to the case for fluctuations around a de Sitter background, a specific background that is conformal to flat.  

Because of the decoupling of the trace,  for conformal to flat backgrounds we can replace (\ref{AP10}) by 
\begin{eqnarray}
W_{\mu\nu}(g_{\mu\nu})= W^{(0)}_{\mu\nu}(g^{(0)}_{\mu\nu})+\delta W_{\mu\nu}(K_{\mu\nu}),\qquad
\bar{W}_{\mu\nu}(\bar{g}_{\mu\nu})=\bar{W}^{(0)}_{\mu\nu}(\bar{g}^{(0)}_{\mu\nu})+\delta\bar{W}_{\mu\nu}(\bar{K}_{\mu\nu}),
\label{AP14}
\end{eqnarray}
where
\begin{eqnarray}
\bar{g}^{(0)}_{\mu\nu}(x)=\Omega^2(x)g^{(0)}_{\mu\nu}(x),
\label{AP15}
\end{eqnarray}
\begin{eqnarray}
\bar{K}_{\mu\nu}(x)=\Omega^2(x)K_{\mu\nu}(x).
\label{AP16}
\end{eqnarray}
In the following then, to construct the fluctuations in a $\bar{g}^{(0)}_{\mu\nu}$ background from the fluctuations in a $g^{(0)}_{\mu\nu}$ background that is conformal to flat, we shall need to utilize (\ref{AP16}) rather than (\ref{AP12}). 

We emphasize that for fluctuations around a conformal to flat background, we are able to reduce the theory to a dependence on the traceless $K_{\mu\nu}$ without needing to make any reference to the fluctuation equations at all. Since one also has the freedom to make four general coordinate transformations, on using them one can reduce  the nine-component $K_{\mu\nu}$ to five independent components, again without needing to make any reference to the fluctuation equations. Any further reduction in the number of independent components of $K_{\mu\nu}$ could only be achieved through use of residual gauge invariances or the structure of the fluctuation equations themselves. The key step in this paper will be in finding the right gauge to make the reduction from nine components to five, with a view to finding fluctuation equations in which there is no mixing of any of the components of $K_{\mu\nu}$ with each other.

\subsection{Implications of Coordinate Invariance}

In general in order to impose a coordinate gauge condition, we recall that since $h^{\mu\nu}$ and $h_{\mu\nu}$ transform into $h^{\mu\nu}-\nabla^{\nu}\epsilon^{\mu}-\nabla^{\mu}\epsilon^{\nu}$ and $h_{\mu\nu}-\nabla_{\nu}\epsilon_{\mu}-\nabla_{\mu}\epsilon_{\nu}$ under a perturbative coordinate gauge transformation of the form $x^{\mu}\rightarrow x^{\mu}+\epsilon^{\mu}(x)$ (all covariant derivatives being taken with respect to the background $g_{(0)}^{\mu\nu}$), we see that under the same transformation $K^{\mu\nu}$ transforms as  
\begin{eqnarray}
K^{\mu\nu}\rightarrow K^{\mu\nu}-\nabla^{\nu}\epsilon^{\mu}-\nabla^{\mu}\epsilon^{\nu}+\frac{1}{2}g_{(0)}^{\mu\nu}\nabla_{\alpha}\epsilon^{\alpha}.
\label{AP17}
\end{eqnarray}
With the covariant derivative of the fluctuation being given as  
\begin{eqnarray}
\nabla_{\nu}K^{\mu\nu}=
\partial_{\nu}K^{\mu\nu}
+K^{\nu\sigma}g_{(0)}^{\mu\rho}\partial_{\nu}g^{(0)}_{\rho\sigma}
-\frac{1}{2}K^{\nu\sigma}g_{(0)}^{\mu\rho}\partial_{\rho}g^{(0)}_{\nu\sigma}
+\frac{1}{2}K^{\mu\sigma}g_{(0)}^{\nu\rho}\partial_{\sigma}g^{(0)}_{\rho\nu},
\label{AP18}
\end{eqnarray}
and recalling that $K^{\nu\sigma}g^{(0)}_{\nu\sigma}=0$, we find that under a conformal transformation $\nabla_{\nu}K^{\mu\nu}$ transforms as
\begin{eqnarray}
\nabla_{\nu}K^{\mu\nu}\rightarrow \Omega^{-2}\nabla_{\nu}K^{\mu\nu}+4\Omega^{-3}K^{\mu\sigma}\partial_{\sigma}\Omega,
\label{AP19}
\end{eqnarray}
with a transverse gauge condition $\nabla_{\nu}K^{\mu\nu}=0$ not being conformal invariant. To identify a coordinate gauge condition that is conformal invariant, we note that under a conformal transformation the quantity $K^{\mu\nu}g_{(0)}^{\alpha\beta}\partial_{\nu}g^{(0)}_{\alpha\beta}$ transforms as
\begin{eqnarray}
K^{\mu\nu}g_{(0)}^{\alpha\beta}\partial_{\nu}g^{(0)}_{\alpha\beta}
\rightarrow \Omega^{-2}K^{\mu\nu}g_{(0)}^{\alpha\beta}\partial_{\nu}g^{(0)}_{\alpha\beta}
+8\Omega^{-3}K^{\mu\nu}\partial_{\nu}\Omega.
\label{AP20}
\end{eqnarray}
Consequently, we obtain
\begin{eqnarray}
\nabla_{\nu}K^{\mu\nu}-\frac{1}{2} K^{\mu\nu}g_{(0)}^{\alpha\beta}\partial_{\nu}g^{(0)}_{\alpha\beta}\rightarrow \Omega^{-2}\left[\nabla_{\nu}K^{\mu\nu}
-\frac{1}{2} K^{\mu\nu}g_{(0)}^{\alpha\beta}\partial_{\nu}g^{(0)}_{\alpha\beta}\right]
=\overline{\nabla_{\nu}K^{\mu\nu}}-\frac{1}{2} \bar{K}^{\mu\nu}\bar{g}_{(0)}^{\alpha\beta}\partial_{\nu}\bar{g}^{(0)}_{\alpha\beta},
\label{AP21}
\end{eqnarray}
where $\overline{\nabla_{\nu}K^{\mu\nu}}$ is evaluated in a geometry with metric $\bar{g}^{(0)}_{\mu\nu}$ according to 
\begin{eqnarray}
\overline{\nabla_{\nu}K^{\mu\nu}}=
\partial_{\nu}\bar{K}^{\mu\nu}
+\bar{K}^{\nu\sigma}\bar{g}_{(0)}^{\mu\rho}\partial_{\nu}\bar{g}^{(0)}_{\rho\sigma}
-\frac{1}{2}\bar{K}^{\nu\sigma}\bar{g}_{(0)}^{\mu\rho}\partial_{\rho}\bar{g}^{(0)}_{\nu\sigma}
+\frac{1}{2}\bar{K}^{\mu\sigma}\bar{g}_{(0)}^{\nu\rho}\partial_{\sigma}\bar{g}^{(0)}_{\rho\nu}.
\label{AP22}
\end{eqnarray}
The quantity $\nabla_{\nu}K^{\mu\nu}- K^{\mu\nu}g_{(0)}^{\alpha\beta}\partial_{\nu}g^{(0)}_{\alpha\beta}/2$ thus transforms into itself under a conformal transformation, and we shall refer to the condition
\begin{eqnarray}
&&\nabla_{\nu}K^{\mu\nu}=\frac{1}{2}K^{\mu\nu}g_{(0)}^{\alpha\beta}\partial_{\nu}g^{(0)}_{\alpha\beta},
\nonumber\\
&&\partial_{\nu}K^{\mu\nu}+\Gamma^{\mu(0)}_{\nu\sigma}K^{\sigma\nu}
+\Gamma^{\nu(0)}_{\nu\sigma}K^{\mu\sigma}=K^{\mu\nu}\Gamma^{\alpha(0)}_{\alpha\nu},
\nonumber\\
&&\partial_{\nu}K^{\mu\nu}+\Gamma^{\mu(0)}_{\nu\sigma}K^{\sigma\nu}=0,
\label{AP23}
\end{eqnarray}
as the conformal gauge. (In (\ref{AP23}) we have written the gauge condition in three equivalent forms, forms which will be convenient for use in the following.) 

While (\ref{AP23}) is left invariant under a local conformal transformation, we note that when the background is flat Minkowski ($g^{(0)}_{\alpha\beta}=\eta_{\alpha\beta}$), (\ref{AP23}) reduces to the transverse condition $\partial_{\nu}K^{\mu\nu}=0$. We are thus able to construct fluctuations around a conformal to flat background in the conformal gauge by conformally transforming fluctuations around a flat background in the transverse gauge, a remarkably convenient and straightforward procedure.

\subsection{Fluctuations Around Flat in the Transverse Gauge}

For fluctuations around a flat background that is in flat Minkowski coordinates it was found, without the  imposition of any gauge condition, that $\delta W^{\mu\nu}$ takes the form \cite{Mannheim2006}
\begin{eqnarray}
\delta W_{\mu\nu}=\frac{1}{2}(\eta^{\rho}_{\phantom{\rho} \mu} \partial^{\alpha}\partial_{\alpha}-\partial^{\rho}\partial_{\mu})
(\eta^{\sigma}_{\phantom{\sigma} \nu} \partial^{\beta}\partial_{\beta}-
\partial^{\sigma}\partial_{\nu})K_{\rho \sigma}- 
\frac{1}{6}(\eta_{\mu \nu} \partial^{\gamma}\partial_{\gamma}-
\partial_{\mu}\partial_{\nu})(\eta^{\rho \sigma} \partial^{\delta}\partial_{\delta}-
\partial^{\rho}\partial^{\sigma})K_{\rho\sigma}.
\label{AP24}
\end{eqnarray}
Now in a flat background we obtain $\partial_{\nu}K^{\mu\nu}\rightarrow \partial_{\nu}K^{\mu\nu}-\partial_{\nu}\partial^{\nu}\epsilon^{\mu}-\partial^{\mu}\partial_{\nu}\epsilon^{\nu}/2$ and $\partial_{\mu}\partial_{\nu}K^{\mu\nu}\rightarrow \partial_{\mu}\partial_{\nu}K^{\mu\nu}-3\partial_{\mu}\partial^{\mu}\partial_{\nu}\epsilon^{\nu}/2$ under a coordinate transformation. In a flat background we can thus solve for $\partial_{\nu}\epsilon^{\nu}$ and then for $\epsilon^{\mu}$ in order to bring $\partial_{\nu}K^{\mu\nu}$ to any assigned value. Then, if we now impose the transverse gauge condition $\partial_{\mu}K^{\mu\nu}=0$ (i.e. we impose the conformal gauge condition given in (\ref{AP23}) but with  $g^{(0)}_{\alpha\beta}=\eta_{\alpha\beta}$) (\ref{AP24}) reduces to the remarkably simple form
\begin{eqnarray}
\delta W_{\mu\nu}=\frac{1}{2}\eta^{\sigma\rho}\eta^{\alpha\beta}\partial_{\sigma}\partial_{\rho} \partial_{\alpha}\partial_{\beta}K_{\mu \nu},
\label{AP25}
\end{eqnarray}
with all the components of $K_{\mu\nu}$ that were coupled in (\ref{AP24}) having decoupled completely in (\ref{AP25}). (In fluctuations around conformal to flat in the Einstein gravity case there would not appear to be any gauge in which an analogous such complete decoupling occurs for the fluctuation $\delta(R_{\mu\nu}-g_{\mu\nu}R^{\alpha}_{\phantom{\alpha}\alpha}/2)$ in the Einstein tensor, though in Appendix \ref{SD} we shall present some that are relevant to cosmology in which such a complete decoupling is only prevented by the presence of the fluctuation trace $h$, while also presenting one in which the equation for $h$ is nothing other than a readily integrable standard flat space free massless particle wave equation.) In terms of the fourth-order derivative Green's function that obeys
\begin{eqnarray}
\partial_{\alpha}\partial^{\alpha} \partial_{\beta}\partial^{\beta}D^{(FO)}(x-x^{\prime})=\delta^4(x-x^{\prime}),
\label{AP26}
\end{eqnarray}
($FO$ denotes fourth order) the solution to $\delta W_{\mu\nu}(K_{\mu\nu})=\delta T_{\mu\nu}/4\alpha_g$ is thus given by 
\begin{eqnarray}
K_{\mu\nu}(x)=\frac{1}{2\alpha_g}\int d^4x^{\prime}D^{(FO)}(x-x^{\prime})\delta T_{\mu\nu}(x^{\prime})
\label{AP27}
\end{eqnarray}
in the conformal gauge.

The retarded Green's function  solution to (\ref{AP26}) is given in \cite{Mannheim2007}, and is of the form
\begin{eqnarray}
D^{(FO)}(x-x^{\prime})=\frac{1}{8\pi}\theta (t-t^{\prime}-|{\bf x}-{\bf x}^{\prime}|),
\label{AP28}
\end{eqnarray}
and as required of a retarded Green's function, $\theta (t-t^{\prime}-|{\bf x}-{\bf x}^{\prime}|)$ does not take support outside the light cone. In addition, momentum-eigenstate solutions to the wave equation $\partial_{\alpha}\partial^{\alpha} \partial_{\beta}\partial^{\beta}K_{\mu \nu}=0$ are given by \cite{Riegert1984a,Mannheim2012b}
\begin{eqnarray}
K_{\mu\nu}=A_{\mu\nu}e^{ik\cdot x}+(n\cdot x)B_{\mu\nu}e^{ik\cdot x}+A^*_{\mu\nu}e^{-ik\cdot x}+(n\cdot x)B^*_{\mu\nu}e^{-ik\cdot x},
\label{AP29}
\end{eqnarray}
where $k_0^2={\bf k}^2$, where $A_{\mu\nu}$ and $B_{\mu\nu}$ are polarization tensors, and where $n^{\mu}=(1,0,0,0)$ is a unit timelike vector. For a given assigned $\delta T_{\mu\nu}$ (\ref{AP27}) can be solved completely, and for a localized $\delta T_{\mu\nu}$ the asymptotic solution for $K_{\mu\nu}$ is given by (\ref{AP29}). Then with the $n\cdot x=t$ term, fluctuations around a flat background grow linearly in time.

\subsection{Fluctuations Around Conformal to Flat}

Since  the transverse gauge condition $\nabla_{\nu}K^{\mu\nu}=0$ and the conformal gauge condition $\nabla_{\nu} K^{\mu\nu}- K^{\mu\nu}g_{(0)}^{\alpha\beta}\partial_{\nu}g^{(0)}_{\alpha\beta}/2=0$ coincide for a flat Minkowski background, it thus follows that around any background  geometry that is conformal to a flat Minkowski $\eta^{\mu\nu}$ metric  (cf. (\ref{AP6})), the fluctuating $\delta\bar{W}_{\mu\nu}(\bar{K}_{\mu\nu})$ given in (\ref{AP14}) must take the form 
\begin{eqnarray}
\delta \bar{W}_{\mu\nu}=\frac{1}{2}\Omega^{-2}(x)\eta^{\sigma\rho}\eta^{\alpha\beta}\partial_{\sigma}\partial_{\rho} \partial_{\alpha}\partial_{\beta}[\Omega^{-2}(x)\bar{K}_{\mu \nu}],
\label{AP30}
\end{eqnarray}
in the gauge
\begin{eqnarray}
\nabla_{\nu}\bar{K}^{\mu\nu}- \frac{1}{2}\bar{K}^{\mu\nu}\bar{g}_{(0)}^{\alpha\beta}\partial_{\nu}\bar{g}^{(0)}_{\alpha\beta}=0.
\label{AP31}
\end{eqnarray}

Thus by working in the conformal gauge,  we are able to completely decouple the components of $\bar{K}^{\mu\nu}$ in the fluctuation equations. And not only that, the only derivative that appears in (\ref{AP30}) is the ordinary flat space derivative $\eta^{\sigma\rho}\eta^{\alpha\beta}\partial_{\sigma}\partial_{\rho} \partial_{\alpha}\partial_{\beta}$ and not some covariant generalization of it. Eq. (\ref{AP30}) thus represents a remarkable simplification of the fluctuation dynamics. To confirm that this is in fact the case, below we will actually calculate $\delta \bar{W}_{\mu\nu}$ directly in any background metric that is conformal to flat (viz. (\ref{AP6}) with arbitrary $\Omega(x)$),  and show that it reduces to (\ref{AP30}) when (\ref{AP31}) is imposed. Moreover, in order to be as general as possible, we shall also determine $\delta \bar{W}_{\mu\nu}$ in an arbitrary background, one that is not required to be conformal to flat at all.

\subsection{The Nature of Fluctuations in the Energy-Momentum Tensor}

The discussion of the matter field $T^{\mu\nu}$ and fluctuations in it in conformal gravity is quite different than in standard gravity. While any $T^{\mu\nu}$ must be conformal invariant in the conformal gravity theory, given that $4\alpha_gW^{\mu\nu}=T^{\mu\nu}$ in (\ref{AP3}), the  background $T^{\mu\nu}$ that is associated with a cosmological background must be zero identically since $W^{\mu\nu}$ vanishes in any geometry that is conformal to flat. However, while the background $T^{\mu\nu}$ must vanish, that does not mean that it has to vanish trivially. In the literature two ways in which it could vanish non-trivially have been identified, one involving a conformally coupled scalar field \cite{Mannheim1990}, and the other involving a conformal perfect fluid \cite{Mannheim2000}.

For a conformally coupled scalar field $S(x)$ the matter action is
\begin{eqnarray}
I_S&=&-\int d^4x(-g)^{1/2}\left[\frac{1}{2}\nabla_{\mu}S
\nabla^{\mu}S-\frac{1}{12}S^2R^\mu_{\phantom         
{\mu}\mu}+\lambda_S S^4\right]
\nonumber\\
&=&\int d^4x(-g)^{1/2}\left[\frac{1}{2}\dot{S}^2-\frac{1}{2}(\vec{\nabla} S)^2
+\frac{1}{12}S^2R^\mu_{\phantom         
{\mu}\mu}-\lambda_S S^4\right],
\label{AP32}
\end{eqnarray}                                 
where  $\lambda_S$ is a dimensionless coupling constant. (Since we use the convention given in  \cite{Weinberg1972} where $g_{00}$ is taken to have negative signature, and where the proper time is written as $ds^2=-g_{\mu\nu}dx^{\mu}dx^{\nu}$, (\ref{AP32}) thus corresponds to a scalar field with a normal positive signatured kinetic energy.) As such, the $I_{\rm S}$ action is the most general curved space matter action for the $S(x)$ field that is invariant under both general coordinate transformations and local conformal transformations of the form
$S(x)\rightarrow e^{-\alpha(x)}S(x)$,  $g_{\mu\nu}(x)\rightarrow e^{2\alpha(x)}g_{\mu\nu}(x)$. Variation of the $I_S$ action with respect to  $S(x)$ yields the scalar field equation of motion
\begin{eqnarray}
\nabla_{\mu}\nabla^{\mu}S+\frac{1}{6}SR^\mu_{\phantom{\mu}\mu}
-4\lambda_S S^3=0,
\label{AP33}
\end{eqnarray}                                 
while variation with respect to the metric yields a matter field  energy-momentum tensor 
\begin{eqnarray}
T_{\rm S}^{\mu \nu}&=&\frac{2}{3}\nabla^{\mu} \nabla^{\nu} S
-\frac{1}{6}g^{\mu\nu}\nabla_{\alpha}S\nabla^{\alpha}S
-\frac{1}{3}S\nabla^{\mu}\nabla^{\nu}S
\nonumber \\             
&+&\frac{1}{3}g^{\mu\nu}S\nabla_{\alpha}\nabla^{\alpha}S                           
-\frac{1}{6}S^2\left(R^{\mu\nu}
-\frac{1}{2}g^{\mu\nu}R^\alpha_{\phantom{\alpha}\alpha}\right)-g^{\mu\nu}\lambda_S S^4. 
\label{AP34}
\end{eqnarray}                                 
Use of the matter field equation of motion then confirms that this energy-momentum tensor obeys the tracelessness condition $g_{\mu\nu}T_{\rm S}^{\mu\nu}=0$, just as it should do in  a conformal invariant theory.

In the presence of a spontaneously broken non-zero constant expectation
value $S_0$ for the scalar field, the scalar field wave equation and the energy-momentum tensor are then found to simplify to  
\begin{eqnarray}
R^\alpha_{\phantom{\alpha}\alpha}&=&24\lambda_S S_0^2,
\nonumber\\
T_{\rm S}^{\mu \nu}&=& 
-\frac{1}{6} S_0^2\left(R^{\mu\nu}-\frac{1}{2}g^{\mu\nu}
R^\alpha_{\phantom{\alpha}\alpha}\right)-g^{\mu\nu}\lambda_S S_0^4=-\frac{1}{6} S_0^2\left(R^{\mu\nu}-\frac{1}{4}g^{\mu\nu}
R^\alpha_{\phantom{\alpha}\alpha}\right).
\label{AP35}
\end{eqnarray}                                 
Since $W^{\mu \nu}$ will vanish identically in a de Sitter geometry in which 
$R^{\lambda\mu\sigma\nu}=K[g^{\mu \sigma}g^{\lambda \nu}-g^{\mu \nu}g^{\lambda \sigma}]$, $R^{\mu\nu}=-3Kg^{\mu\nu}$, $R^\alpha_{\phantom{\alpha}\alpha}=-12K$, $R^{\mu\nu}=(1/4)g^{\mu\nu}
R^\alpha_{\phantom{\alpha}\alpha}$, $T_{\rm S}^{\mu \nu}$ will also  vanish identically in the same geometry, with $K$ being given by $K=-2\lambda_S S_0^2$. Thus even though $W^{\mu\nu}$ and $T^{\mu\nu}$ both vanish identically, as noted in \cite{Mannheim1990}, the conformal cosmology governed by $4\alpha_gW^{\mu\nu}=T^{\mu\nu}$ admits of a non-trivial de Sitter geometry solution, with a non-vanishing four-curvature $K=-2\lambda_S S_0^2$.

A second way in which $T^{\mu\nu}$ can vanish non-trivially was given in \cite{Mannheim2000}. If we drop the $\lambda_S$-dependent term in $I_S$, then in a generic Robertson-Walker geometry with metric
\begin{eqnarray}
ds^2=dt^2-a^2(t)\left[\frac{dr^2}{1-kr^2}+r^2d\theta^2+r^2\sin^2\theta d\phi^2\right]
=dt^2-a^2(t)\gamma_{ij}dx^idx^j,
\label{AP36}
\end{eqnarray}
solutions to the scalar field wave equation (\ref{AP33})  obey \cite{Mannheim2000}
\begin{eqnarray}
\frac{1}{f(p)}\left[\frac{d^2f}{dp^2}+kf(p)\right]=\frac{1}{g(r,\theta,\phi)}\gamma^{-1/2}\partial_i[\gamma^{1/2}\gamma^{ij}\partial_jg(r,\theta,\phi)]=-\lambda^2,
\label{AP37}
\end{eqnarray}
where $p=\int dt/a(t)$, $S=f(p)g(r,\theta,\phi)/a(t)$, $\gamma^{ij}$ is the metric of the spatial part of the Robertson-Walker metric, and $\lambda^2$ is a separation constant. From (\ref{AP37}) we see that $f(p)$ is harmonic with frequencies that obey $\omega^2=\lambda^2+k$, while we can set $g(r,\theta,\phi)=g^{\ell}_{\lambda}(r)Y^m_{\ell}(\theta,\phi)$, where $g^{\ell}_{\lambda}(r)$ obeys
\begin{eqnarray}
\left[(1-kr^2)\frac{\partial^2}{\partial r^2}+\frac{(2-3kr^2)}{r}\frac{\partial}{\partial r}-\frac{\ell(\ell+1)}{r^2}+\lambda^2\right]g^{\ell}_{\lambda}(r)=0.
\label{AP38}
\end{eqnarray}

To form a perfect fluid energy-momentum tensor, in $T^{\mu\nu}_S$ we make an incoherent averaging over all allowed spatial modes associated with a given $\omega$  (this is equivalent to calculating statistical averages using a density matrix that is proportional to the unit matrix and normalized to one). And on doing the sum over all modes,  for each $\omega$ we obtain  \cite{Mannheim2000} the automatically traceless
\begin{eqnarray}
T_S^{\mu\nu}=\frac{\omega^2(g^{\mu\nu}+4U^{\mu}U^{\nu})}{6\pi^2a^4(t)}=
\frac{(\lambda^2+k^2)(g^{\mu\nu}+4U^{\mu}U^{\nu})}{6\pi^2a^4(t)},
\label{AP39}
\end{eqnarray}
where $U^{\mu}$ is a unit timelike vector. This $T^{\mu\nu}_S$ vanishes if $\omega^2=0$, and with $\omega^2=\lambda^2+k$, we can thus satisfy $T^{\mu\nu}_S=0$ non-trivially  if and only if $k$ is negative. In doing the incoherent averaging when $\omega=0$, for $T^{00}_S$ we obtain
\begin{eqnarray}
T_S^{00}=\frac{1}{6}\sum_{\ell,m}\left[\sum _{i=1}^3\gamma^{ii}|\partial_i(g^{\ell}_{(-k)^{1/2}}Y^{m}_{\ell}(\theta,\phi))|^2+k|g^{\ell}_{(-k)^{1/2}}Y^{m}_{\ell}(\theta,\phi)|^2\right]
\label{AP40}
\end{eqnarray}
when $k$ is negative, with it being shown in \cite{Mannheim2000} that the sum in (\ref{AP40}) vanishes identically. Essentially what happens is that a positive contribution to $T^{\mu\nu}_S$ by the scalar field modes is cancelled by a negative contribution from the gravitational field due to its negative spatial 3-curvature. With negative $k$, solutions to (\ref{AP38}) are associated Legendre functions, and even though we have now fixed $\lambda^2$ to $-k$, (\ref{AP38}) still possesses an infinite number of solutions labelled by $\ell$ and $m$. An incoherent averaging over all of these solutions than causes $T^{\mu\nu}_S$ to vanish non-trivially.

In applications of conformal gravity to astrophysical and cosmological data it has been found that phenomenologically $k$ should be negative. In conformal cosmology very good non-fine-tuned, negative $k$  fits to the accelerating universe Hubble plot data have been presented in \cite{Mannheim2006,Mannheim2017},  with very good negative  $k$ conformal gravity fits to galactic rotation curves having been presented in \cite{Mannheim2006,Mannheim2017}.  Now standard gravity inflationary universe fits to the anisotropy of the cosmic microwave background lead to a spatially flat 3-geometry. However, with fluctuations growing at a different rate in the conformal case (as noted above, already around flat we have linear in time growth -- and as we show in  Appendix \ref{SB}  around an expressly negative $k$ Robertson-Walker background we have early universe $t^4$ fluctuation growth), the size of a standard ruler at recombination will be different than in the standard case. It is thus paramount to determine the conformal gravity expectations for the anisotropy to see if it could support $k<0$, and the objective of this paper is to prepare some of the needed groundwork.

Now despite the fact that the background $T^{\mu\nu}$ is zero, that does not mean that it will remain so if it is perturbed, and in fact it could not remain zero if the geometric side of (\ref{AP3}) is perturbed so that $W^{\mu\nu}$ would become non-zero. However, something unusual happens if we do perturb a non-trivially vanishing background $T^{\mu\nu}$, something that does not happen in the standard case. In the standard Einstein case where (\ref{AP5}) holds, with the background $T^{\mu\nu}$ being non-zero, neither the fluctuation in the background Einstein tensor or the fluctuation in the background $T^{\mu\nu}$ will separately be gauge invariant, only the perturbation of the entire $R^{\mu\nu} -g^{\mu\nu}R^{\alpha}_{\phantom{\alpha}\alpha}/2+8\pi GT^{\mu\nu}$ will be gauge invariant.  However, in the conformal case the change in $W^{\mu\nu}$ has the same functional form independent of whether the background $T^{\mu\nu}$ is identically zero or only non-trivially zero. But in the case in which the background $T^{\mu\nu}$ is identically zero (i.e. empty), there is no change in it, and thus the change in $W^{\mu\nu}$ must be gauge invariant all on its own. And then, if the background $T^{\mu\nu}$ is only non-trivially zero, the change in it must also be gauge invariant on it its own. Moreover, since the Weyl tensor is zero for any geometry that is conformal to flat, the Weyl tensor will vanish even if the conformal factor $\Omega(x)$ in (\ref{AP6}) is not associated with a maximally 3-symmetric background such as Robertson-Walker or a maximally 4-symmetric background such as de Sitter. And then since the background $T_{\mu\nu}$ would then have to vanish too (since $W_{\mu\nu}$ would vanish), it must be the case that for fluctuations around (\ref{AP6}) $\delta W_{\mu\nu}$ would then be gauge invariant on its own no matter how complicated a function  $\Omega(x)$ might be. Below we shall exhibit this explicitly by working in the gauge invariant scalar, vector, tensor (SVT) basis discussed in \cite{Lifshitz1946,Bardeen1980} (and also in e.g. \cite{Kodama1984,Bertschinger1996}). Specifically, we shall find that in any background that is conformal to flat (i.e. arbitrary $\Omega(x)$) $\delta W^{\mu\nu}$ can be expressed entirely in terms of the gauge invariant components of the SVT basis even though $\delta(R^{\mu\nu} -g^{\mu\nu}R^{\alpha}_{\phantom{\alpha}\alpha}/2)$ cannot be so expressed in  such an arbitrarily conformal to flat background.

\section{Fluctuation Equations Around an Arbitrary Background}
\label{S3}
\subsection{Setting up the Fluctuation Equations}

In order to perturb $W_{\mu\nu}$ we have found it convenient to use the identity
\begin{eqnarray}
\nabla_{\beta}\nabla_{\nu}T_{\lambda \mu}=\nabla_{\nu}\nabla_{\beta}T_{\lambda \mu}+R_{\lambda\sigma\nu\beta}T^{\sigma}_{\phantom{\sigma}\mu}-R_{\sigma\mu\nu\beta}T_{\lambda}^{\phantom{\lambda}\sigma}
\label{AP41}
\end{eqnarray}
obeyed by any rank two tensor, so that we can write $W^{\mu\nu}$ as 
\begin{eqnarray}
W_{\mu \nu}&=&
-\frac{1}{6}g_{\mu\nu}\nabla_{\beta}\nabla^{\beta}R^{\alpha}_{\phantom{\alpha}\alpha}
+\nabla_{\beta}\nabla^{\beta}R_{\mu\nu}                    
 -\frac{1}{3}\nabla_{\mu}\nabla_{\nu}R^{\alpha}_{\phantom{\alpha}\alpha}  
 -R^{\beta\sigma} R_{\sigma\mu\beta\nu}   
  \nonumber\\
 &-&R^{\beta\sigma} R_{\sigma\nu\beta\mu}  
+\frac{1}{2}g_{\mu\nu}R_{\alpha\beta}R^{\alpha\beta}                                            
+\frac{2}{3}R^{\alpha}_{\phantom{\alpha}\alpha}R_{\mu\nu}                              
-\frac{1}{6}g_{\mu\nu}(R^{\alpha}_{\phantom{\alpha}\alpha})^2.
\label{AP42}
\end{eqnarray}                                 
On taking the metric to be the completely general $g_{\mu\nu}+h_{\mu\nu}$, where here we take $g_{\mu\nu}$ to denote any general background metric (i.e. one not necessarily conformal to flat) and $\delta g_{\mu\nu}=h_{\mu\nu}$ to denote any general fluctuation, perturbing $W^{\mu\nu}$ then gives (following a machine calculation)
\begin{eqnarray}
&&\delta W_{\mu\nu}(h_{\mu\nu})=\tfrac{1}{2} h_{\mu \nu} R_{\alpha \beta} R^{\alpha \beta} -  g_{\mu \nu} h^{\alpha \beta} R_{\alpha}{}^{\gamma} R_{\beta \gamma} -  \tfrac{2}{3} h^{\alpha \beta} R_{\alpha \beta} R_{\mu \nu} + \tfrac{1}{3} g_{\mu \nu} h^{\alpha \beta} R_{\alpha \beta} R -  \tfrac{1}{6} h_{\mu \nu} R^2 
+ h^{\alpha \beta} R_{\alpha}{}^{\gamma} R_{\mu \beta \nu \gamma} 
\nonumber\\
&&+ h^{\alpha \beta} R_{\alpha}{}^{\gamma} R_{\mu \gamma \nu \beta} -  \tfrac{1}{6} h_{\mu \nu} \nabla_{\alpha}\nabla^{\alpha}R -  h^{\alpha \beta} \nabla_{\beta}\nabla_{\alpha}R_{\mu \nu} + \tfrac{1}{6} g_{\mu \nu} h^{\alpha \beta} \nabla_{\beta}\nabla_{\alpha}R + \tfrac{1}{6} g_{\mu \nu} h^{\alpha \beta} \nabla_{\gamma}\nabla^{\gamma}R_{\alpha \beta} 
\nonumber\\
&&+ \tfrac{1}{3} h^{\alpha \beta} \nabla_{\mu}\nabla_{\nu}R_{\alpha \beta}
+\tfrac{1}{3} R \nabla_{\alpha}\nabla^{\alpha}h_{\mu \nu} + R_{\mu \beta \nu \gamma} \nabla_{\alpha}\nabla^{\gamma}h^{\alpha \beta} + R_{\mu \gamma \nu \beta} \nabla_{\alpha}\nabla^{\gamma}h^{\alpha \beta} -  \tfrac{1}{3} R \nabla_{\alpha}\nabla_{\mu}h_{\nu}{}^{\alpha} -  \tfrac{1}{3} R \nabla_{\alpha}\nabla_{\nu}h_{\mu}{}^{\alpha} 
\nonumber\\
&&-  \tfrac{1}{6} \nabla_{\alpha}h_{\mu \nu} \nabla^{\alpha}R 
+ \tfrac{1}{6} g_{\mu \nu} \nabla^{\alpha}R \nabla_{\beta}h_{\alpha}{}^{\beta} -  \nabla_{\alpha}h^{\alpha \beta} \nabla_{\beta}R_{\mu \nu} -  \tfrac{2}{3} R_{\mu \nu} \nabla_{\beta}\nabla_{\alpha}h^{\alpha \beta} + \tfrac{1}{3} g_{\mu \nu} R \nabla_{\beta}\nabla_{\alpha}h^{\alpha \beta} + \tfrac{1}{2} R_{\nu}{}^{\alpha} \nabla_{\beta}\nabla_{\alpha}h_{\mu}{}^{\beta} 
\nonumber\\
&&-  R^{\alpha \beta} \nabla_{\beta}\nabla_{\alpha}h_{\mu \nu} 
+ \tfrac{1}{2} R_{\mu}{}^{\alpha} \nabla_{\beta}\nabla_{\alpha}h_{\nu}{}^{\beta} -  \tfrac{1}{2} R_{\nu}{}^{\alpha} \nabla_{\beta}\nabla^{\beta}h_{\mu \alpha} -  \tfrac{1}{2} R_{\mu}{}^{\alpha} \nabla_{\beta}\nabla^{\beta}h_{\nu \alpha} + \tfrac{1}{2} \nabla_{\beta}\nabla^{\beta}\nabla_{\alpha}\nabla^{\alpha}h_{\mu \nu} 
\nonumber\\
&&-  \tfrac{1}{2} \nabla_{\beta}\nabla^{\beta}\nabla_{\alpha}\nabla_{\mu}h_{\nu}{}^{\alpha} 
-  \tfrac{1}{2} \nabla_{\beta}\nabla^{\beta}\nabla_{\alpha}\nabla_{\nu}h_{\mu}{}^{\alpha} -  \tfrac{1}{2} R_{\nu}{}^{\alpha} \nabla_{\beta}\nabla_{\mu}h_{\alpha}{}^{\beta} + R^{\alpha \beta} \nabla_{\beta}\nabla_{\mu}h_{\nu \alpha} -  \tfrac{1}{2} R_{\mu}{}^{\alpha} \nabla_{\beta}\nabla_{\nu}h_{\alpha}{}^{\beta} 
\nonumber\\
&&+ R^{\alpha \beta} \nabla_{\beta}\nabla_{\nu}h_{\mu \alpha} 
+ \nabla_{\alpha}R_{\nu \beta} \nabla^{\beta}h_{\mu}{}^{\alpha} 
-  \nabla_{\beta}R_{\nu \alpha} \nabla^{\beta}h_{\mu}{}^{\alpha} + \nabla_{\alpha}R_{\mu \beta} \nabla^{\beta}h_{\nu}{}^{\alpha} -  \nabla_{\beta}R_{\mu \alpha} \nabla^{\beta}h_{\nu}{}^{\alpha} -  g_{\mu \nu} R^{\alpha \beta} \nabla_{\gamma}\nabla_{\beta}h_{\alpha}{}^{\gamma} 
\nonumber\\
&&+ \tfrac{2}{3} g_{\mu \nu} R^{\alpha \beta} \nabla_{\gamma}\nabla^{\gamma}h_{\alpha \beta} 
-  R_{\mu \alpha \nu \beta} \nabla_{\gamma}\nabla^{\gamma}h^{\alpha \beta} + \tfrac{1}{6} g_{\mu \nu} \nabla_{\gamma}\nabla^{\gamma}\nabla_{\beta}\nabla_{\alpha}h^{\alpha \beta} + \tfrac{1}{3} g_{\mu \nu} \nabla_{\gamma}R_{\alpha \beta} \nabla^{\gamma}h^{\alpha \beta} -  \nabla_{\beta}R_{\nu \alpha} \nabla_{\mu}h^{\alpha \beta} 
\nonumber\\
&&+ \tfrac{1}{6} \nabla^{\alpha}R \nabla_{\mu}h_{\nu \alpha} 
-  \tfrac{1}{6} R^{\alpha \beta} \nabla_{\mu}\nabla_{\nu}h_{\alpha \beta} -  \nabla_{\beta}R_{\mu \alpha} \nabla_{\nu}h^{\alpha \beta} + \tfrac{1}{3} \nabla_{\mu}R_{\alpha \beta} \nabla_{\nu}h^{\alpha \beta} + \tfrac{1}{6} \nabla^{\alpha}R \nabla_{\nu}h_{\mu \alpha} + \tfrac{1}{3} \nabla_{\mu}h^{\alpha \beta} \nabla_{\nu}R_{\alpha \beta} 
\nonumber\\
&&-  \tfrac{1}{2} R^{\alpha \beta} \nabla_{\nu}\nabla_{\mu}h_{\alpha \beta} 
+ \tfrac{1}{3} \nabla_{\nu}\nabla_{\mu}\nabla_{\beta}\nabla_{\alpha}h^{\alpha \beta}
	+\tfrac{2}{3} R_{\mu \nu} \nabla_{\alpha}\nabla^{\alpha}h -  \tfrac{1}{3} g_{\mu \nu} R \nabla_{\alpha}\nabla^{\alpha}h + \tfrac{1}{2} \nabla_{\alpha}\nabla^{\alpha}\nabla_{\nu}\nabla_{\mu}h 
\nonumber\\
&&-  \tfrac{1}{12} g_{\mu \nu} \nabla_{\alpha}h \nabla^{\alpha}R + \tfrac{1}{2} \nabla_{\alpha}R_{\mu \nu} \nabla^{\alpha}h 
	+ \tfrac{1}{2} g_{\mu \nu} R^{\alpha \beta} \nabla_{\beta}\nabla_{\alpha}h -  \tfrac{1}{6} g_{\mu \nu} \nabla_{\beta}\nabla^{\beta}\nabla_{\alpha}\nabla^{\alpha}h -  R_{\mu \alpha \nu \beta} \nabla^{\beta}\nabla^{\alpha}h + \tfrac{1}{3} R \nabla_{\nu}\nabla_{\mu}h 
\nonumber\\
&&-  \tfrac{1}{3} \nabla_{\nu}\nabla_{\mu}\nabla_{\alpha}\nabla^{\alpha}h.
\label{AP43}
\end{eqnarray}
In (\ref{AP43}) all covariant derivatives are evaluated with respect to the background $g_{\mu\nu}$, and $R$ denotes $R^{\alpha}_{\phantom{\alpha}\alpha}$.
Eq. (\ref{AP43}) contains 62 terms, of which 10 depend on the trace $h=g^{\mu\nu}h_{\mu\nu}$, 
On substituting $h_{\mu\nu}=K_{\mu\nu}+(1/4)g_{\mu\nu}h$ in (\ref{AP43}), $\delta W_{\mu\nu}(h_{\mu\nu})$ breaks into two pieces, a $K_{\mu\nu}$-dependent piece with 52 terms and an $h=g_{\mu\nu}h^{\mu\nu}$-dependent piece with 19 terms, and with $\delta W_{\mu\nu}(h_{\mu\nu})=\delta W_{\mu\nu}(K_{\mu\nu})+\delta W_{\mu\nu}(h)$ they are of the form 
\begin{eqnarray}
&&\delta W_{\mu\nu}(K_{\mu\nu})=\tfrac{1}{2} K_{\mu \nu} R_{\alpha \beta} R^{\alpha \beta} -  g_{\mu \nu} K^{\alpha \beta} R_{\alpha}{}^{\gamma} R_{\beta \gamma} -  \tfrac{2}{3} K^{\alpha \beta} R_{\alpha \beta} R_{\mu \nu} + \tfrac{1}{3} g_{\mu \nu} K^{\alpha \beta} R_{\alpha \beta} R -  \tfrac{1}{6} K_{\mu \nu} R^2 
+ K^{\alpha \beta} R_{\alpha}{}^{\gamma} R_{\mu \beta \nu \gamma} 
\nonumber\\
&&+ K^{\alpha \beta} R_{\alpha}{}^{\gamma} R_{\mu \gamma \nu \beta} -  \tfrac{1}{6} K_{\mu \nu} \nabla_{\alpha}\nabla^{\alpha}R -  K^{\alpha \beta} \nabla_{\beta}\nabla_{\alpha}R_{\mu \nu} + \tfrac{1}{6} g_{\mu \nu} K^{\alpha \beta} \nabla_{\beta}\nabla_{\alpha}R + \tfrac{1}{6} g_{\mu \nu} K^{\alpha \beta} \nabla_{\gamma}\nabla^{\gamma}R_{\alpha \beta} 
\nonumber\\
&&+ \tfrac{1}{3} K^{\alpha \beta} \nabla_{\mu}\nabla_{\nu}R_{\alpha \beta}
+\tfrac{1}{3} R \nabla_{\alpha}\nabla^{\alpha}K_{\mu \nu} + R_{\mu \beta \nu \gamma} \nabla_{\alpha}\nabla^{\gamma}K^{\alpha \beta} + R_{\mu \gamma \nu \beta} \nabla_{\alpha}\nabla^{\gamma}K^{\alpha \beta} -  \tfrac{1}{3} R \nabla_{\alpha}\nabla_{\mu}K_{\nu}{}^{\alpha} -  \tfrac{1}{3} R \nabla_{\alpha}\nabla_{\nu}K_{\mu}{}^{\alpha} 
\nonumber\\
&&-  \tfrac{1}{6} \nabla_{\alpha}K_{\mu \nu} \nabla^{\alpha}R 
+ \tfrac{1}{6} g_{\mu \nu} \nabla^{\alpha}R \nabla_{\beta}K_{\alpha}{}^{\beta} -  \nabla_{\alpha}K^{\alpha \beta} \nabla_{\beta}R_{\mu \nu} -  \tfrac{2}{3} R_{\mu \nu} \nabla_{\beta}\nabla_{\alpha}K^{\alpha \beta} + \tfrac{1}{3} g_{\mu \nu} R \nabla_{\beta}\nabla_{\alpha}K^{\alpha \beta} + \tfrac{1}{2} R_{\nu}{}^{\alpha} \nabla_{\beta}\nabla_{\alpha}K_{\mu}{}^{\beta} 
\nonumber\\
&&-  R^{\alpha \beta} \nabla_{\beta}\nabla_{\alpha}K_{\mu \nu} 
+ \tfrac{1}{2} R_{\mu}{}^{\alpha} \nabla_{\beta}\nabla_{\alpha}K_{\nu}{}^{\beta} -  \tfrac{1}{2} R_{\nu}{}^{\alpha} \nabla_{\beta}\nabla^{\beta}K_{\mu \alpha} -  \tfrac{1}{2} R_{\mu}{}^{\alpha} \nabla_{\beta}\nabla^{\beta}K_{\nu \alpha} + \tfrac{1}{2} \nabla_{\beta}\nabla^{\beta}\nabla_{\alpha}\nabla^{\alpha}K_{\mu \nu} 
\nonumber\\
&&-  \tfrac{1}{2} \nabla_{\beta}\nabla^{\beta}\nabla_{\alpha}\nabla_{\mu}K_{\nu}{}^{\alpha} 
-  \tfrac{1}{2} \nabla_{\beta}\nabla^{\beta}\nabla_{\alpha}\nabla_{\nu}K_{\mu}{}^{\alpha} -  \tfrac{1}{2} R_{\nu}{}^{\alpha} \nabla_{\beta}\nabla_{\mu}K_{\alpha}{}^{\beta} + R^{\alpha \beta} \nabla_{\beta}\nabla_{\mu}K_{\nu \alpha} -  \tfrac{1}{2} R_{\mu}{}^{\alpha} \nabla_{\beta}\nabla_{\nu}K_{\alpha}{}^{\beta} 
\nonumber\\
&&+ R^{\alpha \beta} \nabla_{\beta}\nabla_{\nu}K_{\mu \alpha} 
+ \nabla_{\alpha}R_{\nu \beta} \nabla^{\beta}K_{\mu}{}^{\alpha} 
-  \nabla_{\beta}R_{\nu \alpha} \nabla^{\beta}K_{\mu}{}^{\alpha} + \nabla_{\alpha}R_{\mu \beta} \nabla^{\beta}K_{\nu}{}^{\alpha} -  \nabla_{\beta}R_{\mu \alpha} \nabla^{\beta}K_{\nu}{}^{\alpha} -  g_{\mu \nu} R^{\alpha \beta} \nabla_{\gamma}\nabla_{\beta}K_{\alpha}{}^{\gamma} 
\nonumber\\
&&+ \tfrac{2}{3} g_{\mu \nu} R^{\alpha \beta} \nabla_{\gamma}\nabla^{\gamma}K_{\alpha \beta} 
-  R_{\mu \alpha \nu \beta} \nabla_{\gamma}\nabla^{\gamma}K^{\alpha \beta} + \tfrac{1}{6} g_{\mu \nu} \nabla_{\gamma}\nabla^{\gamma}\nabla_{\beta}\nabla_{\alpha}K^{\alpha \beta} + \tfrac{1}{3} g_{\mu \nu} \nabla_{\gamma}R_{\alpha \beta} \nabla^{\gamma}K^{\alpha \beta} -  \nabla_{\beta}R_{\nu \alpha} \nabla_{\mu}K^{\alpha \beta} 
\nonumber\\
&&+ \tfrac{1}{6} \nabla^{\alpha}R \nabla_{\mu}K_{\nu \alpha} 
-  \tfrac{1}{6} R^{\alpha \beta} \nabla_{\mu}\nabla_{\nu}K_{\alpha \beta} -  \nabla_{\beta}R_{\mu \alpha} \nabla_{\nu}K^{\alpha \beta} + \tfrac{1}{3} \nabla_{\mu}R_{\alpha \beta} \nabla_{\nu}K^{\alpha \beta} + \tfrac{1}{6} \nabla^{\alpha}R \nabla_{\nu}K_{\mu \alpha} + \tfrac{1}{3} \nabla_{\mu}K^{\alpha \beta} \nabla_{\nu}R_{\alpha \beta} 
\nonumber\\
&&-  \tfrac{1}{2} R^{\alpha \beta} \nabla_{\nu}\nabla_{\mu}K_{\alpha \beta}+ \tfrac{1}{3} \nabla_{\nu}\nabla_{\mu}\nabla_{\beta}\nabla_{\alpha}K^{\alpha \beta},
\label{AP44}
\end{eqnarray}
\begin{eqnarray}
&&\delta W_{\mu\nu}(h)=- \tfrac{1}{8} g_{\mu \nu} R_{\alpha \beta} R^{\alpha \beta} h -  \tfrac{1}{6} R_{\mu \nu} R h + \tfrac{1}{24} g_{\mu \nu} R^2 h + \tfrac{1}{2} R^{\alpha \beta} R_{\mu \alpha \nu \beta} h -  \tfrac{1}{4} h \nabla_{\alpha}\nabla^{\alpha}R_{\mu \nu} + \tfrac{1}{24} g_{\mu \nu} h \nabla_{\alpha}\nabla^{\alpha}R 
\nonumber\\
&&+ \tfrac{1}{12} h \nabla_{\nu}\nabla_{\mu}R
+\tfrac{1}{4} \nabla_{\alpha}\nabla^{\alpha}\nabla_{\nu}\nabla_{\mu}h -  \tfrac{1}{4} \nabla_{\alpha}R_{\mu \nu} \nabla^{\alpha}h -  \tfrac{1}{2} R_{\mu \alpha \nu \beta} \nabla^{\beta}\nabla^{\alpha}h + \tfrac{1}{4} \nabla_{\mu}R_{\nu \alpha}\nabla^{\alpha}h  -  \tfrac{1}{4} \nabla_{\alpha}R_{\nu}{}^{\alpha} \nabla_{\mu}h + \tfrac{1}{4} R_{\nu}{}^{\alpha} \nabla_{\mu}\nabla_{\alpha}h
\nonumber\\
&& + \tfrac{1}{4} \nabla_{\nu}R_{\mu \alpha} \nabla^{\alpha}h + \tfrac{1}{8} \nabla_{\nu}R\nabla_{\mu}h  -  \tfrac{1}{4} \nabla_{\alpha}R_{\mu}{}^{\alpha} \nabla_{\nu}h + \tfrac{1}{8} \nabla_{\mu}R \nabla_{\nu}h + \tfrac{1}{4} R_{\mu}{}^{\alpha} \nabla_{\nu}\nabla_{\alpha}h  
-  \tfrac{1}{4} \nabla_{\nu}\nabla_{\mu}\nabla_{\alpha}\nabla^{\alpha}h.
\label{AP45}
\end{eqnarray}

\subsection{Decoupling of the Trace of the Fluctuation}

Given the identity 
\begin{eqnarray}
\nabla_{\kappa}\nabla_{\nu}V_{\lambda}-\nabla_{\nu}\nabla_{\kappa}V_{\lambda}=V^{\sigma}R_{\lambda\sigma\nu\kappa}
\label{AP46}
\end{eqnarray}
that is obeyed by any vector, on setting $V_{\lambda}=\nabla_{\lambda}h$ in (\ref{AP46})  and $T_{\lambda\mu}=\nabla_{\lambda}\nabla_{\mu}h$ in (\ref{AP41}) we obtain
\begin{eqnarray}
&&\nabla_{\nu}\nabla_{\mu}\nabla_{\alpha}\nabla^{\alpha}h
=g^{\alpha\beta}\nabla_{\nu}[\nabla_{\alpha}\nabla_{\mu}\nabla_{\beta}h
+R_{\beta\sigma\alpha\mu}\nabla^{\sigma}h]
=g^{\alpha\beta}\nabla_{\nu}[\nabla_{\alpha}\nabla_{\beta}\nabla_{\mu}h
+R_{\beta\sigma\alpha\mu}\nabla^{\sigma}h]
\nonumber\\
&&=g^{\alpha\beta}[\nabla_{\alpha}\nabla_{\nu}\nabla_{\beta}\nabla_{\mu}h
+R_{\beta\sigma\alpha\nu}\nabla^{\sigma}\nabla_{\mu}h
-R_{\sigma\mu\alpha\nu}\nabla_{\beta}\nabla^{\sigma}h
+R_{\beta\sigma\alpha\mu}\nabla_{\nu}\nabla^{\sigma}h
+\nabla_{\nu}R_{\beta\sigma\alpha\mu}\nabla^{\sigma}h]
\nonumber\\
&&=g^{\alpha\beta}[\nabla_{\alpha}[\nabla_{\beta}\nabla_{\nu}\nabla_{\mu}h
+R_{\mu\sigma\beta\nu}\nabla^{\sigma}h]
+R_{\beta\sigma\alpha\nu}\nabla^{\sigma}\nabla_{\mu}h
-R_{\sigma\mu\alpha\nu}\nabla_{\beta}\nabla^{\sigma}h
+R_{\beta\sigma\alpha\mu}\nabla_{\nu}\nabla^{\sigma}h
+\nabla_{\nu}R_{\beta\sigma\alpha\mu}\nabla^{\sigma}h].
\label{AP47}
\end{eqnarray}
On recalling that 
\begin{eqnarray}
 \nabla^{\nu}R_{\nu\mu\kappa\eta}=\nabla_{\kappa}R_{\mu\eta}-\nabla_{\eta}R_{\mu\kappa},
\label{AP48}
\end{eqnarray}
we obtain
\begin{eqnarray}
&&\nabla_{\nu}\nabla_{\mu}\nabla_{\alpha}\nabla^{\alpha}h-\nabla_{\alpha}\nabla^{\alpha}\nabla_{\nu}\nabla_{\mu}h
=R_{\mu\sigma\alpha\nu}\nabla^{\alpha}\nabla^{\sigma}h
+\nabla_{\mu}R_{\nu\sigma}\nabla^{\sigma}h
-\nabla_{\sigma}R_{\nu\mu}
\nabla^{\sigma}h\nonumber\\
&&+R_{\sigma\nu}\nabla^{\sigma}\nabla_{\mu}h
-R_{\sigma\mu\alpha\nu}\nabla^{\alpha}\nabla^{\sigma}h
+R_{\sigma\mu}\nabla_{\nu}\nabla^{\sigma}h
+\nabla_{\nu}R_{\sigma\mu}\nabla^{\sigma}h.
\label{AP49}
\end{eqnarray}
Then with $\nabla^{\alpha}R_{\mu\alpha}=(1/2)\nabla _{\mu}R$ we find that the 12 terms in (\ref{AP45}) that involve a gradient of $h$ all cancel identically. Finally, comparing the remaining 7 terms in (\ref{AP45}) with a background $W_{\mu\nu}$ that is of the form given in (\ref{AP42}), we find that $\delta W_{\mu\nu}(h)$ reduces to the remarkably simple
\begin{eqnarray}
\delta W_{\mu\nu}(h)=-\frac{1}{4}W_{\mu\nu}h.
\label{AP50}
\end{eqnarray}
Now we had noted above that  the condition $\delta W_{\mu\nu}(h)=-\frac{1}{4}W_{\mu\nu}h$ is required on general grounds. We thus recover this condition, and not only do we see that (\ref{AP50}) is generic, its recovery provides a nice internal  check on our calculations. 

To confirm this result it is instructive to also look at the fluctuation in the Weyl tensor itself. About an arbitrary background it is found to evaluate to $\delta C_{\lambda\mu\nu\kappa}=\delta C_{\lambda\mu\nu\kappa}(K_{\mu\nu})+\delta C_{\lambda\mu\nu\kappa}(h)$, where
\begin{eqnarray}
&&\delta C_{\lambda\mu\nu\kappa}(K_{\mu\nu})=- \tfrac{1}{6} g_{\kappa \mu} g_{\lambda \nu} K^{\alpha \beta} R_{\alpha \beta} + \tfrac{1}{6} g_{\kappa \lambda} g_{\mu \nu} K^{\alpha \beta} R_{\alpha \beta} + \tfrac{1}{2} K_{\mu \nu} R_{\kappa \lambda} -  \tfrac{1}{2} K_{\lambda \nu} R_{\kappa \mu} -  \tfrac{1}{2} K_{\kappa \mu} R_{\lambda \nu} 
+ \tfrac{1}{2} K_{\kappa \lambda} R_{\mu \nu} 
\nonumber\\
&&-  \tfrac{1}{6} g_{\mu \nu} K_{\kappa \lambda} R 
+ \tfrac{1}{6} g_{\lambda \nu} K_{\kappa \mu} R + \tfrac{1}{6} g_{\kappa \mu} K_{\lambda \nu} R -  \tfrac{1}{6} g_{\kappa \lambda} K_{\mu \nu} R + K_{\lambda}{}^{\alpha} R_{\kappa \nu \mu \alpha} + \tfrac{1}{4} g_{\mu \nu} \nabla_{\alpha}\nabla^{\alpha}K_{\kappa \lambda} -  \tfrac{1}{4} g_{\lambda \nu} \nabla_{\alpha}\nabla^{\alpha}K_{\kappa \mu} 
\nonumber\\
&&-  \tfrac{1}{4} g_{\kappa \mu} \nabla_{\alpha}\nabla^{\alpha}K_{\lambda \nu} 
+ \tfrac{1}{4} g_{\kappa \lambda} \nabla_{\alpha}\nabla^{\alpha}K_{\mu \nu} -  \tfrac{1}{4} g_{\mu \nu} \nabla_{\alpha}\nabla_{\kappa}K_{\lambda}{}^{\alpha} + \tfrac{1}{4} g_{\lambda \nu} \nabla_{\alpha}\nabla_{\kappa}K_{\mu}{}^{\alpha} -  \tfrac{1}{4} g_{\mu \nu} \nabla_{\alpha}\nabla_{\lambda}K_{\kappa}{}^{\alpha} + \tfrac{1}{4} g_{\kappa \mu} \nabla_{\alpha}\nabla_{\lambda}K_{\nu}{}^{\alpha} 
\nonumber\\
&&+ \tfrac{1}{4} g_{\lambda \nu} \nabla_{\alpha}\nabla_{\mu}K_{\kappa}{}^{\alpha} -  \tfrac{1}{4} g_{\kappa \lambda} \nabla_{\alpha}\nabla_{\mu}K_{\nu}{}^{\alpha} + \tfrac{1}{4} g_{\kappa \mu} \nabla_{\alpha}\nabla_{\nu}K_{\lambda}{}^{\alpha} -  \tfrac{1}{4} g_{\kappa \lambda} \nabla_{\alpha}\nabla_{\nu}K_{\mu}{}^{\alpha} 
-  \tfrac{1}{6} g_{\kappa \mu} g_{\lambda \nu} \nabla_{\beta}\nabla_{\alpha}K^{\alpha \beta} 
\nonumber\\
&&+ \tfrac{1}{6} g_{\kappa \lambda} g_{\mu \nu} \nabla_{\beta}\nabla_{\alpha}K^{\alpha \beta} 
-  \tfrac{1}{2} \nabla_{\kappa}\nabla_{\lambda}K_{\mu \nu} + \tfrac{1}{2} \nabla_{\kappa}\nabla_{\mu}K_{\lambda \nu} 
+ \tfrac{1}{2} \nabla_{\kappa}\nabla_{\nu}K_{\lambda \mu} - \tfrac{1}{2} \nabla_{\nu}\nabla_{\kappa}K_{\lambda \mu} + \tfrac{1}{2} \nabla_{\nu}\nabla_{\lambda}K_{\kappa \mu} 
-  \tfrac{1}{2} \nabla_{\nu}\nabla_{\mu}K_{\kappa \lambda},
\nonumber\\
\label{AP51}
\end{eqnarray}
\begin{eqnarray}
\delta C_{\lambda\mu\nu\kappa}(h)&=&\left[\tfrac{1}{8} g_{\mu \nu}  R_{\kappa \lambda} -  \tfrac{1}{8} g_{\lambda \nu} R_{\kappa \mu} -  \tfrac{1}{8} g_{\kappa \mu}  R_{\lambda \nu} +\tfrac{1}{8} g_{\kappa \lambda}  R_{\mu \nu} + \tfrac{1}{24} g_{\kappa \mu} g_{\lambda \nu}  R -  \tfrac{1}{24} g_{\kappa \lambda} g_{\mu \nu}  R -  \tfrac{1}{4}  R_{\kappa \nu \lambda \mu}\right]h
\nonumber\\
&=&\frac{1}{4}hC_{\lambda\mu\nu\kappa}.
\label{AP52}
\end{eqnarray}
Thus if the background Weyl tensor is zero, $\delta C_{\lambda\mu\nu\kappa}$ is independent of $h$. However, if the background Weyl tensor is zero, then according to (\ref{AP3}) $\delta W^{\mu\nu}$ is given by 
\begin{eqnarray}
\delta W^{\mu\nu}=2\nabla_{\kappa}\nabla_{\lambda}\delta C^{\mu\lambda\nu\kappa}-
R_{\kappa\lambda}\delta C^{\mu\lambda\nu\kappa},
\label{AP53}
\end{eqnarray}
to thus then also be independent of $h$. Thus when the background Weyl tensor is zero  (in which case the background $W_{\mu\nu}$ is zero too), we confirm that $\delta W_{\mu\nu}$ is  independent of $h$, just as required by (\ref{AP50}). With the dependence of $\delta W_{\mu\nu}$ on $K_{\mu\nu}$ being fully specified in (\ref{AP44}) (in any background), we can now impose the conformal gauge condition and evaluate the structure of $\delta W^{\mu\nu}$ in the conformal to flat background case.

\subsection{Preparing to Implement the Conformal Gauge Condition}

To bring (\ref{AP44}) to a form in which we can apply the conformal gauge condition given in (\ref{AP23}) we need to commute differential operators as per (\ref{AP41}) and (\ref{AP46}).  On doing the needed commutations for $\delta W_{\mu\nu}^{}(K_{\mu\nu})$ we obtain the 59 term
\begin{eqnarray}
&&\delta W_{\mu\nu}^{}(K_{\mu\nu})=\tfrac{1}{2} K_{\mu \nu} R_{\alpha \beta} R^{\alpha \beta} -  \tfrac{1}{2} K_{\nu}{}^{\alpha} R_{\alpha \beta} R_{\mu}{}^{\beta} -  \tfrac{2}{3} K^{\alpha \beta} R_{\alpha \beta} R_{\mu \nu} + K^{\alpha \beta} R_{\mu \alpha} R_{\nu \beta} -  \tfrac{1}{2} K_{\mu}{}^{\alpha} R_{\alpha \beta} R_{\nu}{}^{\beta} + \tfrac{1}{3} g_{\mu \nu} K^{\alpha \beta} R_{\alpha \beta} R 
\nonumber\\
&&+ \tfrac{1}{3} K_{\nu}{}^{\alpha} R_{\mu \alpha} R + \tfrac{1}{3} K_{\mu}{}^{\alpha} R_{\nu \alpha} R -  \tfrac{1}{6} K_{\mu \nu} R^2 -  g_{\mu \nu} K^{\alpha \beta} R^{\gamma \kappa} R_{\alpha \gamma \beta \kappa} -  \tfrac{2}{3} K^{\alpha \beta} R R_{\mu \alpha \nu \beta} -  K_{\nu}{}^{\alpha} R^{\beta \gamma} R_{\mu \beta \alpha \gamma} + 2 K^{\alpha \beta} R_{\alpha}{}^{\gamma} R_{\mu \gamma \nu \beta} 
\nonumber\\
&&+ 2 K^{\alpha \beta} R_{\alpha \gamma \beta \kappa} R_{\mu}{}^{\gamma}{}_{\nu}{}^{\kappa} -  K_{\mu}{}^{\alpha} R^{\beta \gamma} R_{\nu \beta \alpha \gamma} + \tfrac{1}{3} R \nabla_{\alpha}\nabla^{\alpha}K_{\mu \nu} -  \tfrac{1}{6} K_{\mu \nu} \nabla_{\alpha}\nabla^{\alpha}R + \tfrac{1}{2} R_{\nu}{}^{\alpha} \nabla_{\alpha}\nabla_{\beta}K_{\mu}{}^{\beta} + \tfrac{1}{2} R_{\mu}{}^{\alpha} \nabla_{\alpha}\nabla_{\beta}K_{\nu}{}^{\beta} 
\nonumber\\
&&-  \tfrac{1}{6} \nabla_{\alpha}K_{\mu \nu} \nabla^{\alpha}R + \tfrac{1}{6} g_{\mu \nu} \nabla^{\alpha}R \nabla_{\beta}K_{\alpha}{}^{\beta} -  \nabla_{\alpha}K^{\alpha \beta} \nabla_{\beta}R_{\mu \nu} -  \tfrac{2}{3} R_{\mu \nu} \nabla_{\beta}\nabla_{\alpha}K^{\alpha \beta} + \tfrac{1}{3} g_{\mu \nu} R \nabla_{\beta}\nabla_{\alpha}K^{\alpha \beta} -  R^{\alpha \beta} \nabla_{\beta}\nabla_{\alpha}K_{\mu \nu} 
\nonumber\\
&&-  K^{\alpha \beta} \nabla_{\beta}\nabla_{\alpha}R_{\mu \nu} + \tfrac{1}{6} g_{\mu \nu} K^{\alpha \beta} \nabla_{\beta}\nabla_{\alpha}R + \tfrac{1}{2} K_{\nu}{}^{\alpha} \nabla_{\beta}\nabla^{\beta}R_{\mu \alpha} + \tfrac{1}{2} K_{\mu}{}^{\alpha} \nabla_{\beta}\nabla^{\beta}R_{\nu \alpha} + \tfrac{1}{2} \nabla_{\beta}\nabla^{\beta}\nabla_{\alpha}\nabla^{\alpha}K_{\mu \nu} 
\nonumber\\
&&-  \tfrac{1}{2} \nabla_{\beta}\nabla^{\beta}\nabla_{\mu}\nabla_{\alpha}K_{\nu}{}^{\alpha} -  \tfrac{1}{2} \nabla_{\beta}\nabla^{\beta}\nabla_{\nu}\nabla_{\alpha}K_{\mu}{}^{\alpha} -  g_{\mu \nu} R^{\alpha \beta} \nabla_{\beta}\nabla_{\gamma}K_{\alpha}{}^{\gamma} + \nabla_{\alpha}R_{\nu \beta} \nabla^{\beta}K_{\mu}{}^{\alpha} + \nabla_{\alpha}R_{\mu \beta} \nabla^{\beta}K_{\nu}{}^{\alpha} 
\nonumber\\
&&+ \tfrac{2}{3} g_{\mu \nu} R^{\alpha \beta} \nabla_{\gamma}\nabla^{\gamma}K_{\alpha \beta} - 2 R_{\mu \alpha \nu \beta} \nabla_{\gamma}\nabla^{\gamma}K^{\alpha \beta} + \tfrac{1}{6} g_{\mu \nu} K^{\alpha \beta} \nabla_{\gamma}\nabla^{\gamma}R_{\alpha \beta} -  K^{\alpha \beta} \nabla_{\gamma}\nabla^{\gamma}R_{\mu \alpha \nu \beta} + \tfrac{1}{6} g_{\mu \nu} \nabla_{\gamma}\nabla^{\gamma}\nabla_{\beta}\nabla_{\alpha}K^{\alpha \beta} 
\nonumber\\
&&+ \tfrac{1}{3} g_{\mu \nu} \nabla_{\gamma}R_{\alpha \beta} \nabla^{\gamma}K^{\alpha \beta} - 2 \nabla_{\gamma}R_{\mu \alpha \nu \beta} \nabla^{\gamma}K^{\alpha \beta} + R_{\mu \beta \nu \gamma} \nabla^{\gamma}\nabla_{\alpha}K^{\alpha \beta} + R_{\mu \gamma \nu \beta} \nabla^{\gamma}\nabla_{\alpha}K^{\alpha \beta} -  \nabla_{\beta}R_{\nu \alpha} \nabla_{\mu}K^{\alpha \beta} 
\nonumber\\
&&+ \tfrac{1}{6} \nabla^{\alpha}R \nabla_{\mu}K_{\nu \alpha} -  \tfrac{1}{3} R \nabla_{\mu}\nabla_{\alpha}K_{\nu}{}^{\alpha} -  \tfrac{1}{2} R_{\nu}{}^{\alpha} \nabla_{\mu}\nabla_{\beta}K_{\alpha}{}^{\beta} + R^{\alpha \beta} \nabla_{\mu}\nabla_{\beta}K_{\nu \alpha} -  \nabla_{\beta}R_{\mu \alpha} \nabla_{\nu}K^{\alpha \beta} + \tfrac{1}{3} \nabla_{\mu}R_{\alpha \beta} \nabla_{\nu}K^{\alpha \beta} 
\nonumber\\
&&+ \tfrac{1}{6} \nabla^{\alpha}R \nabla_{\nu}K_{\mu \alpha} + \tfrac{1}{3} \nabla_{\mu}K^{\alpha \beta} \nabla_{\nu}R_{\alpha \beta} -  \tfrac{1}{3} R \nabla_{\nu}\nabla_{\alpha}K_{\mu}{}^{\alpha} -  \tfrac{1}{2} R_{\mu}{}^{\alpha} \nabla_{\nu}\nabla_{\beta}K_{\alpha}{}^{\beta} + R^{\alpha \beta} \nabla_{\nu}\nabla_{\beta}K_{\mu \alpha} -  \tfrac{2}{3} R^{\alpha \beta} \nabla_{\nu}\nabla_{\mu}K_{\alpha \beta} 
\nonumber\\
&&+ \tfrac{1}{3} K^{\alpha \beta} \nabla_{\nu}\nabla_{\mu}R_{\alpha \beta} + \tfrac{1}{3} \nabla_{\nu}\nabla_{\mu}\nabla_{\beta}\nabla_{\alpha}K^{\alpha \beta}.
\label{AP54}
\end{eqnarray}

As a check on (\ref{AP54}), we note that if we take the background to be flat,  (\ref{AP54}) reduces to $\delta W_{\mu\nu}^{}(K_{\mu\nu})=\tfrac{1}{2} \nabla_{\beta}\nabla^{\beta}\nabla_{\alpha}\nabla^{\alpha}K_{\mu \nu} -  \tfrac{1}{2} \nabla_{\beta}\nabla^{\beta}\nabla_{\mu}\nabla_{\alpha}K_{\nu}{}^{\alpha} -  \tfrac{1}{2} \nabla_{\beta}\nabla^{\beta}\nabla_{\nu}\nabla_{\alpha}K_{\mu}{}^{\alpha} + \tfrac{1}{6} g_{\mu \nu} \nabla_{\gamma}\nabla^{\gamma}\nabla_{\beta}\nabla_{\alpha}K^{\alpha \beta}+ \tfrac{1}{3} \nabla_{\nu}\nabla_{\mu}\nabla_{\beta}\nabla_{\alpha}K^{\alpha \beta}$. With $K_{\mu\nu}$ being traceless, when written in a flat Minkowski coordinate system we recognize this expression as being (\ref{AP24}), just as it should be.

\section{Fluctuations Around a Conformal to Flat Minkowski Background}
\label{S4}

\subsection{Implementing the Conformal Gauge Condition}

While we can obtain great simplification of the 59-term (\ref{AP54}) by imposing a conformal gauge condition, we can also achieve great simplification by evaluating (\ref{AP54}) directly in the metric given in (\ref{AP6}) without introducing any gauge condition at all, and even without restricting $\Omega(x)$ in any way. We do this in Appendix \ref{SF}, and  even though the rewriting of (\ref{AP54}) in the metric given in (\ref{AP6}) initially expands it to 151 terms, we are able to reduce it first to the five-term (\ref{F2}) and then to the one-term (\ref{F3}). In this section we evaluate (\ref{AP54}) in the conformal to flat background given in (\ref{AP6}) with $\Omega(x)$ again arbitrary by implementing the conformal gauge condition $\nabla_{\nu}K^{\mu\nu}=(1/2)K^{\mu\nu}g^{\alpha\beta}\partial_{\nu}g_{\alpha\beta}$ given in (\ref{AP23}). This will also lead to a one-term expression, viz. (\ref{AP61}). In the  $g_{\mu\nu}=\Omega^2(x)\eta_{\mu\nu}$ background  the gauge condition $\nabla_{\nu}K^{\mu\nu}=\frac{1}{2}K^{\mu\nu}g^{\alpha\beta}\partial_{\nu}g_{\alpha\beta}$ takes the form
\begin{eqnarray}
\nabla_{\nu}K^{\mu\nu}&-&\frac{1}{2}K^{\mu\nu}\Omega^{-2}\eta^{\alpha\beta}\eta_{\alpha\beta}\partial_{\nu}\Omega^2=\nabla_{\nu}K^{\mu\nu}-4\Omega^{-1}K^{\mu\nu}\partial_{\nu}\Omega
=\partial_{\nu}K^{\mu\nu}+6\Omega^{-1}K^{\mu\nu}\partial_{\nu}\Omega-4\Omega^{-1}K^{\mu\nu}\partial_{\nu}\Omega
\nonumber\\
&=&\partial_{\nu}K^{\mu\nu}+2\Omega^{-1}K^{\mu\nu}\partial_{\nu}\Omega=\Omega^{-2}\partial_{\nu}(\Omega^{2}K^{\mu\nu})=0. 
\label{AP55}
\end{eqnarray}
If we extract out a factor of $\Omega^2(x)$ from the fluctuation by setting $K^{\mu\nu}=\Omega^{-2}(x)k^{\mu\nu}$,  $K_{\mu\nu}=\Omega^{2}(x)k_{\mu\nu}$ (where indices on $k^{\mu\nu}$ and $k_{\mu\nu}=\eta_{\mu\alpha}\eta_{\nu\beta}k^{\alpha\beta}$ are raised and lowered with $\eta_{\mu\nu}$ alone), (\ref{AP55}) can then be written in the simple transverse form $\partial_{\nu}k^{\mu\nu}=0$, with our gauge condition being such that the conformal factor dependence factors right out. In (\ref{AP55}) we are taking $\Omega(x)$ to be a general function of the coordinates not just in order to be as general as possible but so that we can encompass as a special case Robertson-Walker geometries with general spatial curvature $k$, since as we show in Appendix \ref{SA}, while $\Omega(x)$ will only depend on the time coordinate $t$ if $k$ is zero, for non-zero spatial curvature $\Omega(x)$ will depend on both $t$ and the radial coordinate $r$.

However before evaluating (\ref{AP54}) in  a conformal to flat Minkowski geometry in the conformal gauge given in (\ref{AP55}), we note that the condition 
\begin{eqnarray}
\nabla_{\nu}K^{\mu\nu}=4\Omega^{-1}K^{\mu\nu}\partial_{\nu}\Omega
\label{AP56}
\end{eqnarray}
just happens to have the form of a covariant gauge condition for a background metric $\Omega^2(x) g_{\mu\nu}$ with any $g_{\mu\nu}$. Thus we can proceed covariantly, and following quite a bit of algebra find that when (\ref{AP56}) is imposed in a conformal to flat but not necessarily Minkowski background (the flat background could for instance be the polar coordinate geometry $ds^2=dt^2-dr^2-r^2d\theta^2-r^2\sin^2\theta d\phi^2$) (\ref{AP54}) takes  form
\begin{eqnarray}
&&\delta W_{\mu\nu}=\frac{1}{2}\Omega^{-4}\tilde{\nabla}_{\beta}\tilde{\nabla}^{\beta}\tilde{\nabla}_{\alpha}\tilde{\nabla}^{\alpha}K_{\mu \nu}-  4\Omega^{-5} \tilde{\nabla}_{\beta}\tilde{\nabla}_{\alpha}K_{\mu \nu} \tilde{\nabla}^{\beta}\tilde{\nabla}^{\alpha}\Omega- 2\Omega^{-5}  \tilde{\nabla}_{\alpha}\tilde{\nabla}^{\alpha}\Omega \tilde{\nabla}_{\beta}\tilde{\nabla}^{\beta}K_{\mu \nu}-  4 \Omega^{-5}\tilde{\nabla}^{\alpha}\Omega \tilde{\nabla}_{\beta}\tilde{\nabla}^{\beta}\tilde{\nabla}_{\alpha}K_{\mu \nu}  
\nonumber\\
&&-  \Omega^{-5}K_{\mu \nu} \tilde{\nabla}_{\beta}\tilde{\nabla}^{\beta}\tilde{\nabla}_{\alpha}\tilde{\nabla}^{\alpha}\Omega -  4\Omega^{-5} \tilde{\nabla}_{\alpha}K_{\mu \nu} \tilde{\nabla}_{\beta}\tilde{\nabla}^{\beta}\tilde{\nabla}^{\alpha}\Omega + 6\Omega^{-6} \tilde{\nabla}_{\alpha}\Omega \tilde{\nabla}^{\alpha}\Omega \tilde{\nabla}_{\beta}\tilde{\nabla}^{\beta}K_{\mu \nu} + 12\Omega^{-6} \tilde{\nabla}^{\alpha}\Omega \tilde{\nabla}_{\beta}\tilde{\nabla}_{\alpha}K_{\mu \nu} \tilde{\nabla}^{\beta}\Omega
\nonumber\\
&&+ 3\Omega^{-6} K_{\mu \nu} \tilde{\nabla}_{\alpha}\tilde{\nabla}^{\alpha}\Omega \tilde{\nabla}_{\beta}\tilde{\nabla}^{\beta}\Omega + 12 \Omega^{-6}\tilde{\nabla}_{\alpha}K_{\mu \nu} \tilde{\nabla}^{\alpha}\Omega \tilde{\nabla}_{\beta}\tilde{\nabla}^{\beta}\Omega+ 24\Omega^{-6}  \tilde{\nabla}^{\alpha}\Omega \tilde{\nabla}_{\beta}K_{\mu \nu} \tilde{\nabla}^{\beta}\tilde{\nabla}_{\alpha}\Omega 
\nonumber\\
&&+ 6\Omega^{-6} K_{\mu \nu} \tilde{\nabla}_{\beta}\tilde{\nabla}_{\alpha}\Omega \tilde{\nabla}^{\beta}\tilde{\nabla}^{\alpha}\Omega 
+ 12\Omega^{-6} K_{\mu \nu} \tilde{\nabla}^{\alpha}\Omega \tilde{\nabla}_{\beta}\tilde{\nabla}^{\beta}\tilde{\nabla}_{\alpha}\Omega -  24 \Omega^{-7}K_{\mu \nu} \tilde{\nabla}_{\alpha}\Omega \tilde{\nabla}^{\alpha}\Omega \tilde{\nabla}_{\beta}\tilde{\nabla}^{\beta}\Omega 
\nonumber\\
&&-  48\Omega^{-7}  \tilde{\nabla}_{\alpha}\Omega \tilde{\nabla}^{\alpha}\Omega \tilde{\nabla}_{\beta}K_{\mu \nu} \tilde{\nabla}^{\beta}\Omega
-  48\Omega^{-7}  K_{\mu \nu} \tilde{\nabla}^{\alpha}\Omega \tilde{\nabla}_{\beta}\tilde{\nabla}_{\alpha}\Omega \tilde{\nabla}^{\beta}\Omega+ 60\Omega^{-8} K_{\mu \nu} \tilde{\nabla}_{\alpha}\Omega \tilde{\nabla}^{\alpha}\Omega \tilde{\nabla}_{\beta}\Omega \tilde{\nabla}^{\beta}\Omega
\label{AP57}
\end{eqnarray}
without approximation. In (\ref{AP57}) we have introduced the symbol $\tilde{\nabla}_{\alpha}$ (with Greek index) to denote the covariant derivative with respect to the flat but not necessarily Minkowski background $g_{\mu\nu}$ alone so that $\tilde{\nabla}^{\alpha}$ is equal to $g^{\alpha\beta}\tilde{\nabla}_{\alpha}$ and not to $\Omega^{-2}g^{\alpha\beta}\tilde{\nabla}_{\alpha}$. Quite  remarkably, we find that the 17 terms in (\ref{AP57}) can be factored into just a single term, viz.
\begin{eqnarray}
\delta W_{\mu\nu}(K_{\mu\nu})=\frac{1}{2}\Omega^{-2}\tilde{\nabla}_{\alpha}\tilde{\nabla}^{\alpha}\tilde{\nabla}_{\beta}\tilde{\nabla}^{\beta}(\Omega^{-2}K_{\mu\nu})
=\frac{1}{2}\Omega^{-2}\tilde{\nabla}_{\alpha}\tilde{\nabla}^{\alpha}\tilde{\nabla}_{\beta}\tilde{\nabla}^{\beta}k_{\mu\nu},
\label{AP58}
\end{eqnarray}
where we have set $k_{\mu\nu}=\Omega^{-2}(x)K_{\mu\nu}$. As written, (\ref{AP58}) describes fluctuations around any geometry that is conformal to any flat background metric as written in  the  $\nabla_{\nu}K^{\mu\nu}=4\Omega^{-1}K^{\mu\nu}\partial_{\nu}\Omega$ gauge. If we set $\Omega(x)=1$  (\ref{AP58}) also describes fluctuations around any flat background geometry in the transverse gauge $\nabla_{\nu}K^{\mu\nu}=0$ as per 
\begin{eqnarray}
\delta W_{\mu\nu}(K_{\mu\nu})=\frac{1}{2}\tilde{\nabla}_{\alpha}\tilde{\nabla}^{\alpha}\tilde{\nabla}_{\beta}\tilde{\nabla}^{\beta}K_{\mu\nu}.
\label{AP59}
\end{eqnarray}
As such (\ref{AP59}) would apply to fluctuations around a flat geometry as written in any general but not necessarily flat Minkowski coordinate system.

\subsection{Obtaining the Fluctuation Equations in the Conformal Gauge}

Despite their simple forms neither (\ref{AP58}) nor (\ref{AP59}) are of straightforward use since they involve covariant derivatives that mix the various components of $K_{\mu\nu}$. However, (\ref{AP58}) and (\ref{AP59}) also apply in the gauge given in (\ref{AP55}) that is of interest to us here. Thus for conformal gauge fluctuations around a conformal to flat geometry that is Minkowski the fluctuation equations take the form 
\begin{eqnarray}
&&\delta W_{\mu\nu}(K_{\mu\nu})=\frac{1}{2}\Omega^{-4}\partial_{\beta}\partial^{\beta}\partial_{\alpha}\partial^{\alpha}K_{\mu \nu}-  4\Omega^{-5} \partial_{\beta}\partial_{\alpha}K_{\mu \nu} \partial^{\beta}\partial^{\alpha}\Omega- 2\Omega^{-5}  \partial_{\alpha}\partial^{\alpha}\Omega \partial_{\beta}\partial^{\beta}K_{\mu \nu}-  4 \Omega^{-5}\partial^{\alpha}\Omega \partial_{\beta}\partial^{\beta}\partial_{\alpha}K_{\mu \nu}  
 \nonumber\\
 &&-  \Omega^{-5}K_{\mu \nu} \partial_{\beta}\partial^{\beta}\partial_{\alpha}\partial^{\alpha}\Omega -  4\Omega^{-5} \partial_{\alpha}K_{\mu \nu} \partial_{\beta}\partial^{\beta}\partial^{\alpha}\Omega 
 + 6\Omega^{-6} \partial_{\alpha}\Omega \partial^{\alpha}\Omega \partial_{\beta}\partial^{\beta}K_{\mu \nu} + 12\Omega^{-6} \partial^{\alpha}\Omega \partial_{\beta}\partial_{\alpha}K_{\mu \nu} \partial^{\beta}\Omega 
 \nonumber\\
 &&+ 3\Omega^{-6} K_{\mu \nu} \partial_{\alpha}\partial^{\alpha}\Omega \partial_{\beta}\partial^{\beta}\Omega + 12 \Omega^{-6}\partial_{\alpha}K_{\mu \nu} \partial^{\alpha}\Omega \partial_{\beta}\partial^{\beta}\Omega+ 24\Omega^{-6}  \partial^{\alpha}\Omega \partial_{\beta}K_{\mu \nu} \partial^{\beta}\partial_{\alpha}\Omega + 6\Omega^{-6} K_{\mu \nu} \partial_{\beta}\partial_{\alpha}\Omega \partial^{\beta}\partial^{\alpha}\Omega 
  \nonumber\\
 &&+ 12\Omega^{-6} K_{\mu \nu} \partial^{\alpha}\Omega \partial_{\beta}\partial^{\beta}\partial_{\alpha}\Omega -  24 \Omega^{-7}K_{\mu \nu} \partial_{\alpha}\Omega \partial^{\alpha}\Omega \partial_{\beta}\partial^{\beta}\Omega -  48\Omega^{-7}  \partial_{\alpha}\Omega \partial^{\alpha}\Omega \partial_{\beta}K_{\mu \nu} \partial^{\beta}\Omega-  48\Omega^{-7}  K_{\mu \nu} \partial^{\alpha}\Omega \partial_{\beta}\partial_{\alpha}\Omega \partial^{\beta}\Omega
  \nonumber\\
 &&+ 60\Omega^{-8} K_{\mu \nu} \partial_{\alpha}\Omega \partial^{\alpha}\Omega \partial_{\beta}\Omega \partial^{\beta}\Omega,
\label{AP60}
\end{eqnarray}
with (\ref{AP60})  simplifying to 
\begin{eqnarray}
\delta W_{\mu\nu}(K_{\mu\nu})=\frac{1}{2}\Omega^{-2}\eta^{\sigma\rho}\eta^{\alpha\beta}\partial_{\sigma}\partial_{\rho} \partial_{\alpha}\partial_{\beta}(\Omega^{-2}K_{\mu\nu})
=\frac{1}{2}\Omega^{-2}\eta^{\sigma\rho}\eta^{\alpha\beta}\partial_{\sigma}\partial_{\rho} \partial_{\alpha}\partial_{\beta}k_{\mu\nu}.
\label{AP61}
\end{eqnarray}
We recognize (\ref{AP61}) as precisely being of the form given earlier on general grounds in  (\ref{AP30}). As we see, despite the fact that the $g_{\mu\nu}=\Omega^2(x)\eta_{\mu\nu}$ background is not itself flat, in (\ref{AP60}) and (\ref{AP61}) all derivatives are flat Minkowski, i.e. associated with the metric $ds^2=-\eta_{\alpha\beta}dx^{\alpha}dx^{\beta}=dt^2-dx^2-dy^2-dz^2$. Moreover, and even more significantly, (\ref{AP60}) and (\ref{AP61}) are diagonal in the $(\mu,\nu)$ indices. Thus with our choice of gauge, in a conformal to flat Minkowski background $\delta W_{\mu\nu}$ is diagonalized in the tensor indices. We thus see the power of conformal symmetry since our starting point was the 62 term $\delta W_{\mu\nu}(h_{\mu\nu})$ given in (\ref{AP43}). Eq. (\ref{AP61}) is our main result. It is exact without approximation, and is to be used to monitor cosmological fluctuations in the conformal gravity theory.

\subsection{Summary of the Calculation}

For the benefit of the reader we briefly summarize the steps in our calculation. We start with a general $W_{\mu\nu}$ in the convenient form given in (\ref{AP42}). We perturb $W_{\mu\nu}$ to first order around a general background  with metric $g_{\mu\nu}$ and take the perturbed metric to be of the form  $g_{\mu\nu}+\delta g_{\mu\nu}=g_{\mu\nu}+h_{\mu\nu}$. Recalling  that $\delta R^{\lambda}_{\phantom{\lambda}\mu\nu\kappa}=
\partial \delta\Gamma^{\lambda}_{\mu\nu}/\partial x^{\kappa}
+\Gamma^{\lambda}_{\kappa\sigma}\delta\Gamma^{\sigma}_{\mu\nu}
-\Gamma^{\sigma}_{\mu\kappa}\delta\Gamma^{\lambda}_{\nu\sigma}
-\partial \delta\Gamma^{\lambda}_{\mu\kappa}/\partial x^{\nu}
-\Gamma^{\lambda}_{\nu\sigma}\delta\Gamma^{\sigma}_{\mu\kappa}
+\Gamma^{\sigma}_{\mu\nu}\delta\Gamma^{\lambda}_{\kappa\sigma}=
\nabla_{\kappa}\delta\Gamma^{\lambda}_{\mu\nu}
-\nabla_{\nu}\delta\Gamma^{\lambda}_{\mu\kappa}$, where $\delta\Gamma^{\lambda}_{\mu\nu}=(1/2)g^{\lambda \rho}[\nabla_{\nu}\delta g_{\rho\mu}+\nabla_{\mu}\delta g_{\rho\nu}-\nabla_{\rho}\delta g_{\mu\nu}]$, on evaluating $\delta W_{\mu\nu}$ to lowest order in $\delta g_{\mu\nu}$ we obtain (\ref{AP43}). 

Since our interest is in traceless fluctuations, in (\ref{AP43}) we set  $h_{\mu\nu}=K_{\mu\nu}+(1/4)g_{\mu\nu}h$ where $h=g^{\mu\nu}h_{\mu\nu}$ and $g^{\mu\nu}K_{\mu\nu}=0$. This leads us to two contributions to $\delta W_{\mu\nu}$, viz. $\delta W_{\mu\nu}(K_{\mu\nu})$ as given in (\ref{AP44}) and $\delta W_{\mu\nu}(h)$ as given in (\ref{AP45}). Using properties of manipulations of covariant derivatives we find that $\delta W_{\mu\nu}(h)=-(1/4)W_{\mu\nu}h$ as exhibited in (\ref{AP50}). This allows to establish that $\delta W_{\mu\nu}(h)$ vanishes if the background $W_{\mu\nu}$ vanishes, just as is the case in background Robertson-Walker and de Sitter cosmologies. 

For background cosmologies in which the background $W_{\mu\nu}$ vanishes, $\delta W_{\mu\nu}$ reduces to $\delta W_{\mu\nu}(K_{\mu\nu})$, with $\delta W_{\mu\nu}$ thus only being dependent on the traceless fluctuation $K_{\mu\nu}$ as per (\ref{AP44}). Using further properties of manipulations of covariant derivatives we rewrite (\ref{AP44}) in the form given in (\ref{AP54}), a form in which we can readily implement the conformal gauge condition given in (\ref{AP23}). On implementing this gauge condition we find that for fluctuations around background geometries that are conformal to flat (such as Robertson-Walker and de Sitter) $\delta W_{\mu\nu}(K_{\mu\nu})$ reduces to (\ref{AP61}), our main result. In order to actually implement all of these various steps we had to quite extensively adapt the $xAct$ tensor calculus software package in order to perform the conformal gravity calculations symbolically on a computer.

\section{SVT Decomposition of the Fluctuations}
\label{S5}

\subsection{The SVT Decomposition}

In studying cosmological fluctuations it is very convenient to use the SVT decomposition of the fluctuations because it readily incorporates gauge  invariance  \cite{Bardeen1980}. For cosmological fluctuations around a conformal to flat Minkowski background geometry of the form $ds^2=\Omega^2(x)[dt^2-\delta_{ij}dx^idx^j]$, one introduces the following $3+1$ decomposition of the background plus fluctuation line element 
\begin{eqnarray}
ds^2 = \Omega^2(x) \left[ (1+2\phi) dt^2 -2(\tilde{\nabla}_i B +B_i)dt dx^i - [(1-2\psi)\delta_{ij} +2\tilde{\nabla}_i\tilde{\nabla}_j E + \tilde{\nabla}_i E_j + \tilde{\nabla}_j E_i + 2E_{ij}]dx^i dx^j\right],
\label{AP62}
\end{eqnarray}
where $\Omega(x)$ is an arbitrary function of the coordinates, where $\tilde{\nabla}_i=\partial/\partial x^i$ (with Latin index) and  $\tilde{\nabla}^i=\delta^{ij}\tilde{\nabla}_j$ (i.e. not $\Omega^{-2}\delta^{ij}\tilde{\nabla}_j$) are defined with respect to the background 3-space metric $\delta_{ij}$, and where the elements of (\ref{AP62}) obey
\begin{eqnarray}
\delta^{ij}\tilde{\nabla}_j B_i = 0,\qquad \delta^{ij}\tilde{\nabla}_j E_i = 0, \qquad E_{ij}=E_{ji},\qquad \delta^{jk}\tilde{\nabla}_kE_{ij} = 0, \qquad \delta^{ij}E_{ij} = 0.
\label{AP63}
\end{eqnarray}
In Appendix \ref{SE} we provide a derivation of (\ref{AP62}) using transverse and transverse-traceless projection techniques, and show that even though the form of (\ref{AP62}) is not manifestly covariant, it is nonetheless covariant (just as it would have to be if we are to be able to provide the gauge invariant  combinations of its coefficients that we give below). In (\ref{AP62}) we note that we have incorporated an explicit factor of $\Omega^2(x)$ not just in the background part of the line element but in the fluctuation part as well. This will prove to be very convenient below. As written, (\ref{AP62}) contains ten elements, whose transformations are defined with respect to the background spatial sector as four 3-dimensional scalars ($\phi$, $B$, $\psi$, $E$), two transverse 3-dimensional vectors ($B_i$, $E_i$) each with two independent degrees of freedom, and one symmetric 3-dimensional transverse-traceless tensor ($E_{ij}$) with two degrees of freedom.

\subsection{SVT in Terms of $h_{\mu\nu}$}

One can express these ten degrees of freedom in terms of the original fluctuations $h_{\mu\nu}$. If one introduces the 3-dimensional Green's function that obeys
\begin{eqnarray}
\delta^{ij}\tilde{\nabla}_i\tilde{\nabla}_jD^{(3)}(\mathbf{x}-\mathbf{y})=\delta^3(\mathbf{x}-\mathbf{y}),
\label{AP64}
\end{eqnarray}
and if one for convenience sets $h_{\mu\nu}=\Omega^2(x)f_{\mu\nu}$, then from the form of the (\ref{AP62}) line element one obtains 
\begin{eqnarray}
ds^2&=&-[\Omega^2(x)\eta_{\alpha\beta}+h_{\alpha\beta}]dx^{\alpha}dx^{\beta}
\nonumber\\
&=&-\Omega^2(x)[\eta_{\alpha\beta}+f_{\alpha\beta}]dx^{\alpha}dx^{\beta}=\Omega^2(x)\left[dt^2-\delta_{ij}dx^idx^j-f_{00}dt^2-2f_{0i}dtdx^i-f_{ij}dx^idx^j\right],
\nonumber\\
\delta^{ij}f_{ij}&=&-6\psi+2\tilde{\nabla}_i\tilde{\nabla}^iE,\qquad
 \tilde{\nabla}^jf_{ij}=-2\tilde{\nabla}_i\psi+2\tilde{\nabla}_i\tilde{\nabla}_k\tilde{\nabla}^kE+\tilde{\nabla}_k\tilde{\nabla}^kE_{i},
 \nonumber\\
 \tilde{\nabla}^i \tilde{\nabla}^jf_{ij}&=&-2\tilde{\nabla}_k\tilde{\nabla}^k\psi+2\tilde{\nabla}_k\tilde{\nabla}^k\tilde{\nabla}_{\ell}\tilde{\nabla}^{\ell}E=\frac{4}{3}\tilde{\nabla}_k\tilde{\nabla}^k\tilde{\nabla}_{\ell}\tilde{\nabla}^{\ell}E+\frac{1}{3}\tilde{\nabla}_k\tilde{\nabla}^k\delta^{ij}f_{ij}=4\tilde{\nabla}_k\tilde{\nabla}^k\psi+\tilde{\nabla}_k\tilde{\nabla}^k(\delta^{ij}f_{ij}),
 \nonumber\\
2\phi&=&-f_{00},\qquad
B=\int d^3yD^{(3)}(\mathbf{x}-\mathbf{y})\tilde{\nabla}_y^if_{0i},\qquad B_i=f_{0i}-\tilde{\nabla}_iB,
\nonumber\\
\psi&=&\frac{1}{4}\int d^3yD^{(3)}(\mathbf{x}-\mathbf{y})\tilde{\nabla}_y^k\tilde{\nabla}_y^{\ell}f_{k\ell}-\frac{1}{4}\delta^{k\ell}f_{k\ell},
\nonumber\\
\qquad
E&=&\int d^3yD^{(3)}(\mathbf{x}-\mathbf{y})\left[\frac{3}{4}\int d^3zD^{(3)}(\mathbf{y}-\mathbf{z})\tilde{\nabla}_z^k\tilde{\nabla}_z^{\ell}f_{k\ell}-\frac{1}{4}\delta^{k\ell}f_{k\ell}\right],
\nonumber\\
E_i&=&\int d^3yD^{(3)}(\mathbf{x}-\mathbf{y})\bigg{[}\tilde{\nabla}_y^jf_{ij}
-\tilde{\nabla}^y_i\int d^3zD^{(3)}(\mathbf{y}-\mathbf{z})\tilde{\nabla}_z^k\tilde{\nabla}_z^{\ell}f_{k\ell}\bigg{]},
\nonumber\\
2E_{ij}&=&f_{ij}+2\psi\delta_{ij} -2\tilde{\nabla}_i\tilde{\nabla}_j E - \tilde{\nabla}_i E_j - \tilde{\nabla}_j E_i, 
\label{AP65}
\end{eqnarray}
with $B_i$,  $E_i$ and $E_{ij}$ then obeying (\ref{AP63}). (In (\ref{AP65}) in a symbol such as $\tilde{\nabla}_y^i$ for instance the $y$ indicates that the derivative is taken with respect to the $y$ coordinate.) 

\subsection{SVT in Terms of the Traceless $k_{\mu\nu}$}

It is also instructive to reexpress the SVT components in terms of the traceless part of $f_{\mu\nu}$ to the degree possible. Since $K_{\mu\nu}=\Omega^2k_{\mu\nu}$ is defined in (\ref{AP13}) as $K_{\mu\nu}=h_{\mu\nu}-(1/4)\Omega^2\eta_{\mu\nu}\Omega^{-2}\eta^{\alpha\beta}h_{\alpha\beta}=h_{\mu\nu}-(1/4)\eta_{\mu\nu}\eta^{\alpha\beta}h_{\alpha\beta}$, we can set $k_{\mu\nu}=f_{\mu\nu}-(1/4)\eta_{\mu\nu}[-f_{00}+\delta^{ij}f_{ij}]$. From (\ref{AP65}) we thus obtain
\begin{eqnarray}
k_{00}&=&\frac{3}{4}f_{00}+\frac{1}{4}\delta^{k\ell}f_{k\ell},\qquad k_{0i}=f_{0i},\qquad k_{ij}=f_{ij}+\frac{1}{4}\delta_{ij}f_{00}-\frac{1}{4}\delta_{ij}\delta^{k\ell}f_{k\ell},
\nonumber\\
\phi&=&-\frac{1}{2}f_{00},\qquad
B=\int d^3yD^{(3)}(\mathbf{x}-\mathbf{y})\tilde{\nabla}_y^ik_{0i},\qquad B_i=k_{0i}-\tilde{\nabla}_iB,
\nonumber\\
\psi&=&\frac{1}{4}\int d^3yD^{(3)}(\mathbf{x}-\mathbf{y})\tilde{\nabla}_y^k\tilde{\nabla}_y^{\ell}k_{k\ell}-\frac{3}{4}k_{00}+\frac{1}{2}f_{00},
\nonumber\\
\qquad
E&=&\int d^3yD^{(3)}(\mathbf{x}-\mathbf{y})\left[\frac{3}{4}\int d^3zD^{(3)}(\mathbf{y}-\mathbf{z})\tilde{\nabla}_z^k\tilde{\nabla}_z^{\ell}k_{k\ell}-\frac{1}{4}k_{00}\right],
\nonumber\\
E_i&=&\int d^3yD^{(3)}(\mathbf{x}-\mathbf{y})\bigg{[}\tilde{\nabla}_y^jk_{ij}
-\tilde{\nabla}^y_i\int d^3zD^{(3)}(\mathbf{y}-\mathbf{z})\tilde{\nabla}_z^k\tilde{\nabla}_z^{\ell}k_{k\ell}\bigg{]},
\nonumber\\
2E_{ij}&+&2\tilde{\nabla}_i\tilde{\nabla}_j E +\tilde{\nabla}_i E_j +\tilde{\nabla}_j E_i
=k_{ij}-\frac{1}{2}\delta_{ij}k_{00}+\frac{1}{2}\delta_{ij}\int d^3yD^{(3)}(\mathbf{x}-\mathbf{y})\tilde{\nabla}_y^k\tilde{\nabla}_y^{\ell}k_{k\ell}.
\label{AP66}
\end{eqnarray}
As we see, everything can be expressed in terms of $k_{\mu\nu}$ together with  just one component of $f_{\mu\nu}=\Omega^{-2}(x)h_{\mu\nu}$, namely $f_{00}$.  Also we note that the combination $\phi+\psi$ is independent of $f_{00}$ and only depends on the components of $k_{\mu\nu}$. As we will see below this is not accidental.

\subsection{Gauge Structure of the SVT Decomposition}

In order to explore the gauge structure of the SVT decomposition we implement an infinitesimal coordinate transformation $\bar{x}_{\mu}=x_{\mu}+\epsilon_{\mu}(x)$ on the proper time $ds^2=\Omega^2(x)[dt^2-\delta_{ij}dx^idx^j]-h_{\mu\nu}dx^{\mu}dx^{\nu}$. It is convenient to write $\epsilon_{\mu}$ in the scalar, vector form associated with the background metric $ds^2=-\Omega^2(x)\eta_{\mu\nu}dx^{\mu}dx^{\nu}=\Omega^2(x)[dt^2-\delta_{ij}dx^idx^j]$, viz.  
\begin{eqnarray}
\epsilon_{\mu}=\Omega^2(x)f_{\mu},\qquad f_{0}=-T,\qquad f_i=L_i+\tilde{\nabla}_iL\qquad \delta^{ij}\tilde{\nabla}_jL_i=\tilde{\nabla}^iL_i=0.
\label{AP67}
\end{eqnarray}
With a general coordinate transformation being of the form $\bar{g}^{\mu\nu}=(\partial \bar{x}^{\mu}/\partial x^{\sigma})(\partial \bar{x}^{\nu}/\partial x^{\tau})g^{\sigma\tau}$,  the proper time $ds^2$ and the fluctuations $h_{\mu\nu}$,  $f_{\mu\nu}=\Omega^{-2}(x)h_{\mu\nu}$, and $k_{\mu\nu}=f_{\mu\nu}-(1/4)\eta_{\mu\nu}[-f_{00}+\delta^{ij}f_{ij}]$ transform into 
\begin{eqnarray}
ds^2&=&\bar{\Omega}^2(\bar{x}) \left[ (1+2\bar{\phi}) d\bar{t}^2 -2(\tilde{\nabla}_i \bar{B} +\bar{B}_i)d\bar{t}d\bar{x}^i - [(1-2\bar{\psi})\bar{\delta}_{ij} +2\tilde{\nabla}_i\tilde{\nabla}_j \bar{E} + \tilde{\nabla}_i \bar{E}_j + \tilde{\nabla}_j \bar{E}_i + 2\bar{E}_{ij}]d\bar{x}^i d\bar{x}^j\right],
\nonumber\\
\bar{h}_{\mu\nu}&=&h_{\mu\nu}-\partial_{\nu}\epsilon_{\mu}-\partial_{\mu}\epsilon_{\nu}+2\Gamma^{\lambda}_{\mu\nu}\epsilon_{\lambda}
\nonumber\\
&=&h_{\mu\nu}-\partial_{\nu}\epsilon_{\mu}-\partial_{\mu}\epsilon_{\nu}+\Omega^{-2}(x)[\epsilon_{\mu}\partial_{\nu}+\epsilon_{\nu}\partial_{\mu}
-\epsilon_{\lambda}\eta_{\mu\nu}\eta^{\lambda\sigma}\partial_{\sigma}]\Omega^2(x),
\nonumber\\
\bar{f}_{\mu\nu}&=&f_{\mu\nu}-\partial_{\nu}f_{\mu}-\partial_{\mu}f_{\nu}-\Omega^{-2}(x)f_{\lambda}\eta_{\mu\nu}\eta^{\lambda\sigma}\partial_{\sigma}\Omega^2(x),
\nonumber\\
\bar{f}_{00}&=&f_{00}+2\dot{T}+\Omega^{-2}(x)[T\partial_0+(L_i+\tilde{\nabla}_iL)\delta^{ij}\partial_j]\Omega^2(x),
\nonumber\\
\bar{f}_{0i}&=&f_{0i}+\partial_iT-\dot{L}_i-\tilde{\nabla}_i\dot{L},
\nonumber\\
\bar{f}_{ij}&=&f_{ij}-\partial_{i}(L_j+\tilde{\nabla}_jL)-\partial_{j}(L_i+\tilde{\nabla}_iL)
-\delta_{ij}\Omega^{-2}(x)[T\partial_0+(L_i+\tilde{\nabla}_iL)\delta^{ij}\partial_j]\Omega^2(x),
\nonumber\\
\bar{k}_{00}&=&k_{00}+\frac{3}{2}\dot{T}-\frac{1}{2}\delta^{ij}\partial_i\partial_jL,
\nonumber\\
\bar{k}_{0i}&=&k_{0i}+\partial_iT-\dot{L}_i-\tilde{\nabla}_i\dot{L},
\nonumber\\
\bar{k}_{ij}&=&k_{ij}-\partial_{i}(L_j+\tilde{\nabla}_jL)-\partial_{j}(L_i+\tilde{\nabla}_iL)
+\frac{1}{2}\delta_{ij}[\dot{T}+\delta^{k\ell}\partial_k\partial_{\ell}L],
\label{AP68}
\end{eqnarray}
where the dot denotes derivative with respect to $t$. Expressing the barred SVT basis in terms of the components of $\bar{f}_{\mu\nu}$ by using the same format as in (\ref{AP65}), and using the relations between the components of $\bar{f}_{\mu\nu}$ and $f_{\mu\nu}$ given in (\ref{AP68}) then yields
\begin{eqnarray}
\bar{\phi}&=&\phi-\dot{T}-\Omega^{-1}[T\partial_0+(L_i+\tilde{\nabla}_iL)\delta^{ij}\partial_j]\Omega,\qquad \bar{B}=B+T-\dot{L},\qquad \bar{\psi}=\psi+\Omega^{-1}[T\partial_0+(L_i+\tilde{\nabla}_iL)\delta^{ij}\partial_j]\Omega,
\nonumber\\
\bar{E}&=&E-L,\qquad \bar{B}_i=B_i-\dot{L}_i,\qquad \bar{E}_i=E_i-L_i,\ \qquad \bar{E}_{ij}=E_{ij},
\label{AP69}
\end{eqnarray}
to lowest order in $\epsilon_{\mu}$. 

In deriving (\ref{AP69}) for $B$ and $B_i$ for instance, we set $\bar{B}=\int d^3yD^{(3)}(\mathbf{x}-\mathbf{y})\tilde{\nabla}_y^i\bar{f}_{0i}=\int d^3yD^{(3)}(\mathbf{x}-\mathbf{y})\tilde{\nabla}_y^i[f_{0i}+\partial_iT-\dot{L}_i-\tilde{\nabla}_i\dot{L}]=B+\int d^3yD^{(3)}(\mathbf{x}-\mathbf{y})\tilde{\nabla}_y^i\partial_i(T-\dot{L})$, and using an integration by parts, one which we justify in Appendix \ref{SE} where we show that through appropriate gauge transformations the $B$ and $\bar{B}$ integrals can be made to exist, obtain $\bar{B}=B+T-\dot{L}$. From this relation we can then set $\bar{B}_i=\bar{f}_{0i}-\partial_i\bar{B}=B_i+\partial_iB+\partial_iT-\dot{L}_i-\partial_i\dot{L}-\partial_i(B+T-\dot{L})=B_i-\dot{L}_i$. An alternative procedure that one could consider employing is to start with $\bar{f}_{0i}=\bar{B}_i+\partial_i\bar{B}=B_i+\partial_iB+\partial_iT-\dot{L}_i-\partial_i\dot{L}$, and then apply $\partial^i$ to it, to  obtain $\nabla^2(\bar{B}-B-T+\dot{L})=0$. However, this latter relation  does not require that $\bar{B}-B-T+\dot{L}$ vanish, as it could be of the form $f(t)+{\bf n}\cdot {\bf x} g(t)$ where ${\bf n}$ is an arbitrary spatially-independent 3-vector and $f(t)$ and $g(t)$ are arbitrary functions of time. In our procedure this issue does not arise since no derivative condition is met for any component of $f_{\mu\nu}$. In Appendix  \ref{SE} we discuss the point further. 

Another procedure that is employed in the literature is to break up the relations between the components of $\bar{f}_{\mu\nu}$ and $f_{\mu\nu}$ by equating scalars with scalars, vectors with vectors, tensors with tensors. While valid if the fluctuation is associated with a single momentum state, it is not valid if the fluctuation is a superposition of momentum states (a localized source would necessarily be in a superposition of momentum states), since it is not the case that a vector that is transverse to a given momentum vector is transverse to some other momentum vector. However, by utilizing the decomposition given in (\ref{AP65}) we are able to accommodate arbitrary superpositions of momenta in an approach which is completely general.

On now eliminating the $T$, $L$ and $L_i$ terms from (\ref{AP69}), we obtain four gauge invariant combinations 
\begin{eqnarray}
&&\bar{\phi}+\bar{\psi}+\dot{\bar{B}}-\ddot{\bar{E}}=\phi+\psi+\dot{B}-\ddot{E},\qquad \bar{B}_i-\dot{\bar{E}}_i=B_i-\dot{E}_i,\qquad \bar{E}_{ij}=E_{ij},
\nonumber\\
&&\bar{\phi}-\bar{\psi}+\dot{\bar{B}}-\ddot{\bar{E}}+2\Omega^{-1}\left[(\bar{B}-\dot{\bar{E}})\dot{\Omega}
-(\bar{E}_i+\partial_i\bar{E})\delta^{ij}\partial_j\Omega\right]
\nonumber\\
&&=\phi-\psi+\dot{B}-\ddot{E}+2\Omega^{-1}\left[(B-\dot{E})\dot{\Omega}
-(E_i+\partial_iE)\delta^{ij}\partial_j\Omega\right],
\label{AP70}
\end{eqnarray}
which together contain a total of six degrees of freedom (one scalar, one two-component transverse vector, one two-component transverse-traceless tensor, and a second scalar).  Since a general ten-component $h_{\mu\nu}$ is reduced to six degrees of freedom by imposing four coordinate invariance conditons, the SVT decomposition precisely generates for us six gauge invariant quantities, just as needed. And other than the last one,  none of the other combinations contains $\Omega(x)$, to thus be gauge invariant no matter how arbitrary $\Omega(x)$ might be. We note that we would not have been able to obtain these particular $\Omega$-independent expressions had we defined the fluctuations in (\ref{AP62}) without an overall multiplying factor of $\Omega^2(x)$. Now in analyzing the gauge structure of (\ref{AP62}) we had not imposed any conformal invariance requirements on it (indeed one uses (\ref{AP62}) for fluctuations in Einstein gravity), and yet we see the value in giving the fluctuations the same overall $\Omega^2(x)$ factor as the background. 

In addition, we note that with the structure given in  (\ref{AP70}) we can see why  a procedure that equates scalars with scalars, vectors with vectors, tensors with tensors could have a shortcoming. Specifically, if we were to apply such a procedure to the last combination in (\ref{AP70}) we would obtain two separate conditions: one for the $\phi$, $\psi$, $B$ and $E$ scalars and the other for the $E_i$ vector. Then we would have seven gauge invariant combinations, but we are only allowed six. An interplay between the scalar and vector sectors is thus needed in order to maintain gauge invariance.

In the event that $\Omega(x)$ is only a function of $t$ (\ref{AP69}) reduces to 
\begin{eqnarray}
\bar{\phi}&=&\phi-\dot{T}-\Omega^{-1}\dot{\Omega}T,\qquad \bar{B}=B+T-\dot{L},\qquad \bar{\psi}=\psi+\Omega^{-1}\dot{\Omega}T,\qquad \bar{E}=E-L,
\nonumber\\
\bar{B}_i&=&B_i-\dot{L}_i,\qquad \bar{E}_i=E_i-L_i,\ \qquad \bar{E}_{ij}=E_{ij},
\label{AP71}
\end{eqnarray}
and gauge invariant combinations  take the form
\begin{eqnarray}
&&\bar{\phi}+\bar{\psi}+\dot{\bar{B}}-\ddot{\bar{E}}=\phi+\psi+\dot{B}-\ddot{E},\qquad \bar{B}_i-\dot{\bar{E}}_i=B_i-\dot{E}_i,\qquad \bar{E}_{ij}=E_{ij},
\nonumber\\
&&\bar{\psi}-\Omega^{-1}\dot{\Omega}(\bar{B}-\dot{\bar{E}})=\psi-\Omega^{-1}\dot{\Omega}(B-\dot{E}).
\label{AP72}
\end{eqnarray}
Finally, when $\Omega(x)=1$ (\ref{AP71}) reduces to 
\begin{eqnarray}
\bar{\phi}&=&\phi-\dot{T},\qquad \bar{B}=B+T-\dot{L},\qquad \bar{\psi}=\psi ,\qquad \bar{E}=E-L,
\nonumber\\
\bar{B}_i&=&B_i-\dot{L}_i,\qquad \bar{E}_i=E_i-L_i,\ \qquad \bar{E}_{ij}=E_{ij},
\label{AP73}
\end{eqnarray}
and the gauge invariant combinations are then 
\begin{eqnarray}
\bar{\psi}=\psi,\qquad \bar{\phi}+\dot{\bar{B}}-\ddot{\bar{E}}=\phi+\dot{B}-\ddot{E},\qquad \bar{B}_i-\dot{\bar{E}}_i=B_i-\dot{E}_i,\qquad \bar{E}_{ij}=E_{ij}.
\label{AP74}
\end{eqnarray}

In addition, we note the transformations of the traceless $k_{\mu\nu}$ that are also given in (\ref{AP68}) do not involve the quantity $\Omega^{-2}(x)[T\partial_0+(L_i+\tilde{\nabla}_iL)\delta^{ij}\partial_j]\Omega^2(x)$. In consequence, the last relation given in (\ref{AP70}) does not apply and in the traceless sector one only has a total of five degrees of freedom (one scalar, one two-component transverse vector, one two-component transverse-traceless tensor), with a general ten-component $h_{\mu\nu}$ being reduced to five degrees of freedom by imposing four coordinate invariance conditions and one tracelessness condition. Thus while a general Einstein gravity $\delta G_{\mu\nu}+8\pi G \delta T_{\mu\nu}$ fluctuation will only depend on all of the gauge invariant combinations given in (\ref{AP70}), the conformal gravity $\delta W_{\mu\nu}$ will not depend on the last combination given in (\ref{AP70}). But since none of the other combinations given in (\ref{AP70}) involve derivatives of $\Omega(x)$,  the dependence on $\Omega(x)$ can only appear as an overall multiplying factor in $\delta W_{\mu\nu}$ and only the $\Omega$-independent combinations given in (\ref{AP70}) can appear. In regard to these various gauge invariant combinations, we note that inspection of (\ref{AP66}) shows that none of the three gauge invariant combinations given in (\ref{AP70}) (combinations that only involve $\phi+\psi$ and not $\phi$ or $\psi$ separately) depends on the trace of $h_{\mu\nu}$ but only on the components of $k_{\mu\nu}$ alone. Since $\delta W_{\mu\nu}$ also does not depend on the trace of $h_{\mu\nu}$ for fluctuations around a conformal to  flat background (cf. (\ref{AP50})), it must be the case that when written in the SVT basis,  such a $\delta W_{\mu\nu}$ must only depend on these same three combinations. We now check this directly.

\section{Conformal Gravity SVT Fluctuations in a Conformal to Flat Minkowski Background}
\label{S6}

Evaluating $\delta W^{\mu\nu}$ for the fluctuations around the conformal to flat Minkowski background geometry as given in (\ref{AP62}) yields (following a machine calculation) the remarkably  compact
\begin{eqnarray}
\delta W_{00}  &=& -\frac{2}{3\Omega^2} \delta^{mn}\delta^{\ell k}\tilde{\nabla}_m\tilde{\nabla}_n\tilde{\nabla}_{\ell}\tilde{\nabla}_k (\phi + \psi +\dot{B}-\ddot{E}),
\nonumber\\	
\delta W_{0i} &=&  -\frac{2}{3\Omega^2} \delta^{mn}\tilde{\nabla}_i\tilde{\nabla}_m\tilde{\nabla}_n\partial_t(\phi +\psi +\dot{B}-\ddot{E})
	+\frac{1}{2\Omega^2}\left[\delta^{mn}\delta^{\ell k}\tilde{\nabla}_m\tilde{\nabla}_n\tilde{\nabla}_{\ell}\tilde{\nabla}_k(B_i - \dot{E}_i) -  \delta^{\ell k}\tilde{\nabla}_{\ell}\tilde{\nabla}_k \partial_t^2(B_i - \dot{E}_i)\right],
\nonumber\\	
\delta W_{ij}  &=& \frac{1}{3\Omega^2}\bigg{[} \delta_{ij}\delta^{\ell k}\tilde{\nabla}_{\ell}\tilde{\nabla}_k  \partial_t^2(\phi+ \psi+\dot{B}-\ddot{E}) + \delta^{\ell k}\tilde{\nabla}_{\ell}\tilde{\nabla}_k \tilde{\nabla}_i\tilde{\nabla}_j (\phi + \psi +\dot{B}-\ddot{E}) 
\nonumber\\
&&- \delta_{ij} \delta^{mn}\delta^{\ell k}\tilde{\nabla}_m\tilde{\nabla}_n\tilde{\nabla}_{\ell}\tilde{\nabla}_k(\phi + \psi +\dot{B}-\ddot{E}) -3\tilde{\nabla}_i\tilde{\nabla}_j \partial_t^2(\phi + \psi +\dot{B}-\ddot{E})\bigg{] }
\nonumber\\
&&+\frac{1}{2\Omega^2}\left[ \delta^{\ell k}\tilde{\nabla}_{\ell}\tilde{\nabla}_k \tilde{\nabla}_i   \partial_t(B_j - \dot{E}_j)+ \delta^{\ell k}\tilde{\nabla}_{\ell}\tilde{\nabla}_k \tilde{\nabla}_j \partial_t(B_i - \dot{E}_i) - \tilde{\nabla}_i\partial_t^3(B_j - \dot{E}_j)-\tilde{\nabla}_j\partial_t^3(B_i - \dot{E}_i)\right]
\nonumber\\
&&+\frac{1}{\Omega^2}\left[\delta^{mn}\tilde{\nabla}_m\tilde{\nabla}_n-\partial_t^2\right]^2E_{ij}.
\label{AP75}
\end{eqnarray}
In (\ref{AP75}) we note that the 3 plus 1 structure is apparent just as it should be, with $\delta W_{00}$, $\delta W_{0i}$,  and $\delta W_{ij}$  respectively being written entirely in terms of quantities that transform as 3-dimensional scalars, 3-dimensional vectors, and 3-dimensional rank two tensors. Also each term in $\delta W_{\mu\nu}$ is given as four derivatives of a gauge invariant combination, with only the $E_{ij}$ term in $\delta W_{ij}$ involving four time derivatives. As we will show in Appendix \ref{SB} this will cause the $E_{ij}$ modes to be leading at large times.

Despite the fact that we have taken $\Omega(x)$ to be completely arbitrary, (\ref{AP75}) contains no derivatives of $\Omega(x)$. This is because in a conformal invariant theory $\delta W_{\mu\nu}$ transforms as $\delta W_{\mu\nu}\rightarrow \Omega^{-2}(x)\delta W_{\mu\nu}$ (cf. (\ref{AP12})), with the $\delta W_{\mu\nu}$ associated with the metric of (\ref{AP62}) being given by $\Omega^{-2}(x)$ times the $\delta W_{\mu\nu}$ that would be associated with a (\ref{AP62}) metric without any overall $\Omega^2(x)$ factor or without any dependence on $\Omega(x)$ at all. Because of this, the only gauge invariant combinations that can appear in the gauge invariant $\delta W_{\mu\nu}$ have to be the ones that are totally independent of $\Omega^2(x)$, just as we have found. This then is the value of defining the fluctuations in (\ref{AP62}) so that they expressly have an overall $\Omega^2(x)$ factor. And in addition we note that our having obtained an overall factor of $\Omega^{-2}(x)$ in (\ref{AP75}) provides a nice internal check on our calculation.

We also note that the gauge invariance of $\delta W_{\mu\nu}$ is immediately apparent since $\delta W_{\mu\nu}$ only depends on the three gauge invariant, five degree of freedom  $\phi+\psi+\dot{B}-\ddot{E}$, $B_i-\dot{E}_i$, and $E_{ij}$ combinations given in (\ref{AP70}).  In fact this would have to be the case, since as we had noted above, $\delta W_{\mu\nu}$ is independent of  the trace of $h_{\mu\nu}$, just as are these three gauge invariant SVT combinations. Moreover, as we had noted earlier and as we now explicitly see, for conformal to flat backgrounds $\delta W_{\mu\nu}$ must be gauge invariant on its own no matter what form $\delta T_{\mu\nu}$ might take and no matter how complicated a function $\Omega(x)$ might be. Because of this the conformal gravity $\delta T_{\mu\nu}$ must also be gauge invariant on its own too, and thus it can also be developed in terms of  SVT decomposition analogs of the three gauge invariant quantities given in (\ref{AP70}), with a structure of this form being the most general form for  $\delta T_{\mu\nu}$ that is allowed in conformal cosmology.

It is of interest to compare the SVT expression for $\delta W_{\mu\nu}$ given in  (\ref{AP75}) with  the conformal gauge calculation expression given in (\ref{AP61}). As we see, the structure of the $E_{ij}$ term in (\ref{AP75}) is of the exactly the same fourth-order derivative form as found in (\ref{AP61}). Now $E_{ij}$ is gauge invariant all on its own. Thus its contribution to $\delta W_{\mu\nu}$ does not depend on the choice of gauge, and thus its contributions to (\ref{AP61}) and (\ref{AP75}) must be identical. While  we can identify $2E_{ij}$ with the transverse ($T$) traceless ($\theta$) quantity $h_{ij}^{T\theta}$ that we introduce in Appendix \ref{SE}, our interest here is  identifying $2E_{ij}$ with the relevant components of $k_{ij}$ that appear in (\ref{AP61}) as constrained by  $\partial_{\nu}k^{\mu\nu}=0$, and we shall explore this issue in Appendix \ref{SB}. 

Now unlike conformal gravity, Einstein gravity does not have an underlying  conformal structure, and thus the simple SVT form obtained for $\delta W_{\mu\nu}$ will not have an analog in $\delta G_{\mu\nu}$, since in Einstein gravity only the entire $\delta G_{\mu\nu}+8\pi G \delta T_{\mu\nu}$ is gauge invariant, and thus only $\delta G_{\mu\nu}+8\pi G \delta T_{\mu\nu}$ can be written in terms of the SVT gauge invariant functions given in (\ref{AP69}) - (\ref{AP74}). (An explicit example of this may for instance be found in \cite{Ellis2012}, where the SVT decomposition of the fluctuation Einstein equations is carried out in a simple case.) Moreover, for the metric given in (\ref{AP62}) the SVT structure given in Appendix \ref{SC} for $\delta G_{\mu\nu}$ when $\Omega(x)$ is arbitrary is not just not gauge invariant, it is extremely complicated. Moreover, as we now show, even for an $\Omega(x)$ that only depends on $t$, the relevant $\delta G_{\mu\nu}$ is not only not gauge invariant, it is also more complicated that the $\delta W_{\mu\nu}$ associated with the same $\Omega(t)$ background (viz. (\ref{AP75}), which holds not just for $\Omega(t)$, but for any $\Omega(x)$).

\section{SVT Decomposition of  $\delta G_{\mu\nu}$ with Conformal factor $\Omega(t)$}
\label{S7}

With  background plus fluctuation metric of the form $ds^2=\Omega^2(x)[dt^2-\delta_{ij}dx^idx^j]-\Omega^2(x)f_{\mu\nu}dx^{\mu}dx^{\nu}$, the fluctuation in the Einstein tensor is given by 
\begin{eqnarray}
&&\delta G_{\mu\nu}=- \tfrac{1}{2}\tilde{\nabla}_{\alpha}\tilde{\nabla}_{\mu}f_{\nu}{}^{\alpha} -  \tfrac{1}{2} \tilde{\nabla}_{\alpha}\tilde{\nabla}_{\nu}f_{\mu}{}^{\alpha} -  \eta^{\alpha \beta} \eta_{\mu \nu} \Omega^{-1}\tilde{\nabla}_{\alpha}f \tilde{\nabla}_{\beta}\Omega + \eta^{\alpha \beta} \Omega^{-1} \tilde{\nabla}_{\alpha}f_{\mu \nu} \tilde{\nabla}_{\beta}\Omega + \eta^{\beta \alpha} f_{\mu \nu} \Omega^{-2}\tilde{\nabla}_{\alpha}\Omega \tilde{\nabla}_{\beta}\Omega 
\nonumber\\
&&-  \tfrac{1}{2} \eta^{\alpha \beta} \eta_{\mu \nu} \tilde{\nabla}_{\beta}\tilde{\nabla}_{\alpha}f + \tfrac{1}{2} \eta_{\mu \nu}\tilde{\nabla}_{\beta}\tilde{\nabla}_{\alpha}f^{\alpha \beta} + \tfrac{1}{2} \eta^{\alpha \beta} \tilde{\nabla}_{\beta}\tilde{\nabla}_{\alpha}f_{\mu \nu} + 2 \eta_{\mu \nu} f^{\alpha \beta} \Omega^{-1} \tilde{\nabla}_{\beta}\tilde{\nabla}_{\alpha}\Omega - 2 \eta^{\alpha \beta} f_{\mu \nu} \Omega^{-1} \tilde{\nabla}_{\beta}\tilde{\nabla}_{\alpha}\Omega 
\nonumber\\
&&+ 2 \eta^{\alpha \gamma} \eta_{\mu \nu} \Omega^{-1} \tilde{\nabla}_{\beta}f_{\alpha}{}^{\beta} \tilde{\nabla}_{\gamma}\Omega -  \eta^{\alpha \gamma} \eta^{\beta \kappa} \eta_{\mu \nu} f_{\alpha \beta} \Omega^{-2} \tilde{\nabla}_{\gamma}\Omega\tilde{\nabla}_{\kappa}\Omega -  \eta^{\alpha \beta} \Omega^{-1}\tilde{\nabla}_{\beta}\Omega \tilde{\nabla}_{\mu}f_{\nu \alpha} -  \eta^{\alpha \beta} \Omega^{-1}\tilde{\nabla}_{\beta}\Omega \tilde{\nabla}_{\nu}f_{\mu \alpha} 
\nonumber\\
&&+ \tfrac{1}{2} \tilde{\nabla}_{\nu}\tilde{\nabla}_{\mu}f,
\label{AP76}
\end{eqnarray}
where $f$ denotes $\eta^{\mu\nu}f_{\mu\nu}$,  $f_{\alpha}{}^{\beta}$ denotes $\eta^{\beta\gamma}f_{\alpha\gamma}$, $f^{\alpha\beta}$ denotes $\eta^{\alpha\mu}\eta^{\beta\nu}f_{\mu\nu}$, and the covariant derivative $\tilde{\nabla}_{\mu}$ is evaluated with respect to $\eta_{\mu\nu}$. Evaluating this $\delta G_{\mu\nu}$ in the SVT basis given in (\ref{AP62}) yields, on restricting $\Omega$ to a function of $t$,
\begin{eqnarray}
&&\delta G_{00}=6 \Omega^{-1} \dot{\Omega} \dot{\psi} + 2 \delta^{ab} \Omega^{-1} \dot{\Omega} \tilde{\nabla}_{b}\tilde{\nabla}_{a}B - 2 \delta^{ab} \Omega^{-1} \dot{\Omega} \tilde{\nabla}_{b}\tilde{\nabla}_{a}\dot{E} - 2 \delta^{ab} \tilde{\nabla}_{b}\tilde{\nabla}_{a}\psi,
\nonumber\\
&&\delta G_{0i}=- \Omega^{-2} \dot{\Omega}^2 \tilde{\nabla}_{i}B + 2 \Omega^{-1} \ddot{\Omega} \tilde{\nabla}_{i}B - 2 \Omega^{-1} \dot{\Omega} \tilde{\nabla}_{i}\phi - 2 \tilde{\nabla}_{i}\dot{\psi}
- B_{i} \Omega^{-2} \dot{\Omega}^2 + 2 B_{i} \Omega^{-1} \ddot{\Omega} + \tfrac{1}{2} \delta^{ab} \tilde{\nabla}_{b}\tilde{\nabla}_{a}B_{i} -  \tfrac{1}{2} \delta^{ab} \tilde{\nabla}_{b}\tilde{\nabla}_{a}\dot{E}_{i},
\nonumber\\
&&\delta G_{ij}=2 \delta_{ij} \Omega^{-2} \dot{\Omega}^2 \phi - 4 \delta_{ij} \Omega^{-1} \ddot{\Omega} \phi - 2 \delta_{ij} \Omega^{-1} \dot{\Omega} \dot{\phi} + 2 \delta_{ij} \Omega^{-2} \dot{\Omega}^2 \psi - 4 \delta_{ij} \Omega^{-1} \ddot{\Omega} \psi - 4 \delta_{ij} \Omega^{-1} \dot{\Omega} \dot{\psi} - 2 \delta_{ij} \ddot{\psi} 
\nonumber\\
&&- 2 \delta^{ab} \delta_{ij} \Omega^{-1} \dot{\Omega} \tilde{\nabla}_{b}\tilde{\nabla}_{a}B -  \delta^{ab} \delta_{ij} \tilde{\nabla}_{b}\tilde{\nabla}_{a}\dot{B} + 2 \delta^{ab} \delta_{ij} \Omega^{-1} \dot{\Omega} \tilde{\nabla}_{b}\tilde{\nabla}_{a}\dot{E} + \delta^{ab} \delta_{ij} \tilde{\nabla}_{b}\tilde{\nabla}_{a}\ddot{E} -  \delta^{ab} \delta_{ij} \tilde{\nabla}_{b}\tilde{\nabla}_{a}\phi + \delta^{ab} \delta_{ij} \tilde{\nabla}_{b}\tilde{\nabla}_{a}\psi 
\nonumber\\
&&+ 2 \Omega^{-1} \dot{\Omega} \tilde{\nabla}_{j}\tilde{\nabla}_{i}B + \tilde{\nabla}_{j}\tilde{\nabla}_{i}\dot{B} - 2 \Omega^{-2} \dot{\Omega}^2 \tilde{\nabla}_{j}\tilde{\nabla}_{i}E + 4 \Omega^{-1} \ddot{\Omega} \tilde{\nabla}_{j}\tilde{\nabla}_{i}E - 2 \Omega^{-1} \dot{\Omega} \tilde{\nabla}_{j}\tilde{\nabla}_{i}\dot{E} -  \tilde{\nabla}_{j}\tilde{\nabla}_{i}\ddot{E} + \tilde{\nabla}_{j}\tilde{\nabla}_{i}\phi -  \tilde{\nabla}_{j}\tilde{\nabla}_{i}\psi
\nonumber\\
&&+\Omega^{-1} \dot{\Omega} \tilde{\nabla}_{i}B_{j} + \tfrac{1}{2} \tilde{\nabla}_{i}\dot{B}_{j} -  \Omega^{-2} \dot{\Omega}^2 \tilde{\nabla}_{i}E_{j} + 2 \Omega^{-1} \ddot{\Omega} \tilde{\nabla}_{i}E_{j} -  \Omega^{-1} \dot{\Omega} \tilde{\nabla}_{i}\dot{E}_{j} -  \tfrac{1}{2} \tilde{\nabla}_{i}\ddot{E}_{j} + \Omega^{-1} \dot{\Omega} \tilde{\nabla}_{j}B_{i} + \tfrac{1}{2} \tilde{\nabla}_{j}\dot{B}_{i} 
\nonumber\\
&&-  \Omega^{-2} \dot{\Omega}^2 \tilde{\nabla}_{j}E_{i} 
+ 2 \Omega^{-1} \ddot{\Omega} \tilde{\nabla}_{j}E_{i} -  \Omega^{-1} \dot{\Omega} \tilde{\nabla}_{j}\dot{E}_{i} -  \tfrac{1}{2} \tilde{\nabla}_{j}\ddot{E}_{i}- \ddot{E}_{ij} - 2 \dot{E}_{ij} \Omega^{-1} \dot{\Omega} - 2 E_{ij} \Omega^{-2} \dot{\Omega}^2 + 4 E_{ij} \Omega^{-1} \ddot{\Omega} 
\nonumber\\
&&+ \delta^{ab} \tilde{\nabla}_{b}\tilde{\nabla}_{a}E_{ij},
\label{AP77}
\end{eqnarray}
where as before $\tilde{\nabla}_{i}$ denotes the covariant derivative with respect to  $\delta_{ij}$.

If we set $\Omega(x)=1$ with the line element then being of the form $ds^2=dt^2-\delta_{ij}dx^idx^j -f_{\mu\nu}dx^{\mu}dx^{\nu}$, (\ref{AP77})  reduces to 
\begin{eqnarray}
\delta G_{00}&=&- 2 \delta^{ab} \tilde{\nabla}_{b}\tilde{\nabla}_{a}\psi,
\nonumber\\
\delta G_{0i}&=&- 2 \tilde{\nabla}_{i}\dot{\psi}+ \tfrac{1}{2} \delta^{ab} \tilde{\nabla}_{b}\tilde{\nabla}_{a}(B_{i} -  \dot{E}_{i}),
\nonumber\\
\delta G_{ij}&=&- 2 \delta_{ij} \ddot{\psi} -  \delta^{ab} \delta_{ij} \tilde{\nabla}_{b}\tilde{\nabla}_{a}(\phi+\dot{B}  -\ddot{E})+ \delta^{ab} \delta_{ij} \tilde{\nabla}_{b}\tilde{\nabla}_{a}\psi 
 + \tilde{\nabla}_{j}\tilde{\nabla}_{i}(\phi+\dot{B} -  \ddot{E})  -  \tilde{\nabla}_{j}\tilde{\nabla}_{i}\psi
\nonumber\\
&& + \tfrac{1}{2} \tilde{\nabla}_{i}(\dot{B}_{j} - \ddot{E}_{j}) + \tfrac{1}{2} \tilde{\nabla}_{j}(\dot{B}_{i}  
- \ddot{E}_{i})- \ddot{E}_{ij} + \delta^{ab} \tilde{\nabla}_{b}\tilde{\nabla}_{a}E_{ij}.
\label{AP78}
\end{eqnarray}
Thus as had been noted earlier, the only situation in which $\delta G_{\mu\nu}$ is gauge invariant is for fluctuations around a flat background (compare (\ref{AP78}) with (\ref{AP74})) since it is only in that case that the Einstein gravity background $T_{\mu\nu}$ is zero. Even though $\delta G_{\mu\nu}$ is not gauge invariant when there is an $\Omega(t)$, the very lack of gauge invariance of (\ref{AP77}) shows exactly what the gauge structure of $\delta T_{\mu\nu}$ must be since the total $\delta G_{\mu\nu}+8\pi G \delta T_{\mu\nu}$ is gauge invariant. As we see from (\ref{AP78}), for fluctuations around flat $\delta G_{\mu\nu}$ depends on all four of the gauge invariant combinations listed in (\ref{AP74}). $\delta G_{\mu\nu}$ thus depends on a total of six degrees of freedom, just as is appropriate to a theory in which coordinate invariance reduces an initial ten component $h_{\mu\nu}$ to six physical components.

\appendix
\setcounter{equation}{0}
\section{Conformal to Flat Minkowski Cosmological Backgrounds}
\label{SA}

\subsection{Robertson-Walker Metric with $k=0$ Written in Conformal to Flat Form}

In order to apply (\ref{AP61}) to cosmology we need to write the Robertson-Walker and de Sitter background geometries in a conformal to flat Minkowski form. For a $k=0$ Robertson-Walker background the comoving coordinate system metric takes the form
\begin{eqnarray}
ds^2({\rm comoving})=dt^2-a^2(t)[dx^2+dy^2+dz^2].
\label{A1}
\end{eqnarray}
The straightforward introduction of the conformal time
\begin{eqnarray}
d\tau=\int \frac{dt}{a(t)}
\label{A2}
\end{eqnarray}
then allows us to write the conformal time metric as
\begin{eqnarray}
ds^2({\rm conformal~time})=a^2(\tau)[d\tau^2-dx^2-dy^2-dz^2].
\label{A3}
\end{eqnarray}

\subsection{Robertson-Walker Metric with $k>0$  Written in Conformal to Flat Form}

For a $k>0$ or a $k<0$ Robertson-Walker background the comoving and conformal time coordinate system metrics take the form
\begin{eqnarray}
ds^2({\rm comoving})&=&dt^2-a^2(t)\left[\frac{dr^2}{1-kr^2}+r^2d\theta^2+r^2\sin^2\theta d\phi^2\right],
\nonumber\\
ds^2({\rm conformal~time})&=&a^2(\tau)\left[d\tau^2-\frac{dr^2}{1-kr^2}-r^2d\theta^2-r^2\sin^2\theta d\phi^2\right].
\label{A4}
\end{eqnarray}

To bring the RW geometries with non-zero $k$ to a conformal to flat form requires coordinate transformations that involve both $\tau$ and $r$. For the $k>0$ case first, it is convenient to set $k=1/L^2$, and introduce $\sin \chi=r/L$, with the conformal time metric given in (\ref{A4}) then taking the form
\begin{eqnarray}
ds^2=L^2a^2(p)\left[dp^2-d\chi^2 -\sin^2\chi d\theta^2-\sin^2\chi \sin^2\theta d\phi^2\right],
\label{A5}
\end{eqnarray}
where $p=\tau/L$. Following e.g. \cite{Mannheim2012a} we introduce
\begin{eqnarray}
p^{\prime}+r^{\prime}=\tan[(p+\chi)/2],\qquad p^{\prime}-r^{\prime}=\tan[(p-\chi)/2],\qquad p^{\prime}=\frac{\sin p}{\cos p+\cos \chi},\qquad r^{\prime}=\frac{\sin \chi}{\cos p+\cos \chi},
\label{A6}
\end{eqnarray}
so that
\begin{eqnarray}
dp^{\prime 2}-dr^{\prime 2}&=&\frac{1}{4}[dp^2-d\chi^2]\sec^2[(p+\chi)/2]\sec^2[(p-\chi)/2],
\nonumber\\
\frac{1}{4}(\cos p +\cos \chi)^2&=&\cos^2[(p+\chi)/2]\cos^2[(p-\chi)/2]=\frac{1}{[1+(p^{\prime}+r^{\prime})^2][1+(p^{\prime}-r^{\prime})^2]}.
\label{A7}
\end{eqnarray}
With these transformations the $k>0$ line element then takes the conformal to flat form
\begin{eqnarray}
ds^2=\frac{4L^2a^2(p)}{[1+(p^{\prime}+r^{\prime})^2][1+(p^{\prime}-r^{\prime})^2]}\left[dp^{\prime 2}-dr^{\prime 2} -r^{\prime 2}d\theta^2-r^{\prime 2} \sin^2\theta d\phi^2\right].
\label{A8}
\end{eqnarray}
To bring the spatial sector  of (\ref{A8}) to Cartesian coordinates we set  $x^{\prime}=r^{\prime}\sin\theta\cos\phi$, $y^{\prime}=r^{\prime}\sin\theta\sin\phi$, $z^{\prime}=r^{\prime}\cos\theta$ and thus bring the line element to the form  
\begin{eqnarray}
ds^2=L^2a^2(p)(\cos p+\cos \chi)^2\left[dp^{\prime 2}-dx^{\prime 2} -dy^{\prime 2} -dz^{\prime 2} \right],
\label{A9}
\end{eqnarray}
where now $r^{\prime}=(x^{\prime 2}+ y^{\prime 2}+z^{\prime 2})^{1/2}$. With these transformations (\ref{A9}) is now in the form given in (\ref{AP6}).

\subsection{Robertson-Walker Metric with $k<0$  Written in Conformal to Flat Form}

For the $k<0$ case, it is convenient to set $k=-1/L^2$, and introduce ${\rm sinh} \chi=r/L$, with the conformal time metric given in (\ref{A4}) then taking the form
\begin{eqnarray}
ds^2=L^2a^2(p)\left[dp^2-d\chi^2 -{\rm sinh}^2\chi d\theta^2-{\rm sinh}^2\chi \sin^2\theta d\phi^2\right],
\label{A10}
\end{eqnarray}
where $p=\tau/L$. Next we introduce
\begin{eqnarray}
p^{\prime}+r^{\prime}=\tanh[(p+\chi)/2],\qquad p^{\prime}-r^{\prime}=\tanh[(p-\chi)/2],\qquad p^{\prime}=\frac{\sinh p}{\cosh p+\cosh \chi},\qquad r^{\prime}=\frac{\sinh \chi}{\cosh p+\cosh \chi},
\label{A11}
\end{eqnarray}
so that
\begin{eqnarray}
dp^{\prime 2}-dr^{\prime 2}&=&\frac{1}{4}[dp^2-d\chi^2]{\rm sech}^2[(p+\chi)/2]{\rm sech}^2[(p-\chi)/2],
\nonumber\\
\frac{1}{4}(\cosh p+\cosh \chi)^2&=&{\rm \cosh}^2[(p+\chi)/2]{\rm \cosh}^2[(p-\chi)/2]=\frac{1}{[1-(p^{\prime}+r^{\prime})^2][1-(p^{\prime}-r^{\prime})^2]}.
\label{A12}
\end{eqnarray}
With these transformations the line element takes the conformal to flat form
\begin{eqnarray}
ds^2=\frac{4L^2a^2(p)}{[1-(p^{\prime}+r^{\prime})^2][1-(p^{\prime}-r^{\prime})^2]}\left[dp^{\prime 2}-dr^{\prime 2} -r^{\prime 2}d\theta^2-r^{\prime 2} \sin^2\theta d\phi^2\right].
\label{A13}
\end{eqnarray}
The spatial sector can then be written in Cartesian form
\begin{eqnarray}
ds^2=L^2a^2(p)(\cosh p+\cosh \chi)^2\left[dp^{\prime 2}-dx^{\prime 2} -dy^{\prime 2} -dz^{\prime 2}\right],
\label{A14}
\end{eqnarray}
where again $r^{\prime}=(x^{\prime 2}+ y^{\prime 2}+z^{\prime 2})^{1/2}$.  We note that in transforming from (\ref{A4}) to (\ref{A9}) or to (\ref{A14}) we have only made coordinate transformations and not made any conformal transformation. 

\subsection{Conformal Cosmological Background Solutions with a Cosmological Constant}

While the conformal to flat Minkowski structures given in (\ref{A3}), (\ref{A9}) and (\ref{A14}) are purely kinematical, the explicit form of $a(t)$ can be determined once a dynamics has been specified. Thus in regard to a de Sitter or anti-de Sitter cosmology, a de Sitter or an anti-de Sitter geometry is  just a particular case of a Robertson-Walker geometry in which $a(t)$ has a specific assigned value for each possible choice of  spatial 3-curvature $k$. On writing the  maximally 4-symmetric geometry condition $R_{\mu\nu}=-3\alpha g_{\mu\nu}$ in Robertson-Walker form one obtains 
\begin{eqnarray} 
\dot{a}^2(t) +k=\alpha  a^2(t).
\label{A15}
\end{eqnarray}
(In terms of the scalar field model described in (\ref{AP32}) -- (\ref{AP35}) we have $K=\alpha =-2\lambda_{S}S^2_0$.) Here $\alpha$ is positive for de Sitter and negative for anti-de Sitter. Allowable solutions to (\ref{A15}) depend on the values of $\alpha$ and $k$, and are of the form (see e.g. \cite{Mannheim2006})
\begin{eqnarray}
a(t,\alpha>0,k<0)&=&\left(-\frac{k}{\alpha}\right)^{1/2}
\sinh(\alpha^{1/2}t),
\nonumber \\
a(t,\alpha>0,k=0)&=&a(t=0)\exp(\alpha^{1/2}t),
\nonumber \\
a(t,\alpha>0,k>0)&=&\left(\frac{k}{\alpha}\right)^{1/2}\cosh(\alpha^{1/2}t),
\nonumber \\
a(t,\alpha=0,k<0)&=&(-k)^{1/2}t,
\nonumber \\
a(t,\alpha<0,k<0)&=&\left(\frac{k}{\alpha}\right)^{1/2}\sin((-\alpha)^{1/2}t).
\label{A16}
\end{eqnarray}
In these solutions (\ref{A3}), (\ref{A9}), and (\ref{A14}) all apply  to a de Sitter or an anti-de Sitter cosmology.

\subsection{Conformal Cosmological Background Solutions with a Cosmological Constant and a Radiation Fluid}

For Robertson-Walker cosmologies we note that with slight modification we can extend the scalar field model given above to include a perfect fluid, with the energy-momentum tensor then being given by \cite{Mannheim1998}
\begin{eqnarray}
T_{\rm S}^{\mu \nu}&=&(\rho+p)U_{\mu}U_{\nu}+pg_{\mu\nu} 
-\frac{1}{6} S_0^2\left(R^{\mu\nu}-\frac{1}{2}g^{\mu\nu}
R^\alpha_{\phantom{\alpha}\alpha}\right)-g^{\mu\nu}\lambda_S S_0^4,
\label{A17}
\end{eqnarray}                                 
with the background conformal cosmology still obeying $T_{\rm S}^{\mu \nu}=0$ since the background  Robertson-Walker geometry continues to obey $W_{\mu\nu}=0$.  On taking the perfect fluid energy-momentum tensor to be traceless radiation  (viz. $\rho=3p$, $\rho=A/a^4(t)$, $A>0$) as needed in the early universe, and with $\alpha =-2\lambda_{S}S^2_0$ as before, the evolution equation takes the form
\begin{eqnarray}
\dot{a}^2+k&=&\alpha a^2-\frac{2A}{S_0^2a^2},
\label{A18}
\end{eqnarray}                                 
with allowed solutions to the cosmology being given by  \cite{Mannheim1998} 
\begin{eqnarray}
a(t,\alpha>0,k<0,A>0)&=&\left(-\frac{k(\beta-1)}{2\alpha}-\frac{k\beta}{\alpha}\sinh^2(\alpha^{1/2}t)\right)^{1/2},
\nonumber \\
a(t,\alpha>0,k=0,A>0)&=&\left(-\frac{A}{\lambda_S S_0^4}\right)^{1/4}\cosh^{1/2}(2\alpha^{1/2}t),
\nonumber \\
a(t,\alpha>0,k>0,A>0)&=&\left(\frac{-k(\beta-1)}{2\alpha}+\frac{k\beta}{\alpha}\cosh^2(\alpha^{1/2}t)\right)^{1/2},
\nonumber\\
a(t,\alpha=0,k<0,A>0)&=&\left(-\frac{2A}{kS_0^2}-kt^2\right)^{1/2},
\nonumber \\
a(t,\alpha<0,k<0,A>0)&=&\left(-\frac{k(\beta-1)}{2\alpha}+\frac{k\beta}{\alpha}\sin^2((-\alpha)^{1/2}t)\right)^{1/2},
\label{A19}
\end{eqnarray}
where $\beta=(1+8A\alpha/k^2S_0^2)^{1/2}$.

\section{Early Universe Fluctuations around Conformal to Flat Minkowski Backgrounds}
\label{SB}

\subsection{Transforming Solutions from Conformal to Flat to Comoving}

In terms of practical applications of the conformal gravity theory we note that we have studied the current era conformal cosmology associated with (\ref{A17}), with very good non-fine-tuned, negative $k$  driven fits to the accelerating universe Hubble plot data having been presented in \cite{Mannheim2006,Mannheim2017}. Similarly, very good non-fine-tuned, negative $k$ driven fits  to the galactic rotation curves of 138 spiral galaxies have been presented in \cite{Mannheim2011b,Mannheim2012c,OBrien2012}, fits in which no galactic dark matter is needed at all. (The essence of these rotation curve fits is that in the conformal gravity theory a test particle in a galaxy is affected by both the local galactic field and the global cosmological field, with the systematics of galactic rotation curves being found to be sensitive to the spatial curvature of the Universe -- in fact it is this global cosmological effect that removes the need for the presence of dark matter within galaxies in the conformal theory.) We shall thus consider conformal gravity fluctuations in Robertson-Walker cosmologies with negative $k$.

In conformal gravity fitting to the current era Hubble plot the cosmological constant term is found to dominate over the perfect fluid contribution. However in the early universe it has to be the other way round with a not-yet-red-shifted radiation era perfect fluid being the dominant contribution since $a(t)$ is small and $A/a^4(t)$ is large. Moreover, if the $A/a^4(t)$ radiation contribution is dominant, then since $k$ is given by $k=-\dot{a}^2-2A/a^2S_0^2$ when the $\alpha$ contribution in (\ref{A18}) is negligible, we are led right back to our observationally  preferred negative $k$. (This line of reasoning for  fixing the sign of $k$ is essentially the same as the one  used in (\ref{AP40}) above.) Thus for studying fluctuation growth in the early universe the only relevant solution for $a(t)$ is the $a(t,\alpha=0,k<0,A>0)$ one. For this solution, on setting $k=-1/L^2$, $d^2=2AL^4/S_0^2$, and $L^2a^2(t)=(d^2+t^2)$, we obtain
\begin{eqnarray}
\tau=L\int_0^t \frac{dt}{(d^2+t^2)^{1/2}}=L~{\rm arc sinh}\left(\frac{t}{d}\right),\qquad t=d\sinh p,
\label{B1}
\end{eqnarray}
where $p=\tau/L$. Now according to (\ref{AP29}) fluctuations around a flat background would grow linearly in the relevant time variable, which according to the $k<0$ (\ref{A14}) is $p^{\prime}$. If we define $\Omega(p,\chi)=La(p)(\cosh p+\cosh \chi)$, we can write (\ref{A14}) in the form
\begin{eqnarray}
ds^2=\Omega^2(p,\chi)\left[dp^{\prime 2}-dx^{\prime 2} -dy^{\prime 2} -dz^{\prime 2}\right].
\label{B2}
\end{eqnarray}
And with fluctuation $\delta W_{\mu\nu}$ being given in (\ref{AP61}) as $\delta W_{\mu\nu}=(1/2)\Omega^{-2}\eta^{\sigma\rho}\eta^{\alpha\beta}\partial_{\sigma}\partial_{\rho} \partial_{\alpha}\partial_{\beta}k_{\mu\nu}$ where $k_{\mu\nu}=\Omega^{-2}(x)K_{\mu\nu}$, following
(\ref{AP29}) we can write the solution to $\delta W_{\mu\nu}=0$ in the primed-variable form
\begin{eqnarray}
k_{\mu\nu}=A^{\prime}_{\mu\nu}e^{ik^{\prime}\cdot x^{\prime}}+(n^{\prime}\cdot x^{\prime})B^{\prime}_{\mu\nu}e^{ik^{\prime}\cdot x^{\prime}}+A^{\prime *}_{\mu\nu}e^{-ik^{\prime}\cdot x^{\prime}}+(n^{\prime}\cdot x^{\prime})B^{\prime *}_{\mu\nu}e^{-ik^{\prime}\cdot x^{\prime}},
\label{B3}
\end{eqnarray}
where $\eta^{\mu\nu}k^{\prime}_{\mu}k^{\prime}_{\nu}=0$. Then with the conformal gauge condition being of the form $\partial^{\prime}_{\nu}k^{\mu\nu}=0$ in this case (where $k^{\mu\nu}=\eta^{\mu\alpha}\eta^{\nu\beta}k_{\alpha\beta}$), we obtain 
\begin{eqnarray}
&&ik^{\prime \nu}\left[A^{\prime}_{\mu\nu}e^{ik^{\prime}\cdot x^{\prime}}+(n^{\prime}\cdot x^{\prime})B^{\prime}_{\mu\nu}e^{ik^{\prime}\cdot x^{\prime}}-A^{\prime *}_{\mu\nu}e^{-ik^{\prime}\cdot x^{\prime}}-(n^{\prime}\cdot x^{\prime})B^{\prime *}_{\mu\nu}e^{-ik^{\prime}\cdot x^{\prime}}\right]
\nonumber\\
&&+n^{\prime\nu}\left[B^{\prime}_{\mu\nu}e^{ik^{\prime}\cdot x^{\prime}}+B^{\prime *}_{\mu\nu}e^{-ik^{\prime}\cdot x^{\prime}}\right]=0.
\label{B4}
\end{eqnarray}
With this relation holding for all $x^{\prime}$ we obtain
\begin{eqnarray}
&&ik^{\prime \nu}A^{\prime}_{\mu\nu}+n^{ \prime\nu}B^{\prime}_{\mu\nu}=0,\quad ik^{\prime \nu}B^{\prime}_{\mu\nu}=0,
\quad -ik^{\prime \nu}A^{\prime *}_{\mu\nu}+n^{ \prime\nu}B^{\prime *}_{\mu\nu}=0,\quad -ik^{\prime \nu}B^{\prime *}_{\mu\nu}=0.
\label{B5}
\end{eqnarray}
With the $B^{\prime}_{\mu\nu}$ term being leading in (\ref{B3}) at large $n^{\prime}\cdot x^{\prime}$, we can ignore the non-leading $A^{\prime}_{\mu\nu}$ modes, and treat the $B^{\prime}_{\mu\nu}$ modes as obeying the transverse momentum space condition  $ik^{\prime \nu}B^{\prime}_{\mu\nu}=0$, $-ik^{\prime \nu}B^{\prime *}_{\mu\nu}=0$. Since the gauge condition $\partial^{\prime}_{\nu}k^{\mu\nu}=0$ takes this momentum space form, it means that all of the components of the $B^{\prime}_{\mu\nu}$ modes have the same $(n^{\prime}\cdot x^{\prime})e^{ik^{\prime}\cdot x^{\prime}}$ leading behavior in coordinate space. And with $n^{\prime \mu}=(1,0,0,0)$, the modes grow linearly in the relevant time variable $p^{\prime}$ associated with (\ref{B2}).

To determine the structure of these solutions in the comoving coordinate system we need to both reexpress $\Omega(p,\chi)$ in terms of the comoving coordinates $(t,r)$ and transform the components of the fluctuation $K_{\mu\nu}$ to the comoving coordinate system. For the $\Omega(p,\chi)$ term first  we note that fluctuations around the conformal to flat, $k<0$ Robertson-Walker geometry grow as $\Omega^2(p,\chi)p^{\prime}$ as per (\ref{AP61}). Thus  from (\ref{A14}), (\ref{A11}), and (\ref{B1}) we find that $\Omega^2(p,\chi)p^{\prime}$ grows as
\begin{eqnarray}
\Omega^2(p,\chi)p^{\prime}&=&L^2a^2(p)(\cosh p+\cosh \chi)^2p^{\prime}=L^2a^2(p)\sinh p (\cosh p+\cosh \chi)
\nonumber\\
&=&(d^2+t^2)\frac{t}{d}\left[\left(1+\frac{t^2}{d^2}\right)^{1/2}+\left(1+\frac{r^2}{L^2}\right)^{1/2}\right].
\label{B6}
\end{eqnarray}
 $\Omega^2(p,\chi)p^{\prime}$ thus starts off linearly in $t$ when $t\ll d$, and subsequently then grows as $t^4$ when $t\gg d$. Thus in the conformal to flat Minkowski coordinate system given in (\ref{A14}) all the components of $K_{\mu\nu}$ grow as $t^4$ when $t \gg d$.

To transform the fluctuation itself we note that transforming from (\ref{A14}) to (\ref{A13}) has no effect on the $p^{\prime}$ behavior. However transforming  the spatial coordinates from the Cartesian (\ref{A14}) to the polar (\ref{A13}) does introduce a dependence on $r^{\prime}$ wherever there is an angular component, and thus it does introduce a dependence on the comoving $t$. Specifically, in terms of the $K_{x^{\prime}x^{\prime}}$ type fluctuations in the (\ref{A14}) coordinate system, in the (\ref{A13}) coordinate system we can set 
\begin{align}
K_{\theta\theta}&=(r^{\prime}\cos\theta\cos\phi)^2K_{x^{\prime}x^{\prime}}+(r^{\prime}\cos\theta\sin\phi)^2K_{y^{\prime}y^{\prime}}+(r^{\prime}\sin\theta)^2K_{z^{\prime}z^{\prime}}
\nonumber\\
&=\frac{\sinh^2\chi}{(\cosh p+\cosh \chi)^2}[\cos^2\theta\cos^2\phi K_{x^{\prime}x^{\prime}}+\cos^2\theta\sin^2\phi K_{y^{\prime}y^{\prime}}+\sin^2\theta K_{z^{\prime}z^{\prime}}]
\nonumber\\
&=\frac{r^2d^2}{[L(d^2+t^2)^{1/2}+d(L^2+r^2)^{1/2}]^2}[\cos^2\theta\cos^2\phi K_{x^{\prime}x^{\prime}}+\cos^2\theta\sin^2\phi K_{y^{\prime}y^{\prime}}+\sin^2\theta K_{z^{\prime}z^{\prime}}],
\nonumber\\
K_{r^{\prime}\theta}&=r^{\prime}\cos\theta\sin\theta\cos^2\phi K_{x^{\prime}x^{\prime}}+r^{\prime}\cos\theta\sin\theta\sin^2\phi K_{y^{\prime}y^{\prime}}-r^{\prime}\sin\theta\cos\theta K_{z^{\prime}z^{\prime}}
\nonumber\\
&=\frac{\sinh\chi}{(\cosh p+\cosh \chi)}[\cos\theta\sin\theta\cos^2\phi K_{x^{\prime}x^{\prime}}+\cos\theta\sin\theta\sin^2\phi K_{y^{\prime}y^{\prime}}-\sin\theta\cos\theta K_{z^{\prime}z^{\prime}}],
\nonumber\\
K_{p^{\prime}\theta}&=r^{\prime}\cos\theta\cos\phi K_{p^{\prime}x^{\prime}}+r^{\prime}\cos\theta\sin\phi K_{p^{\prime}y^{\prime}}-r^{\prime}\sin\theta K_{p^{\prime}z^{\prime}}
\nonumber\\
&=\frac{\sinh\chi}{(\cosh p+\cosh \chi)}[\cos\theta\cos\phi K_{p^{\prime}x^{\prime}}+\cos\theta\sin\phi K_{p^{\prime}y^{\prime}}-\sin\theta K_{p^{\prime}z^{\prime}}],
\label{B7}
\end{align}
with analogous expressions for $K_{\theta\phi}$, $K_{\phi\phi}$, $K_{r^{\prime}\phi}$ and $K_{p^{\prime}\phi}$. The $r^{\prime}=\sinh \chi/(\cosh p +\cosh \chi)$ prefactor in (\ref{B7}) has the property that at large $t$ it behaves  as $t^0$ if $p=\chi$, viz. $t=r$ with both $t$ and $r$ large (lightlike case), and as $t^{-1}$ if $p\gg \chi$, viz. $t \gg r$ (timelike case).

To transform from (\ref{A13}) to (\ref{A10}) we need to transform from $(p^{\prime}, r^{\prime})$ to $(p,\chi)$. To transform from (\ref{A10}) to the comoving Robertson-Walker metric given (\ref{A4}) we need to transform from $(p,\chi)$ to $(t,r)$. Since the angular sector is unaffected by the transformation from (\ref{A13}) to (\ref{A4}), the angular sector fluctuations $K_{\theta\theta}$, $K_{\theta  \phi}$, $K_{\phi\phi}$ associated with the comoving Robertson-Walker geometry given in (\ref{A4}) will thus grow as $t^4$ itself as modified by the prefactor in (\ref{B7}), and thus as $t^4$ for the lightlike case (the solutions to $\eta^{\sigma\rho}\eta^{\alpha\beta}\partial_{\sigma}\partial_{\rho} \partial_{\alpha}\partial_{\beta}[\Omega^{-2}(x)K_{\mu\nu}]=0$ as given in  (\ref{B3}) are lightlike), and as $t^2$ for the timelike case. With the $ds^2=0$ light cone being both general coordinate invariant and conformal invariant, lightlike modes associated with the (\ref{A14}) metric will transform into lightlike modes associated with the metric (\ref{A4}). A $t^4$ growth for lightlike modes is a quite substantial growth rate, a growth rate that is not achievable in standard Einstein gravity if one uses the same radiation matter source. 

Since in transforming from one coordinate system to another the transformation is effected by 
\begin{eqnarray}
K_{\mu\nu}=\frac{\partial x^{\prime \alpha}}{\partial x^{\mu}}\frac{\partial x^{\prime\beta}}{\partial x^{\nu}}K^{\prime}_{\alpha\beta},
\label{B8}
\end{eqnarray}
the transformations between the $(p^{\prime}, r^{\prime})$, $(p,\chi)$ and $(t,r)$ coordinate systems are effected by
\begin{eqnarray}
\frac{\partial p^{\prime }}{\partial p}=\frac{\partial r^{\prime }}{\partial \chi}=\frac{1+\cosh p\cosh\chi}{[\cosh p+\cosh\chi]^2},\qquad
\frac{\partial p^{\prime }}{\partial \chi}=\frac{\partial r^{\prime }}{\partial p}=-\frac{\sinh p\sinh\chi}{[\cosh p+\cosh\chi]^2},\qquad \frac{\partial p}{\partial t}=\frac{1}{La(t)},\qquad \frac{\partial \chi}{\partial r}=\frac{1}{L\cosh\chi}.
\label{B9}
\end{eqnarray}

Discussion of (\ref{B9}) depends on whether $p=\chi$ or $p\gg \chi$. Since according to (\ref{B1}) $t=d\sinh p$ when $La(t)=(d^2+t^2)^{1/2}$, at large $t$ we see that  when $p=\chi$ (i.e. both $p$ and $\chi$ then being large) we have 
\begin{eqnarray}
\frac{\partial p^{\prime }}{\partial p}=\frac{\partial r^{\prime }}{\partial \chi}=1,\qquad
\frac{\partial p^{\prime }}{\partial \chi}=\frac{\partial r^{\prime }}{\partial p}=1,\qquad \frac{\partial p}{\partial t}=\frac{1}{t},\qquad \frac{\partial \chi}{\partial r}=\frac{d}{Lt}.
\label{B10}
\end{eqnarray}
Thus going from (\ref{A13}) to (\ref{A10}) ($(p^{\prime},r^{\prime})$ to $(p,\chi)$) will involve no suppression. Then in going from (\ref{A10}) to the comoving (\ref{A4}) ($(p,\chi)$ to $(t,r)$) we will get a $1/t^2$ suppression in the $K_{tt}$, $K_{tr}$ and $K_{rr}$ sectors, a $1/t$ suppression for $K_{t\theta}$, $K_{t\phi}$, $K_{r\theta}$ and $K_{r\phi}$, and no suppression for $K_{\theta\theta}$, $K_{\theta\phi}$ and $K_{\phi\phi}$ since in this case the prefactor in (\ref{B7}) behaves as $t^0$. Finally, incorporating the $t^4$ dependence of $\Omega^2(p,\chi)p^{\prime}$, we find that overall $K_{tt}$, $K_{tr}$ and $K_{rr}$  grow as $t^2$, $K_{t\theta}$, $K_{t\phi}$, $K_{r\theta}$ and $K_{r\phi}$ grow as $t^3$, and $K_{\theta\theta}$, $K_{\theta\phi}$ and $K_{\phi\phi}$ grow as $t^4$. Thus all seven of the $K_{tt}$, $K_{tr}$, $K_{rr}$, $K_{t\theta}$, $K_{t\phi}$, $K_{r\theta}$ and $K_{r\phi}$ are suppressed with respect to $K_{\theta\theta}$, $K_{\theta\phi}$ and $K_{\phi\phi}$, so the leading growth will be the $t^4$ growth associated with the angular $K_{\theta\theta}$, $K_{\theta\phi}$ and $K_{\phi\phi}$.

Similarly,  at large $p\gg \chi$ the transformations behave as 
\begin{eqnarray}
\frac{\partial p^{\prime }}{\partial p}=\frac{\partial r^{\prime }}{\partial \chi}=\frac{d\cosh\chi}{t},\qquad
\frac{\partial p^{\prime }}{\partial \chi}=\frac{\partial r^{\prime }}{\partial p}=-\frac{d\sinh\chi}{t},\qquad \frac{\partial p}{\partial t}=\frac{1}{t},\qquad \frac{\partial \chi}{\partial r}=\frac{1}{L\cosh\chi}.
\label{B11}
\end{eqnarray}
Thus in going from (\ref{A13}) to (\ref{A10}) we will get a $1/t^2$ suppression in the $K_{pp}$, $K_{p\chi}$ and $K_{\chi\chi}$ sectors, a $1/t$ suppression in the $K_{p\theta}$, $K_{p\phi}$, $K_{\chi\theta}$, and $K_{\chi\phi}$ sectors, and no suppression in the $K_{\theta\theta}$, $K_{\theta  \phi}$, $K_{\phi\phi}$ sectors. Then in going from (\ref{A10}) to the comoving (\ref{A4}) we will get an additional $1/t^2$ suppression in the $K_{tt}$ sector,  and an additional  $1/t$ suppression in the $K_{tr}$, $K_{t\theta}$, $K_{t\phi}$ sectors. Thus in going from (\ref{A13}) to (\ref{A4}) we get a net $1/t^4$ suppression for  $K_{tt}$,  a net $1/t^3$ suppression for  $K_{tr}$, a net $1/t^2$ suppression for  $K_{rr}$, $K_{t\theta}$, $K_{t\phi}$, a net $1/t$ suppression for  $K_{r\theta}$ and $K_{r\phi}$, and no suppression for $K_{\theta\theta}$, $K_{\theta\phi}$ and $K_{\phi\phi}$. Finally, incorporating the $t^4$ dependence of $\Omega^2(p,\chi)p^{\prime}$ and including the prefactor in (\ref{B7}), we find that overall $K_{tt}\sim t^0$, $K_{tr}\sim t^1$, $K_{t\theta}\sim t^1$, $K_{t\phi}\sim t^1$, while $K_{rr}\sim t^2$, $K_{r\theta}\sim t^2$, $K_{r\phi}\sim t^2$,  $K_{\theta\theta}\sim t^2$, $K_{\theta\phi}\sim t^2$ and $K_{\phi\phi}\sim t^2$. Thus the leading growth will grow as $t^2$. To understand this pattern we note that when $\chi$ is negligible then so is $r$, and the spatial part of the metric in (\ref{A4}) effectively becomes flat. Consequently,  $K_{tr}$, $K_{t\theta}$ and $K_{t\phi}$ all then have the same time behavior (viz. $t^1$), and $K_{rr}$, $K_{r\theta}$, $K_{r\phi}$,  $K_{\theta\theta}$, $K_{\theta\phi}$ and $K_{\phi\phi}$  all have the same time behavior (viz. $t^2$), with $t^2$ being the leading growth.

To compare with the results obtained in \cite{Mannheim2012a}, we note that there we imposed a  transverse gauge condition and had restricted to modes that obeyed the synchronous condition $K_{0\mu}=0$. We had worked in a spherical polar coordinate basis for the modes, and in the angular sector had obtained 
\begin{eqnarray}
K_{\theta\theta}(t,r,\theta,\phi)&=&L^2a^2(t)\cosh^2[(p+\chi)/2]\cosh^2[(p-\chi)/2]
\nonumber \\
&\times&\bigg{[}2\alpha^{(-)}_{\theta\theta}+\beta^{(-)}_{\theta\theta}\tanh[(p+\chi)/2]+\beta^{(-)}_{\theta\theta}{\rm tanh}[(p-\chi)/2]\bigg{]}
\nonumber\\
&\times&\bigg{[}\tanh[(p+\chi)/2]-\tanh[(p-\chi)/2]\bigg{]}\frac{\exp[-iq(p^{\prime}-r^{\prime})]}{\sin^2\theta},
\label{B12}
\end{eqnarray}
where $\alpha^{(-)}_{\theta\theta}$ and $\beta^{(-)}_{\theta\theta}$ are constants and $q$ is the radial momentum of the mode. For our purposes here we note that is more convenient to solve the radial equation not in terms of $e^{iqr^{\prime}}$ functions, but in terms of spherical Bessel functions instead, with the relevant one being $j_0(qr^{\prime})=\sin(qr^{\prime})/qr^{\prime}$. We thus replace (\ref{B12}) by
\begin{eqnarray}
K_{\theta\theta}(t,r,\theta,\phi)&=&L^2a^2(t)\cosh^2[(p+\chi)/2]\cosh^2[(p-\chi)/2]
\nonumber \\
&\times&\bigg{[}2\alpha^{(-)}_{\theta\theta}+\beta^{(-)}_{\theta\theta}\tanh[(p+\chi)/2]+\beta^{(-)}_{\theta\theta}{\rm tanh}[(p-\chi)/2]\bigg{]}
\nonumber\\
&\times&\bigg{[}\tanh[(p+\chi)/2]-\tanh[(p-\chi)/2]\bigg{]}\frac{\exp[-iqp^{\prime}]}{\sin^2\theta}qr^{\prime}j_0(qr^{\prime}),
\label{B13}
\end{eqnarray}
With $r^{\prime}$ being given by $r^{\prime}=(\tanh[(p+\chi)/2]-\tanh[(p-\chi)/2])/2=\sinh\chi/(\cosh p+\cosh \chi)$,  we can thus set: 
\begin{eqnarray}
K_{\theta\theta}(t,r,\theta,\phi)&=&L^2a^2(t)\bigg{[}\alpha^{(-)}_{\theta\theta}
+\beta^{(-)}_{\theta\theta}\frac{\sinh p}{\cosh p +\cosh \chi}\bigg{]}\sinh^2\chi\frac{\exp[-iqp^{\prime}]}{\sin^2\theta}qj_0(qr^{\prime}).
\label{B14}
\end{eqnarray}
Thus, for the lightlike $p=\chi$ case (where $r^{\prime}\rightarrow 1/2$ at large $p$) $K_{\theta\theta}(t,r,\theta,\phi)$ grows like $t^4$, while for the timelike $p\gg \chi$ case (where $r^{\prime}\rightarrow 0$ at large $p$) $K_{\theta\theta}(t,r,\theta,\phi)$ grows like $t^2$, just as found above. (The timelike $t^2$ growth had been incorrectly given in \cite{Mannheim2012a}.) Interestingly, in \cite{Mannheim2012a}  we had only considered a particular class of solutions, namely those with $K_{tt}=0$, $K_{tr}=0$, $K_{t\theta}=0$ and $K_{t\phi}=0$. We now see that for both $p=\chi$ and $p \gg \chi$, all of the modes that obey the synchronous condition $K_{0\mu}=0$  are non-leading, with the angular modes respectively being leading or coleading in the two cases.

\subsection{Comparison with SVT}

It is of interest to see how the behavior in time that we have found  is compatible with the SVT approach. As we see from (\ref{AP75}), the various gauge invariant combinations that appear in it appear with differing orders of time derivatives. $E_{ij}$ is the only one that appears with a fourth-order time derivative, and is the only one that obeys exactly the same equation as $k_{\mu\nu}=\Omega^{-2}K_{\mu\nu}$ does in in (\ref{AP61}). Now according to (\ref{AP66}), the relation between $E_{ij}$ and $k_{ij}$ is quite complicated. However if we work in the gauge $\partial_{\nu}k^{\mu\nu}=0$ and then set $k_{0 \mu}=0$, we find that  $k_{ij}$ obeys the transverse condition $\partial_{j}k^{ij}=0$. Under these conditions we find that $2E_{ij}=k_{ij}$, with both quantities being both transverse and traceless with respect to the spatial part of the metric.  

Now while one is not free to impose  the synchronous gauge condition $k_{0\mu}=0$ in the equations of motion since we have already used up the gauge freedom in  order to set  $\partial_{\nu}k^{\mu\nu}=0$, we can still look for solutions to the equations of motion that obey $k_{0\mu}=0$. Since we have shown that for a radiation-fluid-dominated  early Universe conformal Robertson-Walker cosmology all four  $k_{0\mu}$ components become non-leading at late times, at late times the solution becomes synchronous, and it is legitimate to drop the four $k_{0\mu}$ components of $k_{\mu\nu}$.

Now even with $k_{0\mu}=0$, $k_{ij}$ would still contain six components as constrained by $\eta^{\mu\nu}k_{\mu\nu}=-k_{00}+\delta^{ij}k_{ij}=\delta^{ij}k_{ij}=0$. In comparison, $E_{ij}$ only contains two independent components. However, in polar coordinates $k_{rr}$, $k_{r\theta}$ and $k_{r\phi}$ are also non-leading in the solution in the lightlike $p=\chi$ case. The leading part of the traceless $k_{ij}$ thus reduces to just the two independent components $k_{\theta\phi}$ and $k_{\phi\phi}=-\sin^2\theta k_{\theta\theta}$ in the lightlike case, and they can be identified with the two-component $E_{ij}$, just as required. Since we have introduced an overall factor of $\Omega^2(x)$ in defining the SVT basis in (\ref{AP62}), it is the two independent components  of $\Omega^2E_{ij}$ that are equivalent to the two leading components of $K_{ij}=\Omega^2k_{ij}$ in the lightlike case, as they all grow as a leading $t^4$ in the early Universe.

For completeness we note that if we consider an era that is cosmological constant dominated, then with $k<0$ the relevant solution is $a(t,\alpha>0,k<0)=\sinh(\alpha^{1/2}t)/L\alpha^{1/2}$. For this solution we obtain 
\begin{eqnarray}
Lp=\tau=L\alpha^{1/2}\int_{t_i}^t \frac{dt}{\sinh(\alpha^{1/2}t)}
=L \log \tanh(\alpha^{1/2}t/2)\bigg{|}_{t_i}^t,\qquad e^p=\frac{\tanh(\alpha^{1/2}t/2)}{\tanh(\alpha^{1/2}t_i/2)},
\label{B15}
\end{eqnarray}
as normalized so that $p=0$ at a convenient initial time $t_i$. The $\Omega^2p^{\prime}$ contribution to the fluctuations thus grows as 
\begin{eqnarray}
L^2a^2(p)\sinh p (\cosh p+\cosh \chi)
&=&\frac{1}{4\alpha}\sinh^2(\alpha^{1/2}t)\left[\frac{\tanh(\alpha^{1/2}t/2)}{\tanh(\alpha^{1/2}t_i/2)}-\frac{\tanh(\alpha^{1/2}t_i/2)}{\tanh(\alpha^{1/2}t/2)}\right]
\nonumber\\
&&\times\left[\frac{\tanh(\alpha^{1/2}t/2)}{\tanh(\alpha^{1/2}t_i/2)}+\frac{\tanh(\alpha^{1/2}t_i/2)}{\tanh(\alpha^{1/2}t/2)}+2\left(1+\frac{r^2}{L^2}\right)^{1/2}\right],
\label{B16}
\end{eqnarray}
i.e.  as $\exp(2\alpha^{1/2}t)$ at large $t$, while the transformation factors behave as 
\begin{eqnarray}
\frac{\partial p^{\prime }}{\partial p}=\frac{\partial r^{\prime }}{\partial \chi}=1,\qquad
\frac{\partial p^{\prime }}{\partial \chi}=\frac{\partial r^{\prime }}{\partial p}=-1,\qquad \frac{\partial p}{\partial t}=\frac{\alpha^{1/2}}{\sinh(\alpha^{1/2}t)},\qquad \frac{\partial \chi}{\partial r}=\frac{1}{L\cosh\chi}
\label{B17}
\end{eqnarray}
in the lightlike case. When $t \rightarrow\infty$ the $\Omega^2p^{\prime}$ contribution grows as $\exp(2\alpha^{1/2}t)$. Hence for the overall growth in the lighlike case the components behave as $K_{tt}\sim t^0$, $K_{tr}\sim \exp(\alpha^{1/2}t)$, $K_{t\theta}\sim \exp(\alpha^{1/2}t)$, $K_{t\phi}\sim \exp(\alpha^{1/2}t)$, with the six other  $K_{rr}$, $K_{r\theta}$, $K_{r\phi}$, $K_{\theta\theta}$, $K_{\theta\phi}$ and $K_{\phi\phi}$ components behaving as a leading $\exp(2\alpha^{1/2}t)$ at late times, i.e. growing as rapidly as the background itself, just as is found in standard Einstein gravity for radiative modes in a cosmological constant dominated $k=0$ cosmology.

\section{SVT Decomposition of  $\delta G_{\mu\nu}$ with General $\Omega(x)$}
\label{SC}

With a general conformal to flat Minkowski background plus fluctuation metric being of the form $ds^2=\Omega^2(x)[dt^2-\delta_{ij}dx^idx^j]-\Omega^2(x)f_{\mu\nu}dx^{\mu}dx^{\nu}$, then with $\tilde{\nabla}_{a}$ denoting the covariant derivative with respect to  $\delta_{ij}$, the fluctuation in the Einstein tensor in the SVT basis associated with (\ref{AP62}) is given by 
\begin{eqnarray}
&&\delta G_{00}=6 \Omega^{-1} \dot{\Omega} \dot{\psi} + 4 \Omega^{-1} \tilde{\nabla}_{a}\dot{\Omega} \tilde{\nabla}^{a}B - 2 \Omega^{-2} \dot{\Omega} \tilde{\nabla}_{a}\Omega \tilde{\nabla}^{a}B - 2 \Omega^{-1} \tilde{\nabla}_{a}\Omega \tilde{\nabla}^{a}\psi - 2 \Omega^{-2} \phi \tilde{\nabla}_{a}\Omega \tilde{\nabla}^{a}\Omega - 2 \Omega^{-2} \psi \tilde{\nabla}_{a}\Omega \tilde{\nabla}^{a}\Omega 
\nonumber\\
&&+ 2 \delta^{ab} \Omega^{-1} \dot{\Omega} \tilde{\nabla}_{b}\tilde{\nabla}_{a}B - 2 \delta^{ab} \Omega^{-1} \dot{\Omega} \tilde{\nabla}_{b}\tilde{\nabla}_{a}\dot{E} - 2 \delta^{ab} \tilde{\nabla}_{b}\tilde{\nabla}_{a}\psi + 4 \delta^{ab} \Omega^{-1} \phi \tilde{\nabla}_{b}\tilde{\nabla}_{a}\Omega + 4 \delta^{ab} \Omega^{-1} \psi \tilde{\nabla}_{b}\tilde{\nabla}_{a}\Omega 
\nonumber\\
&&+ 2 \Omega^{-2} \tilde{\nabla}^{a}\Omega \tilde{\nabla}_{b}\tilde{\nabla}_{a}E \tilde{\nabla}^{b}\Omega 
- 2 \delta^{bc} \Omega^{-1} \tilde{\nabla}^{a}\Omega \tilde{\nabla}_{c}\tilde{\nabla}_{b}\tilde{\nabla}_{a}E - 4 \delta^{ab} \delta^{cd} \Omega^{-1} \tilde{\nabla}_{c}\tilde{\nabla}_{a}E \tilde{\nabla}_{d}\tilde{\nabla}_{b}\Omega
+4 B^{a} \Omega^{-1} \tilde{\nabla}_{a}\dot{\Omega} - 2 B^{a} \Omega^{-2} \dot{\Omega} \tilde{\nabla}_{a}\Omega 
\nonumber\\
&&+ 2 \Omega^{-2} \tilde{\nabla}_{a}\Omega \tilde{\nabla}_{b}\Omega \tilde{\nabla}^{b}E^{a} - 4 \Omega^{-1} \tilde{\nabla}_{b}\tilde{\nabla}_{a}\Omega \tilde{\nabla}^{b}E^{a} - 2 \delta^{bc} \Omega^{-1} \tilde{\nabla}^{a}\Omega \tilde{\nabla}_{c}\tilde{\nabla}_{b}E_{a}-4 E^{ab} \Omega^{-1} \tilde{\nabla}_{b}\tilde{\nabla}_{a}\Omega + 2 E_{ab} \Omega^{-2} \tilde{\nabla}^{a}\Omega \tilde{\nabla}^{b}\Omega,
\label{C1}
\end{eqnarray}
\begin{eqnarray}
&&\delta G_{0i}=- \Omega^{-2} \dot{\Omega}^2 \tilde{\nabla}_{i}B + 2 \Omega^{-1} \ddot{\Omega} \tilde{\nabla}_{i}B + \Omega^{-2} \tilde{\nabla}_{a}\Omega \tilde{\nabla}^{a}\Omega \tilde{\nabla}_{i}B - 2 \delta^{ab} \Omega^{-1} \tilde{\nabla}_{b}\tilde{\nabla}_{a}\Omega \tilde{\nabla}_{i}B - 2 \Omega^{-1} \dot{\Omega} \tilde{\nabla}_{i}\phi - 2 \tilde{\nabla}_{i}\dot{\psi} 
\nonumber\\
&&+ 2 \Omega^{-1} \dot{\psi} \tilde{\nabla}_{i}\Omega 
- 2 \Omega^{-1} \tilde{\nabla}^{a}\Omega \tilde{\nabla}_{i}\tilde{\nabla}_{a}\dot{E}
- B_{i} \Omega^{-2} \dot{\Omega}^2 + 2 B_{i} \Omega^{-1} \ddot{\Omega} + \Omega^{-1} \tilde{\nabla}_{a}\Omega \tilde{\nabla}^{a}B_{i} -  \Omega^{-1} \tilde{\nabla}_{a}\Omega \tilde{\nabla}^{a}\dot{E}_{i} + B_{i} \Omega^{-2} \tilde{\nabla}_{a}\Omega \tilde{\nabla}^{a}\Omega 
\nonumber\\
&&+ \tfrac{1}{2} \delta^{ab} \tilde{\nabla}_{b}\tilde{\nabla}_{a}B_{i}
 -  \tfrac{1}{2} \delta^{ab} \tilde{\nabla}_{b}\tilde{\nabla}_{a}\dot{E}_{i} - 2 B_{i} \delta^{ab} \Omega^{-1} \tilde{\nabla}_{b}\tilde{\nabla}_{a}\Omega -  \Omega^{-1} \tilde{\nabla}_{a}\Omega \tilde{\nabla}_{i}B^{a} -  \Omega^{-1} \tilde{\nabla}_{a}\Omega \tilde{\nabla}_{i}\dot{E}^{a}
-2 \dot{E}_{ia} \Omega^{-1} \tilde{\nabla}^{a}\Omega,
\label{C2}
\end{eqnarray}
\begin{eqnarray}
&&\delta G_{ij}=2 \delta_{ij} \Omega^{-2} \dot{\Omega}^2 \phi - 4 \delta_{ij} \Omega^{-1} \ddot{\Omega} \phi - 2 \delta_{ij} \Omega^{-1} \dot{\Omega} \dot{\phi} + 2 \delta_{ij} \Omega^{-2} \dot{\Omega}^2 \psi - 4 \delta_{ij} \Omega^{-1} \ddot{\Omega} \psi - 4 \delta_{ij} \Omega^{-1} \dot{\Omega} \dot{\psi} - 2 \delta_{ij} \ddot{\psi} 
\nonumber\\
&&- 4 \delta_{ij} \Omega^{-1} \tilde{\nabla}_{a}\dot{\Omega} \tilde{\nabla}^{a}B 
+ 2 \delta_{ij} \Omega^{-2} \dot{\Omega} \tilde{\nabla}_{a}\Omega \tilde{\nabla}^{a}B - 2 \delta_{ij} \Omega^{-1} \tilde{\nabla}_{a}\Omega \tilde{\nabla}^{a}\dot{B} - 2 \delta_{ij} \Omega^{-1} \tilde{\nabla}_{a}\Omega \tilde{\nabla}^{a}\phi - 2 \Omega^{-1} \tilde{\nabla}_{a}\tilde{\nabla}_{j}\tilde{\nabla}_{i}E \tilde{\nabla}^{a}\Omega 
\nonumber\\
&&- 2 \delta^{ab} \delta_{ij} \Omega^{-1} \dot{\Omega} \tilde{\nabla}_{b}\tilde{\nabla}_{a}B 
-  \delta^{ab} \delta_{ij} \tilde{\nabla}_{b}\tilde{\nabla}_{a}\dot{B} + 2 \delta^{ab} \delta_{ij} \Omega^{-1} \dot{\Omega} \tilde{\nabla}_{b}\tilde{\nabla}_{a}\dot{E} + \delta^{ab} \delta_{ij} \tilde{\nabla}_{b}\tilde{\nabla}_{a}\ddot{E} -  \delta^{ab} \delta_{ij} \tilde{\nabla}_{b}\tilde{\nabla}_{a}\phi + \delta^{ab} \delta_{ij} \tilde{\nabla}_{b}\tilde{\nabla}_{a}\psi 
\nonumber\\
&&- 2 \delta_{ij} \Omega^{-2} \tilde{\nabla}^{a}\Omega \tilde{\nabla}_{b}\tilde{\nabla}_{a}E \tilde{\nabla}^{b}\Omega 
+ 2 \delta^{bc} \delta_{ij} \Omega^{-1} \tilde{\nabla}^{a}\Omega \tilde{\nabla}_{c}\tilde{\nabla}_{b}\tilde{\nabla}_{a}E + 4 \delta^{ab} \delta^{cd} \delta_{ij} \Omega^{-1} \tilde{\nabla}_{c}\tilde{\nabla}_{a}E \tilde{\nabla}_{d}\tilde{\nabla}_{b}\Omega + 2 \Omega^{-1} \tilde{\nabla}_{i}\Omega \tilde{\nabla}_{j}\psi 
\nonumber\\
&&+ 2 \Omega^{-1} \tilde{\nabla}_{i}\psi \tilde{\nabla}_{j}\Omega + 2 \Omega^{-1} \dot{\Omega} \tilde{\nabla}_{j}\tilde{\nabla}_{i}B 
+ \tilde{\nabla}_{j}\tilde{\nabla}_{i}\dot{B} - 2 \Omega^{-2} \dot{\Omega}^2 \tilde{\nabla}_{j}\tilde{\nabla}_{i}E + 4 \Omega^{-1} \ddot{\Omega} \tilde{\nabla}_{j}\tilde{\nabla}_{i}E + 2 \Omega^{-2} \tilde{\nabla}_{a}\Omega \tilde{\nabla}^{a}\Omega \tilde{\nabla}_{j}\tilde{\nabla}_{i}E 
\nonumber\\
&&- 4 \delta^{ab} \Omega^{-1} \tilde{\nabla}_{b}\tilde{\nabla}_{a}\Omega \tilde{\nabla}_{j}\tilde{\nabla}_{i}E - 2 \Omega^{-1} \dot{\Omega} \tilde{\nabla}_{j}\tilde{\nabla}_{i}\dot{E} 
-  \tilde{\nabla}_{j}\tilde{\nabla}_{i}\ddot{E} 
+ \tilde{\nabla}_{j}\tilde{\nabla}_{i}\phi -  \tilde{\nabla}_{j}\tilde{\nabla}_{i}\psi-4 B^{a} \delta_{ij} \Omega^{-1} \tilde{\nabla}_{a}\dot{\Omega} - 2 \dot{B}^{a} \delta_{ij} \Omega^{-1} \tilde{\nabla}_{a}\Omega 
\nonumber\\
&&+ 2 B^{a} \delta_{ij} \Omega^{-2} \dot{\Omega} \tilde{\nabla}_{a}\Omega - 2 \delta_{ij} \Omega^{-2} \tilde{\nabla}_{a}\Omega \tilde{\nabla}_{b}\Omega \tilde{\nabla}^{b}E^{a}  
+ 4 \delta_{ij} \Omega^{-1} \tilde{\nabla}_{b}\tilde{\nabla}_{a}\Omega \tilde{\nabla}^{b}E^{a} + 2 \delta^{bc} \delta_{ij} \Omega^{-1} \tilde{\nabla}^{a}\Omega \tilde{\nabla}_{c}\tilde{\nabla}_{b}E_{a} + \Omega^{-1} \dot{\Omega} \tilde{\nabla}_{i}B_{j} 
\nonumber\\
&&+ \tfrac{1}{2} \tilde{\nabla}_{i}\dot{B}_{j} -  \Omega^{-2} \dot{\Omega}^2 \tilde{\nabla}_{i}E_{j} + 2 \Omega^{-1} \ddot{\Omega} \tilde{\nabla}_{i}E_{j}  
+ \Omega^{-2} \tilde{\nabla}_{a}\Omega \tilde{\nabla}^{a}\Omega \tilde{\nabla}_{i}E_{j} - 2 \delta^{ab} \Omega^{-1} \tilde{\nabla}_{b}\tilde{\nabla}_{a}\Omega \tilde{\nabla}_{i}E_{j} -  \Omega^{-1} \dot{\Omega} \tilde{\nabla}_{i}\dot{E}_{j} -  \tfrac{1}{2} \tilde{\nabla}_{i}\ddot{E}_{j} 
\nonumber\\
&&+ \Omega^{-1} \dot{\Omega} \tilde{\nabla}_{j}B_{i} + \tfrac{1}{2} \tilde{\nabla}_{j}\dot{B}_{i} -  \Omega^{-2} \dot{\Omega}^2 \tilde{\nabla}_{j}E_{i}  
+ 2 \Omega^{-1} \ddot{\Omega} \tilde{\nabla}_{j}E_{i} + \Omega^{-2} \tilde{\nabla}_{a}\Omega \tilde{\nabla}^{a}\Omega \tilde{\nabla}_{j}E_{i} 
- 2 \delta^{ab} \Omega^{-1} \tilde{\nabla}_{b}\tilde{\nabla}_{a}\Omega \tilde{\nabla}_{j}E_{i} -  \Omega^{-1} \dot{\Omega} \tilde{\nabla}_{j}\dot{E}_{i} 
\nonumber\\
&&-  \tfrac{1}{2} \tilde{\nabla}_{j}\ddot{E}_{i} - 2 \Omega^{-1} \tilde{\nabla}^{a}\Omega \tilde{\nabla}_{j}\tilde{\nabla}_{i}E_{a}
- \ddot{E}_{ij} 
- 2 \dot{E}_{ij} \Omega^{-1} \dot{\Omega} 
- 2 E_{ij} \Omega^{-2} \dot{\Omega}^2 
+ 4 E_{ij} \Omega^{-1} \ddot{\Omega} + 2 \Omega^{-1} \tilde{\nabla}_{a}E_{ij} \tilde{\nabla}^{a}\Omega 
\nonumber\\
&&+ 2 E_{ij} \Omega^{-2} \tilde{\nabla}_{a}\Omega \tilde{\nabla}^{a}\Omega 
+ \delta^{ab} \tilde{\nabla}_{b}\tilde{\nabla}_{a}E_{ij} - 4 E_{ij} \delta^{ab} \Omega^{-1} \tilde{\nabla}_{b}\tilde{\nabla}_{a}\Omega 
+ 4 E^{ab} \delta_{ij} \Omega^{-1} \tilde{\nabla}_{b}\tilde{\nabla}_{a}\Omega - 2 E_{ab} \delta_{ij} \Omega^{-2} \tilde{\nabla}^{a}\Omega \tilde{\nabla}^{b}\Omega 
\nonumber\\
&&- 2 \Omega^{-1} \tilde{\nabla}^{a}\Omega \tilde{\nabla}_{i}E_{ja} - 2 \Omega^{-1} \tilde{\nabla}^{a}\Omega \tilde{\nabla}_{j}E_{ia}.
\label{C3}
\end{eqnarray}
In $\delta G_{00}$ there are 21 terms, in $\delta G_{0i}$ there are 19 terms, and in $\delta G_{ij}$ there are 68 terms. In comparison, as we see from (\ref{AP75}),  in $\delta W_{00}$ there are 4 terms, in $\delta W_{0i}$ there are 8 terms, and in $\delta W_{ij}$ there are 27 terms. And unlike $\delta W_{\mu\nu}$, for any $\Omega(x)$ not equal to one $\delta G_{\mu\nu}$ is not gauge invariant.

\section{Compact Expressions for $\delta G_{\mu\nu}$ in some Specific Gauges}
\label{SD}

For the conformal to flat Minkowski line element  $ds^2=- \Omega^2(x)(\eta_{\mu\nu}+f_{\mu\nu})dx^{\mu}dx^{\nu}$, where $\Omega(x)$ is an arbitrary function of $x_{\mu}$, the Einstein tensor fluctuation  $\delta G_{\mu\nu}$ is given by (\ref{AP76}). We introduce $k_{\mu\nu}=f_{\mu\nu}-(1/4)\eta^{\alpha\beta}f_{\alpha\beta}$ with $\eta^{\mu\nu}k_{\mu\nu}=0$, and rewrite (\ref{AP76}) as 
\begin{eqnarray}
\delta G_{\mu\nu}&=&- \tfrac{1}{4} \eta^{\alpha \beta} \eta_{\mu \nu} \Omega^{-1} \partial_{\alpha}\Omega \partial_{\beta}f + \eta^{\alpha \beta} \Omega^{-1} \partial_{\alpha}k_{\mu \nu} \partial_{\beta}\Omega + \eta^{\beta \alpha} k_{\mu \nu} \Omega^{-2} \partial_{\alpha}\Omega \partial_{\beta}\Omega + \tfrac{1}{2} \eta^{\alpha \beta} \partial_{\beta}\partial_{\alpha}k_{\mu \nu} -  \tfrac{1}{4} \eta^{\alpha \beta} \eta_{\mu \nu} \partial_{\beta}\partial_{\alpha}f 
\nonumber\\
&-& 2 \eta^{\alpha \beta} k_{\mu \nu} \Omega^{-1} \partial_{\beta}\partial_{\alpha}\Omega -  \tfrac{1}{2} \eta^{\alpha \beta} \partial_{\beta}\partial_{\mu}k_{\nu \alpha} -  \tfrac{1}{2} \eta^{\alpha \beta} \partial_{\beta}\partial_{\nu}k_{\mu \alpha} + 2 \eta^{\alpha \beta} \eta^{\gamma \kappa} \eta_{\mu \nu} \Omega^{-1} \partial_{\beta}\Omega \partial_{\kappa}k_{\alpha \gamma} 
\nonumber\\
&-&  \eta^{\alpha \gamma} \eta^{\beta \kappa} \eta_{\mu \nu} k_{\alpha \beta} \Omega^{-2} \partial_{\gamma}\Omega \partial_{\kappa}\Omega + \tfrac{1}{2} \eta^{\alpha \beta} \eta^{\gamma \kappa} \eta_{\mu \nu} \partial_{\kappa}\partial_{\beta}k_{\alpha \gamma} + 2 \eta^{\alpha \beta} \eta^{\gamma \kappa} \eta_{\mu \nu} k_{\alpha \gamma} \Omega^{-1} \partial_{\kappa}\partial_{\beta}\Omega -  \eta^{\alpha \beta} \Omega^{-1} \partial_{\beta}\Omega \partial_{\mu}k_{\nu \alpha} 
\nonumber\\
&-&  \eta^{\alpha \beta} \Omega^{-1} \partial_{\beta}\Omega \partial_{\nu}k_{\mu \alpha} -  \tfrac{1}{4} \Omega^{-1} \partial_{\mu}\Omega \partial_{\nu}f -  \tfrac{1}{4} \Omega^{-1} \partial_{\mu}f \partial_{\nu}\Omega + \tfrac{1}{4} \partial_{\nu}\partial_{\mu}f,
\label{D1}
\end{eqnarray}
where $f$ denotes $\eta^{\alpha\beta}f_{\alpha\beta}$. We have considered gauges of the form:
\begin{eqnarray}
\eta^{\alpha\beta}\partial_{\alpha}k_{\beta\nu} = \Omega^{-1} J \eta^{\alpha\beta}k_{\nu\alpha}\partial_\beta \Omega + P \partial_\nu f+ R \Omega^{-1} f\partial_\nu \Omega,
\label{D2}
\end{eqnarray}
where $J$, $P$, and $R$ are constants. For various choices of these parameters we can simplify the structure of $\delta G_{\mu\nu}$, and on taking  $J = -2$, $P = 1/2$, $R = 0$, viz. on setting
\begin{eqnarray}
\eta^{\alpha\beta}\partial_{\alpha}k_{\beta\nu} = -2 \Omega^{-1}  \eta^{\alpha\beta}k_{\nu\alpha}\partial_\beta \Omega + \tfrac{1}{2} \partial_\nu f,
\label{D3}
\end{eqnarray}
we find that when $\Omega$ is only a function of the conformal time $\tau$, $\delta G_{\mu\nu}$ evaluates  to 
\begin{eqnarray}
\delta G_{00}&=&
(3 \Omega^{-2} \dot{\Omega}^2 -  \Omega^{-1} \ddot{\Omega} + \tfrac{1}{2} \eta^{\mu \nu} \partial_{\mu} \partial_{\nu} -  \Omega^{-1} \dot{\Omega} \partial_{0}) k_{00} - \tfrac{1}{4} ( \Omega^{-1} \dot{\Omega} \partial_{0} + \partial_{0} \partial_{0}) f,
\nonumber\\
\delta G_{0i}&=&
(\Omega^{-1} \ddot{\Omega} + \tfrac{1}{2} \eta^{\mu \nu} \partial_{\mu} \partial_{\nu} -  \Omega^{-1} \dot{\Omega} \partial_{0}) k_{0i} - \tfrac{1}{4}  (\Omega^{-1} \dot{\Omega} \partial_{i} +\partial_{i} \partial_{0}) f,
\nonumber\\
\delta G_{ij}&=&
\delta_{ij}(-2 \Omega^{-2} \dot{\Omega}^2 + \Omega^{-1} \ddot{\Omega}) k_{00} + (- \Omega^{-2} \dot{\Omega}^2 + 2 \Omega^{-1} \ddot{\Omega} + \tfrac{1}{2} \eta^{\mu \nu} \partial_{\mu} \partial_{\nu} -  \Omega^{-1} \dot{\Omega} \partial_{0}) k_{ij}, 
\nonumber\\
&-& \tfrac{1}{4} (\delta_{ij} \Omega^{-1} \dot{\Omega} \partial_{0} + \partial_{i} \partial_{j}) f,
\nonumber\\
\eta^{\mu\nu}\delta G_{\mu\nu}&=&(-10\Omega^{-2} \dot{\Omega}^2 +6  \Omega^{-1} \ddot{\Omega})k_{00}- \tfrac{1}{4}  (2\Omega^{-1} \dot{\Omega} \partial_{0} +\eta^{\mu\nu}\partial_{\mu}\partial_{\nu}) f,
\label{D4}
\end{eqnarray}
where the dot denotes the derivative with respect to $\tau$, and $\partial_0$ denotes $\partial_{\tau}$. While not diagonal in the $(\mu,\nu)$ indices, we note that $\delta G_{\mu\nu}$ is close to being so, since apart from  the $k_{00}$ and $f$ dependence  $\delta G_{\mu\nu}$ otherwise would be. Moreover, the ten fluctuation equations contained in  $\delta G_{\mu\nu}=-8\pi G \delta T_{\mu\nu}$ can be solved completely once $\delta T_{\mu\nu}$ is specified, since from the $\delta G_{00}$ and $\eta^{\mu\nu}\delta G_{\mu\nu}$ equations one can determine $k_{00}$ and $f$, and then from the other $\delta G_{\mu\nu}$ equations one can determine all the other components of $k_{\mu\nu}$.

For completeness we note that if the background is the inflationary universe de Sitter geometry where $\Omega(\tau)=1/H\tau$ with $H$ constant, (\ref{D4}) takes the form
\begin{eqnarray}
\delta G_{00}&=&
(\tau^{-2} + \tfrac{1}{2} \eta^{\mu \nu} \partial_{\mu} \partial_{\nu} + \tau^{-1} \partial_{0}) k_{00} + \tfrac{1}{4} (\tau^{-1} \partial_{0} -   \partial_{0} \partial_{0}) f,
\nonumber\\
\delta G_{0i}&=&
(2 \tau^{-2} + \tfrac{1}{2} \eta^{\mu \nu} \partial_{\mu} \partial_{\nu} + \tau^{-1} \partial_{0}) k_{0i} + \tfrac{1}{4} ( \tau^{-1} \partial_{i} -  \partial_{i} \partial_{0}) f,
\nonumber\\
\delta G_{ij}&=&
(3 \tau^{-2} + \tfrac{1}{2} \eta^{\mu \nu} \partial_{\mu} \partial_{\nu} + \tau^{-1} \partial_{0}) k_{ij} + \tfrac{1}{4} (\delta_{ij} \tau^{-1} \partial_{0} -   \partial_{i} \partial_{j}) f,
\nonumber\\
\eta^{\mu\nu}\delta G_{\mu\nu}&=&2\tau^{-2}k_{00}+ \tfrac{1}{4}  (2\tau^{-1}\partial_{0} -\eta^{\mu\nu}\partial_{\mu}\partial_{\nu}) f.
\label{D5}
\end{eqnarray}
As we see, by incorporating $\Omega(\tau)$ into both the background and the fluctuation we obtain very compact forms for $\delta G_{\mu\nu}$  in (\ref{D4}) and (\ref{D5}), with the $\delta G_{\mu\nu}=-8\pi G \delta T_{\mu\nu}$ fluctuation equations then being completely integrable.

We have found one further convenient decomposition of $\delta G_{\mu\nu}$, namely the gauge choice $J=-4$, $R=2P-3/2$, $P$ arbitrary. For a de Sitter background this gauge choice leads to
\begin{eqnarray}
\delta G_{00}&=&(-2 \tau^{-2}
 + \tfrac{1}{2} \eta^{\mu \nu} \partial_{\mu} \partial_{\nu}
 + 3 \tau^{-1} \partial_{0}) k_{00}
 + \left[(\tfrac{3}{4} -  P) \tau^{-2}
 + \tfrac{1}{4}(1-2P) \eta^{\mu \nu} \partial_{\mu} \partial_{\nu}
 + P \tau^{-1} \partial_{0}
 +( \tfrac{1}{4} -P)\partial^2_{0}\right]f,
\nonumber\\
\delta G_{0i}&=&\tau^{-1} \partial_{i} k_{00}
 + (\tau^{-2}
 + \tfrac{1}{2} \eta^{\mu \nu} \partial_{\mu} \partial_{\nu}
 + 2 \tau^{-1} \partial_{0}) k_{0i}
 + \left[(P- \tfrac{1}{2}) \tau^{-1} \partial_{i}
 + (\tfrac{1}{4}-P) \partial_{i} \partial_{0} \right]f,
\nonumber\\
\delta G_{ij}&=&\delta_{ij}\tau^{-2} k_{00}
 +\tau^{-1} \partial_{j} k_{0i}
 + \tau^{-1} \partial_{i} k_{0j}
 + (3 \tau^{-2}
 + \tfrac{1}{2} \eta^{\mu \nu} \partial_{\mu} \partial_{\nu}
 + \tau^{-1} \partial_{0}) k_{ij}
  \nonumber\\
 &+& \delta_{ij}\left[(\tfrac{3}{4}-P) \tau^{-2}
 + \tfrac{1}{4}(2P-1) \eta^{\mu \nu} \partial_{\mu} \partial_{\nu}
  +(P-1)\tau^{-1} \partial_{0}\right]f
 + (\tfrac{1}{4}-P) \partial_{i} \partial_{j}f.
\nonumber\\
\eta^{\alpha\beta}\delta G_{\alpha\beta}&=&(P-\tfrac{3}{4})( \eta^{\alpha\beta}\partial_{\alpha}\partial_{\beta}f +4\tau^{-1}\partial_0 f-6\tau^{-2}f)=(P-\tfrac{3}{4})\tau^2 \eta^{\alpha\beta}\partial_{\alpha}\partial_{\beta}(\tau^{-2}f).
\label{D6}
\end{eqnarray}
In this gauge $\eta^{\mu\nu}\delta G_{\mu\nu}=\Omega^2g^{\mu\nu}\delta G_{\mu\nu}$ depends on the trace $f$ of the fluctuation alone. One can immediately solve for $f$ and then proceed to the other components of the fluctuation in turn. As we see,  the quantity $\eta^{\alpha\beta}\delta G_{\alpha\beta}$ takes the form of the flat space free massless particle wave operator acting on $\tau^{-2} f$, with the equation $g^{\mu\nu}\delta G_{\mu\nu}=-8\pi G g^{\mu\nu}\delta T_{\mu\nu}$ immediately being integrable with the $D^{(4)}(x-y)$ propagator that obeys $\eta^{\alpha\beta}\partial_{\alpha}\partial_{\beta}D^{(4)}(x-y)=\delta^4(x-y)$ as
\begin{eqnarray}
f=\eta^{\mu\nu}f_{\mu\nu}=-\frac{8\pi G}{(P-\tfrac{3}{4})}\tau^2(x) \int d^4yD^{(4)}(x-y)\tau^{-2}(y)\eta^{\mu\nu}\delta T_{\mu\nu}(y),
\label{D7}
\end{eqnarray}
where $\tau(x)=x^0$, $\tau(y)=y^0$.

\section{First Principles Derivation of the SVT Decomposition by Projection}
\label{SE}

\subsection{The $3+1$ Decomposition}

The standard covariant $3+1$ decomposition of a symmetric rank two tensor $T_{\mu\nu}$ in a 4-dimensional geometry with metric $g_{\mu\nu}$ is constructed by introducing a 4-vector $U^{\mu}$ that obeys $g_{\mu\nu}U^{\mu}U^{\nu}=-1$ and a projector 
\begin{eqnarray}
P_{\mu\nu}=g_{\mu\nu}+U_{\mu}U_{\nu}
\label{E1}
\end{eqnarray}
that obeys
\begin{eqnarray}
U_{\mu}P^{\mu\nu}=0, \qquad P_{\mu\nu}P^{\mu\nu}=g_{\mu\nu}P^{\mu\nu}=3,\qquad P_{\mu\sigma}P^{\sigma}_{\phantom{\sigma}\nu}=P_{\mu\nu}.
\label{E2}
\end{eqnarray}
In terms of the projector we can write
\begin{eqnarray}
T_{\mu\nu}=g_{\mu}^{\phantom{\mu}\sigma}g_{\nu}^{\phantom{\nu}\tau}T_{\sigma\tau}=
P_{\mu}^{\phantom{\mu}\sigma}P_{\nu}^{\phantom{\nu}\tau}T_{\sigma\tau}
-U_{\mu}U^{\sigma}P_{\nu}^{\phantom{\nu}\tau}T_{\sigma\tau}
-P_{\mu}^{\phantom{\mu}\sigma}U_{\nu}U^{\tau}T_{\sigma\tau}
+U_{\mu}U_{\nu}U^{\sigma}U^{\tau}T_{\sigma\tau}.
\label{E3}
\end{eqnarray}
On introducing
\begin{eqnarray}
\rho&=&U^{\sigma}U^{\tau}T_{\sigma\tau},\qquad p=\frac{1}{3}P^{\sigma\tau}T_{\sigma\tau},\qquad 
q_{\mu}=-P_{\mu}^{\phantom{\mu}\sigma}U^{\tau}T_{\sigma\tau},
\nonumber\\
\pi_{\mu\nu}&=&\left[\frac{1}{2}P_{\mu}^{\phantom{\mu}\sigma}P_{\nu}^{\phantom{\nu}\tau}
+\frac{1}{2}P_{\nu}^{\phantom{\nu}\sigma}P_{\mu}^{\phantom{\mu}\tau}
-\frac{1}{3}P_{\mu\nu}P^{\sigma\tau}\right]T_{\sigma\tau},
\label{E4}
\end{eqnarray}
which obey
\begin{eqnarray}
U^{\mu}q_{\mu}=0,\qquad U^{\nu}\pi_{\mu\nu}=0,\qquad \pi_{\mu\nu}=\pi_{\nu\mu},\qquad g^{\mu\nu}\pi_{\mu\nu}=P^{\mu\nu}\pi_{\mu\nu}=0,
\label{E5}
\end{eqnarray}
we can rewrite $T_{\mu\nu}$ as
\begin{eqnarray}
T_{\mu\nu}=(\rho+p)U_{\mu}U_{\nu}+pg_{\mu\nu}+U_{\mu}q_{\nu}+U_{\nu}q_{\mu}+\pi_{\mu\nu},
\label{E6}
\end{eqnarray}
a familiar form that may for instance be found in \cite{Ellis1971}. As constructed, the ten-component $T_{\mu\nu}$ has been covariantly decomposed into two one-component 4-scalars, one three-component  4-vector that is orthogonal to $U_{\mu}$ and one five-component traceless, rank two tensor that is also orthogonal  to $U_{\mu}$.

\subsection{Transverse and Longitudinal Components for the Vector Sector}

While the form of (\ref{E6}) has same $3+1$ structure as (\ref{AP62}), we need to bring $q_{\mu}$ and $\pi_{\mu\nu}$ to a form that involves components that are transverse and longitudinal.  To this end we take the background metric to be flat Minkowski,  we take $U^{\mu}$ to be the unit timelike vector $U^{\mu}=(1,0,0,0)$, we set $g_{\mu\nu}=\eta_{\mu\nu}+h_{\mu\nu}$, and look to put (\ref{E6}) into the form given in (\ref{AP62}). We first consider $\Omega(x)=1$ and then conformally transform afterwards.

For $q_{\mu}$ the form  for $U^{\mu}$ then requires that $q_0$ be zero, so that $q_{\mu}=(0,q_i)$. With the general $h_{0i}$ not obeying $\nabla^{i}h_{0i}=0$, we can immediately introduce 
\begin{eqnarray}
B=\int d^3yD^{(3)}(x-y)\tilde{\nabla}_y^ih_{0i}
\label{E7}
\end{eqnarray}
just as in (\ref{AP65}) and set
\begin{eqnarray}
h_{0i}=q_i=B_i+\tilde{\nabla}_iB,\qquad \tilde{\nabla}^iB_i=0.
\label{E8}
\end{eqnarray}
As noted in \cite{Mannheim2005}, the reason for introducing the Green's function term in (\ref{E7}) is that while any longitudinal vector can be written as the derivative $\tilde{\nabla}_i\phi$ of some scalar $\phi$, not every $\tilde{\nabla}_i\phi$ is in fact longitudinal since $\tilde{\nabla}_i\phi$ would be transverse if $\phi$ obeys $\tilde{\nabla}_i\tilde{\nabla}^i\phi=0$. Thus if we start with some general vector for which $\tilde{\nabla}^iV_i$ is non-zero, and define a scalar that obeys $\tilde{\nabla}_i\tilde{\nabla}^i\phi = \tilde{\nabla}^iV_i$, the quantity $\phi=\int d^3yD^{(3)}(x-y)\tilde{\nabla}_y^iV_i$ could never obey $\tilde{\nabla}_i\tilde{\nabla}^i\phi=0$ and would thus necessarily be longitudinal, while the quantity $V_i-\tilde{\nabla}_i\phi$ would automatically be transverse. That  breaking up a vector into transverse and longitudinal pieces would involve integrals of Green's functions may also be anticipated since in (\ref{AP65}) Green's function integrals appeared automatically when we wrote the SVT components of (\ref{AP62}) in terms of the $h_{ij}$. 

In regard to (\ref{E7}), it had been noted in \cite{Mannheim2005} that  for an arbitrarily  given $\tilde{\nabla}_y^ih_{0i}$ the integral that appears in (\ref{E7}) might not exist. However, by making a gauge transformation  of the form $h_{0i}\rightarrow \partial_0 \epsilon_i+\partial_i\epsilon_0$, $\partial_ih_{0i}\rightarrow \partial_i h_{0i}+\partial_0\partial_i\epsilon_i+\nabla^2\epsilon_0$, we can choose a gauge so that the integral in (\ref{E7}) then is finite.  And when we do this we can integrate by parts and set $\int d^3yD^{(3)}(x-y)\tilde{\nabla}_y^ih_{0i}=-\int d^3y[\tilde{\nabla}_y^iD^{(3)}(x-y)]h_{0i}=
+\tilde{\nabla}_x^i\int d^3yD^{(3)}(x-y)h_{0i}$. In the following then we shall take the vector-sector (\ref{E7}) integral  and its tensor-sector analogs given below to be finite, and can do so since making gauge transformations has no effect on the SVT decomposition of $\delta W_{\mu\nu}$ given in (\ref{AP75}), as it is written entirely in terms of gauge invariant quantities. Now if we had initially set $B=\int d^3yD^{(3)}(x-y)\tilde{\nabla}_y^ih_{0i}+f(t)+{\bf n}\cdot {\bf x} g(t)$ where ${\bf n}$ is an arbitrary spatially-independent 3-vector and $f(t)$ and $g(t)$ are arbitrary functions of time, it would still follow that $\tilde{\nabla}^iB_i=0$. However, in constructing (\ref{E7}) our starting point is that (up to  gauge transformations) $h_{i0}$ is already given, and our task is to decompose it into $\tilde{\nabla}_iB$ and $B_i$ components. This leads us to (\ref{E7}) with no extraneous $f(t)$ or ${\bf n}\cdot {\bf x} g(t)$ terms. 

\subsection{Transverse and Longitudinal Components for the Tensor Sector}

To decompose $\pi_{\mu\nu}$ we follow \cite{Mannheim2005} and introduce a set of transverse and transverse-traceless projectors. We first discuss the projection technique in a Minkowski or Cartesian flat space in an arbitrary $D$ dimensions, where one can define a Green's function that obeys
\begin{eqnarray}
\nabla_{\mu}\nabla^{\mu}D^{(D)}(x-y)=\delta^{D}(x-y).
\label{E9}
\end{eqnarray}
For a fluctuation $h_{\mu\nu}$ around a background $\eta_{\mu\nu}$ we introduce a $D$-dimensional vector $W_{\mu}$ according to 
\begin{eqnarray}
W_{\mu}=\int d^Dy D^{(D)}(x-y)
\nabla_y^{\nu}h_{\mu\nu},
\label{E10}
\end{eqnarray}
and can pick $D$ gauge conditions  so that each $\mu$ component of (\ref{E10}) and the integrals given below that involve $W_{\mu}$ can all  exist. (When $D=3$ there are three $W_i$, which together with $B$ as given in (\ref{E7}) requires exactly four gauge transformations in the four-dimensional spacetime associated with the SVT decomposition given in (\ref{AP62}).) With $\nabla^{\mu}W_{\mu}=\int d^Dy D^{(D)}(x-y)\nabla_y^{\mu}\nabla_y^{\nu}h_{\mu\nu}$ not in general being zero for a general $h_{\mu\nu}$, we construct the longitudinal and transverse
\begin{eqnarray}
h^L_{\mu\nu}=\nabla_{\mu}W_{\nu}+\nabla_{\nu}W_{\mu}-\nabla_{\mu}\nabla_{\nu}\int d^Dy D^{(D)}(x-y)\nabla_y^{\sigma}W_{\sigma},\qquad h^T_{\mu\nu}=h_{\mu\nu}-h^L_{\mu\nu},
\label{E11}
\end{eqnarray}
with $h^{L}_{\mu\nu}$ and $h^{T}_{\mu\nu}$  obeying 
\begin{eqnarray}
\nabla^{\nu}h^L_{\mu\nu}&=&\nabla^{\nu}\nabla_{\nu}W_{\mu}=\nabla^{\nu}h_{\mu\nu}, \qquad \nabla^{\nu}h^T_{\mu\nu}=0,\qquad g^{\mu\nu}h^{L}_{\mu\nu}=\nabla^{\mu}W_{\mu},
\nonumber\\
\nabla^{\mu}\nabla^{\nu}h_{\mu\nu}&=&\nabla^{\mu}\nabla^{\nu}h^{L}_{\mu\nu} =\nabla_{\alpha}\nabla^{\alpha}(g^{\mu\nu}h^{L}_{\mu\nu}).
\label{E12}
\end{eqnarray}

\subsection{Transverse-Traceless  and Longitudinal-Traceless Components for the Tensor Sector}

To extract out the part of $h^L_{\mu\nu}$ that is traceless in a way that does not affect the divergence structure of  $h^L_{\mu\nu}$ we introduce
\begin{eqnarray}
h^{L\theta}_{\mu\nu}= h^{L}_{\mu\nu}-\frac{1}{D-1}g_{\mu\nu}g^{\sigma\tau}h^{L}_{\sigma\tau}
+\frac{1}{D-1}\nabla_{\mu}\nabla_{\nu}\int d^Dy D^{(D)}(x-y)g^{\sigma\tau}h^{L}_{\sigma\tau}
\label{E13}
\end{eqnarray}
($\theta$ denotes zero trace), to find that not only does $h^{L\theta}_{\mu\nu}$ obey $g^{\mu\nu}h^{L\theta}_{\mu\nu}=0$,  it also obeys $\nabla^{\nu}h^{L\theta}_{\mu\nu}=\nabla^{\nu}h^{L}_{\mu\nu}$, since $h^{L\theta}_{\mu\nu}- h^{L}_{\mu\nu}$ is divergenceless.

To simplify $h^{L\theta}_{\mu\nu}$ we introduce a  $V_{\mu}$ that obeys
\begin{eqnarray}
V_{\mu}&=&W_{\mu}-\frac{(D-2)}{2(D-1)}\nabla_{\mu}\int d^Dy D^{(D)}(x-y)\nabla_y^{\nu}W_{\nu},
\qquad
\nabla^{\mu}V_{\mu}=\frac{D}{2(D-1)}\nabla^{\mu}W_{\mu},
\nonumber\\
W_{\mu}&=&V_{\mu}+\frac{(D-2)}{D}\nabla_{\mu}\int d^Dy D^{(D)}(x-y)\nabla_y^{\nu}V_{\nu},
\label{E14}
\end{eqnarray}
on recalling that  $\nabla^{\mu}W_{\mu}=g^{\mu\nu}h^{L}_{\mu\nu}$, we find that we can rewrite $h^{L\theta}_{\mu\nu}$ as
\begin{eqnarray}
h^{L\theta}_{\mu\nu}= \nabla_{\mu}V_{\nu}+\nabla_{\nu}V_{\mu}-\frac{2}{D}g_{\mu\nu}\nabla^{\sigma}V_{\sigma},
\label{E15}
\end{eqnarray}
a form that may also be found in \cite{York1973}. 

A similar analysis can be made for $h^T_{\mu\nu}$, with the quantity 
\begin{eqnarray}
h^{T\theta}_{\mu\nu}= h^{T}_{\mu\nu}-\frac{1}{D-1}g_{\mu\nu}g^{\sigma\tau}h^{T}_{\sigma\tau}
+\frac{1}{D-1}\nabla_{\mu}\nabla_{\nu}\int d^Dy D^{(D)}(x-y)g^{\sigma\tau}h^{T}_{\sigma\tau}
\label{E16}
\end{eqnarray}
being both transverse ($\nabla^{\mu}h^{T\theta}_{\mu\nu}=0$), and traceless ($g^{\mu\nu}h^{T\theta}_{\mu\nu}=0$).
The general $h_{\mu\nu}$ can thus be written as 
\begin{eqnarray}
h_{\mu\nu}=h^{T\theta}_{\mu\nu}+h^{L\theta}_{\mu\nu}+\frac{1}{D-1}g_{\mu\nu}g^{\sigma\tau}h_{\sigma\tau}
-\frac{1}{D-1}\nabla_{\mu}\nabla_{\nu}\int d^Dy D^{(D)}(x-y)g^{\sigma\tau}h_{\sigma\tau}.
\label{E17}
\end{eqnarray}

\subsection{The SVT Basis}

Since we took $W_{\mu}$ to not be divergenceless, $V_{\mu}$ would not be divergenceless either, and thus we can introduce a scalar $F$ that obeys
\begin{eqnarray}
\nabla_{\mu}\nabla^{\mu}F=\nabla^{\mu}V_{\mu},\qquad F=\int d^Dy D^{(D)}(x-y)\nabla_y^{\mu}V_{\mu},
\label{E18}
\end{eqnarray}
and decompose $V_{\mu}$ into $\nabla_{\mu}F$ and a divergenceless $F_{\mu}$ that is  given by
\begin{eqnarray}
F_{\mu}=V_{\mu}-\nabla_{\mu}F=V_{\mu}-\nabla_{\mu}\int d^Dy D^{(D)}(x-y)\nabla_y^{\nu}V_{\nu}=
W_{\mu}-\nabla_{\mu}\int d^Dy D^{(D)}(x-y)\nabla_y^{\nu}W_{\nu},\qquad \nabla^{\mu}F_{\mu}=0.
\label{E19}
\end{eqnarray}
We can thus rewrite $h^{L\theta}_{\mu\nu}$ as 
\begin{eqnarray}
h^{L\theta}_{\mu\nu}= \nabla_{\mu}F_{\nu}+\nabla_{\nu}F_{\mu}+2\nabla_{\mu}\nabla_{\nu}F-\frac{2}{D}g_{\mu\nu}\nabla_{\sigma}\nabla^{\sigma}F,
\label{E20}
\end{eqnarray}
and write $h_{\mu\nu}$ as 
\begin{eqnarray}
h_{\mu\nu}&=&h^{T\theta}_{\mu\nu}+ \nabla_{\mu}F_{\nu}+\nabla_{\nu}F_{\mu}+2\nabla_{\mu}\nabla_{\nu}F-\frac{2}{D}g_{\mu\nu}\nabla_{\sigma}\nabla^{\sigma}F
\nonumber\\
&+&\frac{1}{D-1}g_{\mu\nu}g^{\sigma\tau}h_{\sigma\tau}
-\frac{1}{D-1}\nabla_{\mu}\nabla_{\nu}\int d^Dy D^{(D)}(x-y)g^{\sigma\tau}h_{\sigma\tau}.
\label{E21}
\end{eqnarray}
On defining
\begin{eqnarray}
2\psi&=&\frac{2}{D}\nabla_{\mu}\nabla^{\mu}F-\frac{1}{(D-1)}g^{\sigma\tau}h_{\sigma\tau}
=\frac{1}{D-1}\int d^Dy D^{(D)}(x-y)\nabla_y^{\mu}\nabla_y^{\nu}h_{\mu\nu}-\frac{1}{(D-1)}g^{\sigma\tau}h_{\sigma\tau},
\nonumber\\
\qquad 2E&=&2F-\frac{1}{D-1}\int d^Dy D^{(D)}(x-y)g^{\sigma\tau}h_{\sigma\tau}
\nonumber\\
&=&\frac{D}{D-1}\int d^Dy D^{(D)}(x-y)\int d^Dz D^{(D)}(y-z)\nabla_z^{\mu}\nabla_z^{\nu}h_{\mu\nu} -\frac{1}{D-1}\int d^Dy D^{(D)}(x-y)g^{\sigma\tau}h_{\sigma\tau},
\nonumber\\
 E_{\mu}&=&F_{\mu}=\int d^Dy D^{(D)}(x-y)\nabla_y^{\nu}h_{\mu\nu}
 -\nabla^x_{\mu}\int d^Dy D^{(D)}(x-y)\nabla_y^{\alpha}\int d^Dz D^{(D)}(y-z)\nabla_z^{\beta}h_{\alpha\beta}
 \nonumber\\
 &=&\int d^Dy D^{(D)}(x-y)\nabla_y^{\nu}h_{\mu\nu}
 -\int d^Dy D^{(D)}(x-y)\nabla^y_{\mu}\int d^Dz D^{(D)}(y-z)\nabla_z^{\alpha}\nabla_z^{\beta}h_{\alpha\beta},
\nonumber\\
2E_{\mu\nu}&=&h^{T\theta}_{\mu\nu},
\label{E22}
\end{eqnarray}
we can write (\ref{E21}) in the form
\begin{eqnarray}
h_{\mu\nu}&=&-2g_{\mu\nu}\psi+2\nabla_{\mu}\nabla_{\nu}E
+ \nabla_{\mu}E_{\nu}+\nabla_{\nu}E_{\mu}+2E_{\mu\nu},
\label{E23}
\end{eqnarray}
and can recognize (\ref{E22}) as being of the same form as (\ref{AP65}). On multiplying by the general coordinate scalar $\Omega^2(x)$ and restricting to $D=3$ we recognize (\ref{E23}) as being the spatial part of (\ref{AP62}). In deriving the dependencies of $B$, $B_i$, $E$, $E_i$ on $f_{ij}$ that we had presented in (\ref{AP65}) we had started with the metric in the form given in (\ref{AP62}) and determined everything from it. In this appendix we proceed in the opposite direction by starting with the $3+1$ decomposition and then using projection techniques to derive the form for the metric given in (\ref{E23}), in a thus first principles approach.

In deriving (\ref{E23}) we have proceeded covariantly, and even though the form for the metric given in (\ref{AP62}) is thus necessarily covariant too, is not manifestly covariant. However, noting that $P^{\mu\nu}\nabla_{\nu}=[\eta^{\mu\nu}+U^{\mu}U^{\nu}]\nabla_{\nu}= (0,\eta^{ij}\nabla_j)$, through use of  the $P^{\mu\nu}$ projector the metric in (\ref{AP62}) could then be written in a manifestly covariant form.

\subsection{Projector Algebra}

It is of interest to relate the above formalism to the projector algebra techniques described in \cite{Mannheim2005}. In \cite{Mannheim2005} the following $D$-dimensional flat (Cartesian or Minkowski) spacetime transverse and longitudinal projection operators were introduced: 
\begin{eqnarray}
T_{\mu\nu\sigma\tau}&=&\eta_{\mu\sigma}\eta_{\nu\tau}
-\nabla_{\mu}\int d^DyD^{(D)}(x-y)
\eta_{\nu\tau}\nabla_{\sigma}
-\nabla_{\nu}\int d^DyD^{(D)}(x-y)
\eta_{\mu\sigma}\nabla_{\tau}
\nonumber \\
&+&\nabla_{\mu}\nabla_{\nu}\int
d^DyD^{(D)}(x-y)\nabla_{\sigma}\int
d^DzD^{(D)}(y-z)
\nabla_{\tau},
\label{E24}
\end{eqnarray}
\begin{eqnarray}
L_{\mu\nu\sigma\tau}&=&\nabla_{\mu}\int d^DyD^{(D)}(x-y)
\eta_{\nu\tau}\nabla_{\sigma}
+\nabla_{\nu}\int d^DyD^{(D)}(x-y)
\eta_{\mu\sigma}\nabla_{\tau}
\nonumber \\
&-&\nabla_{\mu}\nabla_{\nu}\int
d^DyD^{(D)}(x-y)\nabla_{\sigma}\int
d^DzD^{(D)}(y-z)
\nabla_{\tau},
\label{E25}
\end{eqnarray}
as they project out the transverse and longitudinal components of symmetric rank two tensors such as the fluctuation $h_{\mu\nu}$ according to 
\begin{eqnarray}
\nabla_{\nu}T^{\mu\nu\sigma\tau}h_{\sigma\tau}
=0,\qquad 
\nabla_{\nu}L^{\mu\nu\sigma\tau}h_{\sigma\tau}=\nabla_{\nu}h^{\mu\nu}.
\label{E26}
\end{eqnarray}
As constructed, these projectors obey the standard projector algebra
\begin{eqnarray}
&&T_{\mu\nu\sigma\tau}T^{\sigma\tau}_{\phantom{\sigma\tau}\alpha\beta}=
T_{\mu\nu\alpha\beta},\qquad
L_{\mu\nu\sigma\tau}L^{\sigma\tau}_{\phantom{\sigma\tau}\alpha\beta}
=L_{\mu\nu\alpha\beta},
\nonumber \\
&&T_{\mu\nu\sigma\tau}L^{\sigma\tau}_{\phantom{\sigma\tau}\alpha\beta}=
0,\qquad
L_{\mu\nu\sigma\tau}T^{\sigma\tau}_{\phantom{\sigma\tau}\alpha\beta}
=0,\qquad L_{\mu\nu\sigma\tau}
+T_{\mu\nu\sigma\tau}
=\eta_{\mu\sigma}\eta_{\nu\tau}.
\label{E27}
\end{eqnarray}
The specific sequencing of derivatives indicated in (\ref{E24}) and (\ref{E25}) was introduced so that we could establish these projector relations without needing to make any  integrations by parts. When we can integrate by parts we find that application of $T_{\mu\nu\sigma\tau}$ and $L_{\mu\nu\sigma\tau}$ precisely yields 
\begin{eqnarray}
T_{\mu\nu\sigma\tau}h^{\sigma\tau}=h^T_{\mu\nu},\qquad 
L_{\mu\nu\sigma\tau}h^{\sigma\tau}
=h^L_{\mu\nu},
\label{E28}
\end{eqnarray}
i.e. precisely yields the  $h^T_{\mu\nu}$ and $h^L_{\mu\nu}$ defined in (\ref{E11}) above. The procedure we developed above is thus recognized as being equivalent to the use of a projector algebra. 

In \cite{Mannheim2005} two further projectors were introduced
\begin{eqnarray}
Q_{\mu\nu\sigma\tau}=\frac{1}{D-1}\left[\eta_{\mu\nu}
-\nabla_{\mu}\nabla_{\nu}\int d^DyD^{(D)}(x-y)\right]
\left[\eta_{\sigma\tau}-\nabla_{\sigma}\int
d^DzD(y-z)\nabla_{\tau}\right],
\label{E29}
\end{eqnarray}
\begin{eqnarray}
P_{\mu\nu\sigma\tau}=T_{\mu\nu\sigma\tau}-Q_{\mu\nu\sigma\tau}.
\label{E30}
\end{eqnarray}
They obey the projector algebra
\begin{eqnarray}
T_{\mu\nu\sigma\tau}Q^{\sigma\tau}_{\phantom{\sigma\tau}\alpha\beta}
&=&Q_{\mu\nu\alpha\beta},\qquad
Q_{\mu\nu\sigma\tau}T^{\sigma\tau}_{\phantom{\sigma\tau}\alpha\beta}
=Q_{\mu\nu\alpha\beta},\qquad
Q_{\mu\nu\sigma\tau}Q^{\sigma\tau}_{\phantom{\sigma\tau}\alpha\beta}
=Q_{\mu\nu\alpha\beta}, 
\nonumber\\
P_{\mu\nu\sigma\tau}Q^{\sigma\tau\alpha\beta}&=&0,\qquad
Q_{\mu\nu\sigma\tau}P^{\sigma\tau\alpha\beta}=0,\qquad
P_{\mu\nu\sigma\tau}P^{\sigma\tau}_{\phantom{\sigma\tau}\alpha\beta}
=P_{\mu\nu\alpha\beta}.
\label{E31}
\end{eqnarray}
The projector $P_{\mu\nu\sigma\tau}$ projects out the traceless piece of $h^T_{\mu\nu}$, while $Q_{\mu\nu\sigma\tau}$ projects out its complement.  Comparison with (\ref{E16}) shows that 
\begin{eqnarray}
P_{\mu\nu}^{\phantom{\mu\nu}\sigma\tau}h^T_{\sigma\tau}=h^{T\theta}_{\mu\nu},\qquad 
Q_{\mu\nu}^{\phantom{\mu\nu}\sigma\tau}h^T_{\sigma\tau}
=h^T_{\mu\nu}-h^{T\theta}_{\mu\nu},
\label{E32}
\end{eqnarray}
with $h^{T\theta}_{\mu\nu}$ being both traceless and transverse. Now with the aid of (\ref{E12}) we find that $Q_{\mu\nu}^{\phantom{\mu\nu}\sigma\tau}h^L_{\sigma\tau}=0$. And thus with $P_{\mu\nu}^{\phantom{\mu\nu}\sigma\tau}h^L_{\sigma\tau}=0$ as well, we obtain 
\begin{eqnarray}
 P_{\mu\nu}^{\phantom{\mu\nu}\sigma\tau}h_{\sigma\tau}=h^{T\theta}_{\mu\nu}.
\label{E33}
\end{eqnarray}
$P_{\mu\nu\sigma\tau}$ is thus a traceless projector not just for the transverse $h_{\mu\nu}^T$ but for the full $h_{\mu\nu}$ as well. We can thus introduce its complementary projector $U_{\mu\nu\sigma\tau}=\eta_{\mu\sigma}\eta_{\nu\tau}-P_{\mu\nu\sigma\tau}$, as it obeys
\begin{eqnarray}
P_{\mu\nu\sigma\tau}U^{\sigma\tau\alpha\beta}&=&0,\qquad
U_{\mu\nu\sigma\tau}P^{\sigma\tau\alpha\beta}=0,\qquad
U_{\mu\nu\sigma\tau}U^{\sigma\tau}_{\phantom{\sigma\tau}\alpha\beta}
=U_{\mu\nu\alpha\beta},
\nonumber\\
U_{\mu\nu}^{\phantom{\mu\nu}\sigma\tau}h_{\sigma\tau}&=&h_{\mu\nu}-h^{T\theta}_{\mu\nu}=
h^{L\theta}_{\mu\nu}+\frac{1}{D-1}g_{\mu\nu}g^{\sigma\tau}h_{\sigma\tau}
-\frac{1}{D-1}\nabla_{\mu}\nabla_{\nu}\int d^Dy D^{(D)}(x-y)g^{\sigma\tau}h_{\sigma\tau},
\label{E34}
\end{eqnarray}
with the last relation following from (\ref{E17}). We can thus develop the SVT decomposition by the use of projection operators.

\subsection{General Form of the Fluctuation Equations}

As a final comment on the various components of $h_{\mu\nu}$ that we have identified, it is instructive to reexpress them in terms of $K_{\mu\nu}$ by setting  $h_{\mu\nu}=K_{\mu\nu}+(1/4)g_{\mu\nu}g^{\sigma\tau}h_{\sigma\tau}$, where we are now restricting to flat $D=4$ Minkowski spacetime. For the transverse $h_{\mu\nu}^T$ given in (\ref{E11}) this yields 
\begin{eqnarray}
h_{\mu\nu}^T&=&K_{\mu\nu}-\int d^4y D^{(4)}(x-y)\nabla_{\mu}\nabla^{\alpha}K_{\alpha\nu}
-\int d^4y D^{(4)}(x-y)\nabla_{\nu}\nabla^{\alpha}K_{\alpha\mu}+\frac{1}{4}g_{\mu\nu}g^{\alpha\beta}h_{\alpha\beta}
\nonumber\\
&+&\nabla_{\mu}\nabla_{\nu}\int d^4y D^{(4)}(x-y)\int d^4z D^{(4)}(y-z)\nabla^{\alpha}\nabla^{\beta}K_{\alpha\beta}
-\frac{1}{4}\nabla_{\mu}\nabla_{\nu}\int d^4y D^{(4)}(x-y)g^{\alpha\beta}h_{\alpha\beta},
\nonumber\\
g^{\mu\nu}h^T_{\mu\nu}&=&-\int d^4y D^{(4)}(x-y)\nabla^{\alpha}\nabla^{\beta}K_{\alpha\beta}+\frac{3}{4}g^{\alpha\beta}h_{\alpha\beta},
\label{E35}
\end{eqnarray}
and for the transverse-traceless $h_{\mu\nu}^{T\theta}$ given in (\ref{E16}) it yields 
\begin{eqnarray}
h_{\mu\nu}^{T\theta}&=&K_{\mu\nu}-\int d^4y D^{(4)}(x-y)\nabla_{\mu}\nabla^{\alpha}K_{\alpha\nu}
-\int d^4y D^{(4)}(x-y)\nabla_{\nu}\nabla^{\alpha}K_{\alpha\mu}
\nonumber\\
&+&\frac{2}{3}\nabla_{\mu}\nabla_{\nu}\int d^4y D^{(4)}(x-y)\int d^4z D^{(4)}(y-z)\nabla^{\alpha}\nabla^{\beta}K_{\alpha\beta}+\frac{1}{3}g_{\mu\nu}\int d^4x D^{(4)}(x-y)\nabla^{\alpha}\nabla^{\beta}K_{\alpha\beta}.
\label{E36}
\end{eqnarray}
As we see, the transverse-traceless $h_{\mu\nu}^{T\theta}$ is independent of the trace of the fluctuation $g^{\alpha\beta}h_{\alpha\beta}$, with it depending solely on the traceless $K_{\mu\nu}$,  just as one would expect. Recalling (cf. (\ref{AP17})) that $K_{\mu\nu}$ transforms as $K^{\mu\nu}\rightarrow K^{\mu\nu}-\nabla^{\nu}\epsilon^{\mu}-\nabla^{\mu}\epsilon^{\nu}+\frac{1}{2}\eta^{\mu\nu}\nabla_{\alpha}\epsilon^{\alpha}$ under a gauge transformation, we can readily check that $h_{\mu\nu}^{T\theta}$ is not only transverse and traceless but  most conveniently it is gauge invariant as well. Finally, applying the flat spacetime four box to $h_{\mu\nu}^{T\theta}$ we obtain
\begin{eqnarray}
\partial_{\sigma}\partial^{\sigma}\partial_{\tau}\partial^{\tau}h_{\mu\nu}^{T\theta}&=&
\partial_{\sigma}\partial^{\sigma}\partial_{\tau}\partial^{\tau}K_{\mu\nu}
-\partial_{\sigma}\partial^{\sigma}\partial_{\mu}\partial^{\alpha}K_{\alpha\nu}
-\partial_{\sigma}\partial^{\sigma}\partial_{\nu}\partial^{\alpha}K_{\alpha\mu}
\nonumber\\
&+&\frac{2}{3}\partial_{\mu}\partial_{\nu}\partial^{\alpha}\partial^{\beta}K_{\alpha\beta}+\frac{1}{3}\eta_{\mu\nu}\partial_{\sigma}\partial^{\sigma}\partial^{\alpha}\partial^{\beta}K_{\alpha\beta}.
\label{E37}
\end{eqnarray}
Comparing with (\ref{AP24}) we see that for fluctuations around flat we can set 
\begin{eqnarray}
\delta W_{\mu\nu}=\frac{1}{2}\eta^{\sigma\rho}\eta^{\alpha\beta}\partial_{\sigma}\partial_{\rho} \partial_{\alpha}\partial_{\beta}h_{\mu\nu}^{T\theta},
\label{E38}
\end{eqnarray}
to thus write $\delta W_{\mu\nu}$ in a very compact form that is manifestly gauge invariant. The significance of (\ref{E38}) is that without our needing to choose a gauge, (\ref{E38}) is already diagonal in the $(\mu,\nu)$ indices. As a check on (\ref{E38}),  we note that if we now work in the transverse gauge in which $\partial^{\alpha}K_{\alpha\mu}=0$ we find that $h_{\mu\nu}^{T\theta}$ becomes $K_{\mu\nu}$, just as would be expected since the already traceless $K^{\mu\nu}$ is transverse in this gauge. And at the same time (\ref{E38}) becomes $\delta W_{\mu\nu}=(1/2)\partial_{\sigma}\partial^{\sigma}\partial_{\tau}\partial^{\tau}K_{\mu\nu}$, an expression we recognize as (\ref{AP25}).

According to (\ref{AP3}) the conformal gravity fluctuation equation takes the form $4\alpha_g\delta W_{\mu\nu}=\delta T_{\mu\nu}$, an equation whose consistency is maintained by the fact that both of its sides are transverse and traceless. (In an arbitrary background  $\delta T_{\mu\nu}$ is covariantly conserved, and thus for fluctuations around a Minkowski background $\partial_{\nu}\delta T^{\mu\nu}$ is zero. Moreover, since $\delta T_{\mu\nu}$ arises in a conformal theory it is traceless.) For fluctuations around a four-dimensional Minkowski background, the fluctuation equation takes the form 
\begin{eqnarray}
\frac{1}{2}\partial_{\sigma}\partial^{\sigma}\partial_{\tau}\partial^{\tau}h_{\mu\nu}^{T\theta}=\frac{1}{4\alpha_g}\delta T_{\mu\nu},
\label{E39}
\end{eqnarray}
and given (\ref{AP28}) has solution 
\begin{eqnarray}
h_{\mu\nu}^{T\theta}(x)=\frac{1}{16\pi\alpha_g}\int d^4x^{\prime}\theta(t-t^{\prime}-|{\bf x}-{\bf x}^{\prime}|)
\delta T_{\mu\nu}(x^{\prime}).
\label{E40}
\end{eqnarray}
This solution is exact without approximation and holds in every gauge, with the dynamics only depending on the transverse-traceless components of the fluctuation. As far as the counting of degrees of freedom is concerned, with $h_{\mu\nu}^{T\theta}$ obeying $\nabla^{\nu}h_{\mu\nu}^{T\theta}=0$, $\eta^{\mu\nu}h_{\mu\nu}^{T\theta}=0$ there are five constraints. With the initial $h_{\mu\nu}$ having ten degrees of freedom $h_{\mu\nu}^{T\theta}$ thus has five. The utility of constructing (\ref{E39}) by projection is that, apart from the tracelessness constraint, by transverse projection one is able to reduce the initial ten-component $h_{\mu\nu}$ by four degrees of freedom, to thus secure the standard four coordinate invariance freedom on $h_{\mu\nu}$ without needing to explicitly impose it.

As noted in \cite{Mannheim2005}, an analogous situation occurs in Einstein gravity. For fluctuations around flat Minkowski the fluctuation in the Einstein tensor takes the form
\begin{eqnarray}
\delta G_{\mu\nu}&=&\delta R_{\mu\nu}-\frac{1}{2}\eta_{\mu\nu}\eta^{\alpha\beta}\delta R_{\alpha\beta}\nonumber\\
&=&
\frac{1}{2}\left[\partial_{\alpha}\partial^{\alpha}h_{\mu\nu}
-\partial_{\mu}\partial^{\alpha}h_{\alpha\nu}
-\partial_{\nu}\partial^{\alpha}h_{\alpha\mu}
+\partial_{\mu}\partial_{\nu}\eta^{\alpha\beta}h_{\alpha\beta}\right]
-\frac{1}{2}\eta_{\mu\nu}\left[\partial_{\alpha}\partial^{\alpha}\eta^{\sigma\tau}h_{\sigma\tau}-
\partial_{\alpha}\partial_{\beta}h^{\alpha\beta}\right].
\label{E41}
\end{eqnarray}
Comparing with (\ref{E11}), we see that we can rewrite this expression as 
\begin{eqnarray}
\delta G_{\mu\nu}=\frac{1}{2}\left[\partial_{\alpha}\partial^{\alpha}h^T_{\mu\nu}
+\partial_{\mu}\partial_{\nu}\eta^{\alpha\beta}h^T_{\alpha\beta}
-\eta_{\mu\nu}\partial_{\alpha}\partial^{\alpha}\eta^{\sigma\tau}h^T_{\sigma\tau}\right],
\label{E42}
\end{eqnarray}
i.e. we can write it entirely in terms of the six degree of freedom transverse $h^T_{\mu\nu}$ without needing to choose a gauge. (Under $h_{\mu\nu}\rightarrow h_{\mu\nu}-\partial_{\mu}\epsilon_{\nu}-\partial_{\nu}\epsilon_{\mu}$ the transverse $h^T_{\mu\nu}$ is invariant, with $T_{\mu\nu\sigma\tau}(\partial^{\sigma}\epsilon^{\tau}+\partial^{\tau}\epsilon^{\sigma})$ being zero identically, where $T_{\mu\nu\sigma\tau}$ is given in (\ref{E24}).) Thus while one can reduce (\ref{E41}) to $\delta G_{\mu\nu}=(1/2)[\partial_{\alpha}\partial^{\alpha}h_{\mu\nu}+\partial_{\mu}\partial_{\nu}\eta^{\alpha\beta}h_{\alpha\beta}-\eta_{\mu\nu}\partial_{\alpha}\partial^{\alpha}\eta^{\sigma\tau}h_{\sigma\tau}]$ by working in the transverse gauge, one can reduce (\ref{E41}) to (\ref{E42}) without needing to specify a gauge at all. Thus again one has secured the standard four coordinate invariance freedom on $h_{\mu\nu}$ by transverse projection without needing to explicitly impose it, with transverse projection reducing the ten-component $h_{\mu\nu}$ to its six physical degrees of freedom. Moreover, just like the conformal gravity (\ref{E38}), (\ref{E42}) is also diagonal in the $(\mu,\nu)$ indices. Thus the fluctuation equations of both Einstein gravity and conformal gravity  can be written in very simple forms if one uses transverse and transverse-traceless projection operators, forms in which they become diagonal in the  $(\mu,\nu)$ indices. For the fluctuation Einstein equations themselves one has 
\begin{eqnarray}
\delta R_{\mu\nu}-\frac{1}{2}\eta_{\mu\nu}\eta^{\alpha\beta}\delta R_{\alpha\beta}=-8\pi G\delta T_{\mu\nu},\qquad \eta^{\alpha\beta}\delta R_{\alpha\beta}=8\pi G \eta^{\alpha\beta}\delta T_{\alpha\beta}=
\partial_{\alpha}\partial^{\alpha}\eta^{\sigma\tau}h^T_{\sigma\tau}.
\label{E43}
\end{eqnarray}
With the retarded second-order derivative theory propagator being given by $\theta(t)\delta(t-r)/4\pi r$, we can thus set
\begin{eqnarray}
\frac{1}{2}\partial_{\alpha}\partial^{\alpha}h^T_{\mu\nu}=-8\pi G\left[\delta T_{\mu\nu}-
\frac{1}{2}\eta_{\mu\nu}\eta^{\sigma\tau}\delta T_{\sigma\tau}
+\frac{1}{8\pi}\partial_{\mu}\partial_{\nu}\int \frac{d^4x^{\prime}}{|{\bf x}-{\bf x}^{\prime}|}
\theta(t-t^{\prime})\delta(t-t^{\prime}-|{\bf x}-{\bf x}^{\prime}|)\eta^{\sigma\tau}\delta T_{\sigma\tau}(x^{\prime})\right],
\label{E44}
\end{eqnarray}
and thus
\begin{eqnarray}
h_{\mu\nu}^{T}(x)&=&-4G\int \frac{d^4x^{\prime}}{|{\bf x}-{\bf x}^{\prime}|}\theta(t-t^{\prime})\delta(t-t^{\prime}-|{\bf x}-{\bf x}^{\prime}|)
\bigg{[}\delta T_{\mu\nu}(x^{\prime})-\frac{1}{2}\eta_{\mu\nu}\eta^{\alpha\beta}\delta T_{\alpha \beta}(x^{\prime})
\nonumber\\
&+&\frac{1}{8\pi}\partial^{\prime}_{\mu}\partial^{\prime}_{\nu}\int \frac{d^4x^{\prime\prime}}{|{\bf x}^{\prime}-{\bf x}^{\prime\prime}|}\theta(t^{\prime}-t^{\prime\prime})\delta(t^{\prime}-t^{\prime\prime}-|{\bf x}^{\prime}-{\bf x}^{\prime\prime}|)\eta^{\sigma\tau}\delta T_{\sigma\tau}(x^{\prime\prime})\bigg{]},
\label{E45}
\end{eqnarray}
a gauge invariant expression that is exact without approximation.

\section{Projection technique for fluctuations around a general conformal to flat Minkowski background}
\label{SF}

Through use of the projection technique, in Appendix \ref{SE} we obtained a very compact one-term expression, (\ref{E38}),  for fluctuations around a flat Minkowski background. We can readily generalize the technique to fluctuations about a general conformal to flat Minkowski background. Since (\ref{AP43}) holds for fluctuations around a completely general and arbitrary background, it in particular holds in conformal to flat Minkowski geometries that are described by the metric given in (\ref{AP6}), with (\ref{AP43}) then reducing to (\ref{AP44}) and (\ref{AP54}). And with indices being raised with $\eta^{\mu\nu}$ (so that $\partial^{\mu}=\eta^{\mu\nu}\partial_{\nu}$) we find that in this case $\delta W_{\mu\nu}$ evaluates to the 151 term
\begin{align}
\delta W_{\mu\nu}&=\Omega^{-5} \partial_{\alpha}\partial_{\nu}\partial^{\alpha}\Omega \partial_{\beta}K_{\mu}{}^{\beta} + \Omega^{-5} \partial_{\alpha}\partial_{\mu}\partial^{\alpha}\Omega \partial_{\beta}K_{\nu}{}^{\beta} + 2 \Omega^{-5} \partial^{\alpha}\partial_{\nu}\Omega \partial_{\beta}\partial_{\alpha}K_{\mu}{}^{\beta}
\nonumber\\
& + 2 \Omega^{-5} \partial^{\alpha}\partial_{\mu}\Omega \partial_{\beta}\partial_{\alpha}K_{\nu}{}^{\beta} + 2 \Omega^{-5} \partial^{\alpha}\Omega \partial_{\beta}\partial_{\alpha}\partial_{\mu}K_{\nu}{}^{\beta} + 2 \Omega^{-5} \partial^{\alpha}\Omega \partial_{\beta}\partial_{\alpha}\partial_{\nu}K_{\mu}{}^{\beta} 
\nonumber\\
&+ \tfrac{1}{3} \Omega^{-4} \partial_{\beta}\partial_{\alpha}\partial_{\nu}\partial_{\mu}K^{\alpha \beta} -  \tfrac{2}{3} K^{\alpha \beta} \Omega^{-5} \partial_{\beta}\partial_{\alpha}\partial_{\nu}\partial_{\mu}\Omega + \Omega^{-5} \partial^{\alpha}\partial_{\nu}\Omega \partial_{\beta}\partial^{\beta}K_{\mu \alpha}
\nonumber\\
& - 2 \Omega^{-5} \partial_{\alpha}\partial^{\alpha}\Omega \partial_{\beta}\partial^{\beta}K_{\mu \nu} + 6 \Omega^{-6} \partial_{\alpha}\Omega \partial^{\alpha}\Omega \partial_{\beta}\partial^{\beta}K_{\mu \nu} + \Omega^{-5} \partial^{\alpha}\partial_{\mu}\Omega \partial_{\beta}\partial^{\beta}K_{\nu \alpha} 
\nonumber\\
&+ 3 K_{\mu \nu} \Omega^{-6} \partial_{\alpha}\partial^{\alpha}\Omega \partial_{\beta}\partial^{\beta}\Omega + 12 \Omega^{-6} \partial_{\alpha}K_{\mu \nu} \partial^{\alpha}\Omega \partial_{\beta}\partial^{\beta}\Omega - 24 K_{\mu \nu} \Omega^{-7} \partial_{\alpha}\Omega \partial^{\alpha}\Omega \partial_{\beta}\partial^{\beta}\Omega 
\nonumber\\
&- 4 \Omega^{-5} \partial^{\alpha}\Omega \partial_{\beta}\partial^{\beta}\partial_{\alpha}K_{\mu \nu} + 12 K_{\mu \nu} \Omega^{-6} \partial^{\alpha}\Omega \partial_{\beta}\partial^{\beta}\partial_{\alpha}\Omega + \tfrac{1}{2} \Omega^{-4} \partial_{\beta}\partial^{\beta}\partial_{\alpha}\partial^{\alpha}K_{\mu \nu}
\nonumber\\
& -  K_{\mu \nu} \Omega^{-5} \partial_{\beta}\partial^{\beta}\partial_{\alpha}\partial^{\alpha}\Omega -  \tfrac{1}{2} \Omega^{-4} \partial_{\beta}\partial^{\beta}\partial_{\alpha}\partial_{\mu}K_{\nu}{}^{\alpha} -  \tfrac{1}{2} \Omega^{-4} \partial_{\beta}\partial^{\beta}\partial_{\alpha}\partial_{\nu}K_{\mu}{}^{\alpha} 
\nonumber\\
&- 4 \Omega^{-5} \partial_{\alpha}K_{\mu \nu} \partial_{\beta}\partial^{\beta}\partial^{\alpha}\Omega + \Omega^{-5} \partial^{\alpha}\Omega \partial_{\beta}\partial^{\beta}\partial_{\mu}K_{\nu \alpha} + \Omega^{-5} \partial^{\alpha}\Omega \partial_{\beta}\partial^{\beta}\partial_{\nu}K_{\mu \alpha} 
\nonumber\\
&-  \tfrac{4}{3} \Omega^{-5} \partial^{\alpha}\partial_{\nu}\Omega \partial_{\beta}\partial_{\mu}K_{\alpha}{}^{\beta} + \Omega^{-5} \partial_{\alpha}\partial^{\alpha}\Omega \partial_{\beta}\partial_{\mu}K_{\nu}{}^{\beta} - 3 \Omega^{-6} \partial_{\alpha}\Omega \partial^{\alpha}\Omega \partial_{\beta}\partial_{\mu}K_{\nu}{}^{\beta} 
\nonumber\\
&- 6 K_{\nu}{}^{\beta} \Omega^{-6} \partial^{\alpha}\Omega \partial_{\beta}\partial_{\mu}\partial_{\alpha}\Omega - 3 K_{\nu \alpha} \Omega^{-6} \partial^{\alpha}\Omega \partial_{\beta}\partial_{\mu}\partial^{\beta}\Omega -  \tfrac{4}{3} \Omega^{-5} \partial^{\alpha}\partial_{\mu}\Omega \partial_{\beta}\partial_{\nu}K_{\alpha}{}^{\beta}
\nonumber\\
& + \Omega^{-5} \partial_{\alpha}\partial^{\alpha}\Omega \partial_{\beta}\partial_{\nu}K_{\mu}{}^{\beta} - 3 \Omega^{-6} \partial_{\alpha}\Omega \partial^{\alpha}\Omega \partial_{\beta}\partial_{\nu}K_{\mu}{}^{\beta} - 6 K_{\mu}{}^{\beta} \Omega^{-6} \partial^{\alpha}\Omega \partial_{\beta}\partial_{\nu}\partial_{\alpha}\Omega 
\nonumber\\
&- 3 K_{\mu \alpha} \Omega^{-6} \partial^{\alpha}\Omega \partial_{\beta}\partial_{\nu}\partial^{\beta}\Omega -  \tfrac{4}{3} \Omega^{-5} \partial^{\alpha}\Omega \partial_{\beta}\partial_{\nu}\partial_{\mu}K_{\alpha}{}^{\beta} -  \tfrac{4}{3} \Omega^{-5} \partial_{\alpha}K^{\alpha \beta} \partial_{\beta}\partial_{\nu}\partial_{\mu}\Omega 
\nonumber\\
&+ 4 K_{\alpha}{}^{\beta} \Omega^{-6} \partial^{\alpha}\Omega \partial_{\beta}\partial_{\nu}\partial_{\mu}\Omega - 48 \Omega^{-7} \partial_{\alpha}\Omega \partial^{\alpha}\Omega \partial_{\beta}K_{\mu \nu} \partial^{\beta}\Omega + 60 K_{\mu \nu} \Omega^{-8} \partial_{\alpha}\Omega \partial^{\alpha}\Omega \partial_{\beta}\Omega \partial^{\beta}\Omega
\nonumber\\
& + 12 \Omega^{-6} \partial^{\alpha}\Omega \partial_{\beta}\partial_{\alpha}K_{\mu \nu} \partial^{\beta}\Omega - 48 K_{\mu \nu} \Omega^{-7} \partial^{\alpha}\Omega \partial_{\beta}\partial_{\alpha}\Omega \partial^{\beta}\Omega - 6 \Omega^{-6} \partial^{\alpha}\Omega \partial_{\beta}\partial_{\mu}K_{\nu \alpha} \partial^{\beta}\Omega
\nonumber\\
& - 6 \Omega^{-6} \partial^{\alpha}\Omega \partial_{\beta}\partial_{\nu}K_{\mu \alpha} \partial^{\beta}\Omega
 + 24 \Omega^{-6} \partial^{\alpha}\Omega \partial_{\beta}K_{\mu \nu} \partial^{\beta}\partial_{\alpha}\Omega + K_{\nu \beta} \Omega^{-5} \partial^{\beta}\partial_{\alpha}\partial_{\mu}\partial^{\alpha}\Omega 
 \nonumber\\
 &+ K_{\mu \beta} \Omega^{-5} \partial^{\beta}\partial_{\alpha}\partial_{\nu}\partial^{\alpha}\Omega + 2 \Omega^{-5} \partial_{\alpha}\partial_{\mu}K_{\nu \beta} \partial^{\beta}\partial^{\alpha}\Omega + 2 \Omega^{-5} \partial_{\alpha}\partial_{\nu}K_{\mu \beta} \partial^{\beta}\partial^{\alpha}\Omega 
 \nonumber\\
 &- 4 \Omega^{-5} \partial_{\beta}\partial_{\alpha}K_{\mu \nu} \partial^{\beta}\partial^{\alpha}\Omega + 6 K_{\mu \nu} \Omega^{-6} \partial_{\beta}\partial_{\alpha}\Omega \partial^{\beta}\partial^{\alpha}\Omega - 6 \Omega^{-6} \partial_{\alpha}K_{\nu \beta} \partial^{\alpha}\Omega \partial^{\beta}\partial_{\mu}\Omega 
 \nonumber\\
 &+ 2 \Omega^{-5} \partial_{\alpha}K_{\nu \beta} \partial^{\beta}\partial_{\mu}\partial^{\alpha}\Omega - 6 \Omega^{-6} \partial_{\alpha}K_{\mu \beta} \partial^{\alpha}\Omega \partial^{\beta}\partial_{\nu}\Omega + 2 \Omega^{-5} \partial_{\alpha}K_{\mu \beta} \partial^{\beta}\partial_{\nu}\partial^{\alpha}\Omega
 \nonumber\\
 & + 2 \eta_{\mu \nu} \Omega^{-6} \partial^{\alpha}\Omega \partial_{\beta}\partial^{\beta}\Omega \partial_{\gamma}K_{\alpha}{}^{\gamma} - 8 \eta_{\mu \nu} \Omega^{-7} \partial_{\alpha}\Omega \partial^{\alpha}\Omega \partial^{\beta}\Omega \partial_{\gamma}K_{\beta}{}^{\gamma} + 4 \eta_{\mu \nu} \Omega^{-6} \partial^{\alpha}\Omega \partial^{\beta}\partial_{\alpha}\Omega \partial_{\gamma}K_{\beta}{}^{\gamma} 
 \nonumber\\
 &-  \tfrac{2}{3} \eta_{\mu \nu} \Omega^{-5} \partial^{\beta}\partial_{\alpha}\partial^{\alpha}\Omega \partial_{\gamma}K_{\beta}{}^{\gamma} + 2 \eta_{\mu \nu} K_{\beta}{}^{\gamma} \Omega^{-6} \partial^{\beta}\partial^{\alpha}\Omega \partial_{\gamma}\partial_{\alpha}\Omega + 4 \eta_{\mu \nu} \Omega^{-6} \partial^{\alpha}\Omega \partial^{\beta}\Omega \partial_{\gamma}\partial_{\beta}K_{\alpha}{}^{\gamma} 
 \nonumber\\
 &-  \tfrac{4}{3} \eta_{\mu \nu} \Omega^{-5} \partial^{\beta}\partial^{\alpha}\Omega \partial_{\gamma}\partial_{\beta}K_{\alpha}{}^{\gamma} -  \tfrac{1}{3} \eta_{\mu \nu} \Omega^{-5} \partial_{\alpha}\partial^{\alpha}\Omega \partial_{\gamma}\partial_{\beta}K^{\beta \gamma} + \eta_{\mu \nu} \Omega^{-6} \partial_{\alpha}\Omega \partial^{\alpha}\Omega \partial_{\gamma}\partial_{\beta}K^{\beta \gamma}
 \nonumber\\
 & + \eta_{\mu \nu} K^{\beta \gamma} \Omega^{-6} \partial_{\alpha}\partial^{\alpha}\Omega \partial_{\gamma}\partial_{\beta}\Omega - 4 \eta_{\mu \nu} K^{\beta \gamma} \Omega^{-7} \partial_{\alpha}\Omega \partial^{\alpha}\Omega \partial_{\gamma}\partial_{\beta}\Omega - 16 \eta_{\mu \nu} K_{\alpha}{}^{\gamma} \Omega^{-7} \partial^{\alpha}\Omega \partial^{\beta}\Omega \partial_{\gamma}\partial_{\beta}\Omega 
 \nonumber\\
 &-  \tfrac{2}{3} \eta_{\mu \nu} \Omega^{-5} \partial^{\alpha}\Omega \partial_{\gamma}\partial_{\beta}\partial_{\alpha}K^{\beta \gamma} + 2 \eta_{\mu \nu} K^{\beta \gamma} \Omega^{-6} \partial^{\alpha}\Omega \partial_{\gamma}\partial_{\beta}\partial_{\alpha}\Omega + \eta_{\mu \nu} \Omega^{-6} \partial^{\alpha}\Omega \partial^{\beta}\Omega \partial_{\gamma}\partial^{\gamma}K_{\alpha \beta}
 \nonumber\\
 & -  \tfrac{1}{3} \eta_{\mu \nu} \Omega^{-5} \partial^{\beta}\partial^{\alpha}\Omega \partial_{\gamma}\partial^{\gamma}K_{\alpha \beta} - 4 \eta_{\mu \nu} K_{\alpha \beta} \Omega^{-7} \partial^{\alpha}\Omega \partial^{\beta}\Omega \partial_{\gamma}\partial^{\gamma}\Omega -  \tfrac{2}{3} \eta_{\mu \nu} \Omega^{-5} \partial^{\alpha}\Omega \partial_{\gamma}\partial^{\gamma}\partial_{\beta}K_{\alpha}{}^{\beta}
\nonumber \\
 & + \tfrac{1}{6} \eta_{\mu \nu} \Omega^{-4} \partial_{\gamma}\partial^{\gamma}\partial_{\beta}\partial_{\alpha}K^{\alpha \beta} + 20 \eta_{\mu \nu} K_{\beta \gamma} \Omega^{-8} \partial_{\alpha}\Omega \partial^{\alpha}\Omega \partial^{\beta}\Omega \partial^{\gamma}\Omega - 8 \eta_{\mu \nu} \Omega^{-7} \partial^{\alpha}\Omega \partial^{\beta}\Omega \partial_{\gamma}K_{\alpha \beta} \partial^{\gamma}\Omega
 \nonumber\\
 & + 2 \eta_{\mu \nu} K_{\alpha \gamma} \Omega^{-6} \partial^{\alpha}\Omega \partial^{\gamma}\partial_{\beta}\partial^{\beta}\Omega + 2 \eta_{\mu \nu} \Omega^{-6} \partial_{\alpha}K_{\beta \gamma} \partial^{\alpha}\Omega \partial^{\gamma}\partial^{\beta}\Omega + 4 \eta_{\mu \nu} \Omega^{-6} \partial^{\alpha}\Omega \partial_{\beta}K_{\alpha \gamma} \partial^{\gamma}\partial^{\beta}\Omega 
 \nonumber\\
 &-  \tfrac{1}{3} \eta_{\mu \nu} K_{\beta \gamma} \Omega^{-5} \partial^{\gamma}\partial^{\beta}\partial_{\alpha}\partial^{\alpha}\Omega -  \tfrac{2}{3} \eta_{\mu \nu} \Omega^{-5} \partial_{\alpha}K_{\beta \gamma} \partial^{\gamma}\partial^{\beta}\partial^{\alpha}\Omega + 4 \Omega^{-6} \partial^{\alpha}\Omega \partial^{\beta}\partial_{\nu}\Omega \partial_{\mu}K_{\alpha \beta} 
 \nonumber\\
 &-  \tfrac{2}{3} \Omega^{-5} \partial_{\beta}\partial_{\nu}\partial_{\alpha}\Omega \partial_{\mu}K^{\alpha \beta} - 3 \Omega^{-6} \partial^{\alpha}\Omega \partial_{\beta}\partial^{\beta}\Omega \partial_{\mu}K_{\nu \alpha} + 12 \Omega^{-7} \partial_{\alpha}\Omega \partial^{\alpha}\Omega \partial^{\beta}\Omega \partial_{\mu}K_{\nu \beta} 
 \nonumber\\
 &- 6 \Omega^{-6} \partial^{\alpha}\Omega \partial^{\beta}\partial_{\alpha}\Omega \partial_{\mu}K_{\nu \beta} + \Omega^{-5} \partial^{\beta}\partial_{\alpha}\partial^{\alpha}\Omega \partial_{\mu}K_{\nu \beta} + 4 \Omega^{-6} \partial^{\alpha}\partial_{\nu}\Omega \partial_{\beta}K_{\alpha}{}^{\beta} \partial_{\mu}\Omega 
 \nonumber\\
 &- 3 \Omega^{-6} \partial_{\alpha}\partial^{\alpha}\Omega \partial_{\beta}K_{\nu}{}^{\beta} \partial_{\mu}\Omega + 12 \Omega^{-7} \partial_{\alpha}\Omega \partial^{\alpha}\Omega \partial_{\beta}K_{\nu}{}^{\beta} \partial_{\mu}\Omega - 6 \Omega^{-6} \partial^{\alpha}\Omega \partial_{\beta}\partial_{\alpha}K_{\nu}{}^{\beta} \partial_{\mu}\Omega
 \nonumber\\
 & + 24 K_{\nu}{}^{\beta} \Omega^{-7} \partial^{\alpha}\Omega \partial_{\beta}\partial_{\alpha}\Omega \partial_{\mu}\Omega -  \tfrac{2}{3} \Omega^{-5} \partial_{\beta}\partial_{\alpha}\partial_{\nu}K^{\alpha \beta} \partial_{\mu}\Omega - 3 \Omega^{-6} \partial^{\alpha}\Omega \partial_{\beta}\partial^{\beta}K_{\nu \alpha} \partial_{\mu}\Omega 
 \nonumber\\
 &+ 12 K_{\nu \alpha} \Omega^{-7} \partial^{\alpha}\Omega \partial_{\beta}\partial^{\beta}\Omega \partial_{\mu}\Omega + \Omega^{-5} \partial_{\beta}\partial^{\beta}\partial_{\alpha}K_{\nu}{}^{\alpha} \partial_{\mu}\Omega + 4 \Omega^{-6} \partial^{\alpha}\Omega \partial_{\beta}\partial_{\nu}K_{\alpha}{}^{\beta} \partial_{\mu}\Omega 
 \nonumber\\
 &+ 2 K^{\alpha \beta} \Omega^{-6} \partial_{\beta}\partial_{\nu}\partial_{\alpha}\Omega \partial_{\mu}\Omega - 60 K_{\nu \beta} \Omega^{-8} \partial_{\alpha}\Omega \partial^{\alpha}\Omega \partial^{\beta}\Omega \partial_{\mu}\Omega + 24 \Omega^{-7} \partial^{\alpha}\Omega \partial_{\beta}K_{\nu \alpha} \partial^{\beta}\Omega \partial_{\mu}\Omega 
 \nonumber\\
 &- 3 K_{\nu \beta} \Omega^{-6} \partial^{\beta}\partial_{\alpha}\partial^{\alpha}\Omega \partial_{\mu}\Omega - 6 \Omega^{-6} \partial_{\alpha}K_{\nu \beta} \partial^{\beta}\partial^{\alpha}\Omega \partial_{\mu}\Omega - 6 \Omega^{-6} \partial^{\alpha}\Omega \partial_{\beta}K_{\nu}{}^{\beta} \partial_{\mu}\partial_{\alpha}\Omega 
 \nonumber\\
 &- 6 K_{\nu \beta} \Omega^{-6} \partial^{\beta}\partial^{\alpha}\Omega \partial_{\mu}\partial_{\alpha}\Omega - 3 K_{\nu}{}^{\beta} \Omega^{-6} \partial_{\alpha}\partial^{\alpha}\Omega \partial_{\mu}\partial_{\beta}\Omega + 12 K_{\nu}{}^{\beta} \Omega^{-7} \partial_{\alpha}\Omega \partial^{\alpha}\Omega \partial_{\mu}\partial_{\beta}\Omega 
\nonumber\\
&+ 24 K_{\nu \alpha} \Omega^{-7} \partial^{\alpha}\Omega \partial^{\beta}\Omega \partial_{\mu}\partial_{\beta}\Omega - 6 \Omega^{-6} \partial^{\alpha}\Omega \partial_{\beta}K_{\nu \alpha} \partial_{\mu}\partial^{\beta}\Omega + 4 \Omega^{-6} \partial^{\alpha}\Omega \partial^{\beta}\partial_{\mu}\Omega \partial_{\nu}K_{\alpha \beta} 
\nonumber\\
&- 8 \Omega^{-7} \partial^{\alpha}\Omega \partial^{\beta}\Omega \partial_{\mu}\Omega \partial_{\nu}K_{\alpha \beta} + 2 \Omega^{-6} \partial^{\beta}\partial^{\alpha}\Omega \partial_{\mu}\Omega \partial_{\nu}K_{\alpha \beta} -  \tfrac{2}{3} \Omega^{-5} \partial_{\beta}\partial_{\mu}\partial_{\alpha}\Omega \partial_{\nu}K^{\alpha \beta} 
\nonumber\\
&- 3 \Omega^{-6} \partial^{\alpha}\Omega \partial_{\beta}\partial^{\beta}\Omega \partial_{\nu}K_{\mu \alpha} + 12 \Omega^{-7} \partial_{\alpha}\Omega \partial^{\alpha}\Omega \partial^{\beta}\Omega \partial_{\nu}K_{\mu \beta} - 6 \Omega^{-6} \partial^{\alpha}\Omega \partial^{\beta}\partial_{\alpha}\Omega \partial_{\nu}K_{\mu \beta} 
\nonumber\\
&+ \Omega^{-5} \partial^{\beta}\partial_{\alpha}\partial^{\alpha}\Omega \partial_{\nu}K_{\mu \beta} + 4 \Omega^{-6} \partial^{\alpha}\partial_{\mu}\Omega \partial_{\beta}K_{\alpha}{}^{\beta} \partial_{\nu}\Omega - 3 \Omega^{-6} \partial_{\alpha}\partial^{\alpha}\Omega \partial_{\beta}K_{\mu}{}^{\beta} \partial_{\nu}\Omega 
\nonumber\\
&+ 12 \Omega^{-7} \partial_{\alpha}\Omega \partial^{\alpha}\Omega \partial_{\beta}K_{\mu}{}^{\beta} \partial_{\nu}\Omega - 6 \Omega^{-6} \partial^{\alpha}\Omega \partial_{\beta}\partial_{\alpha}K_{\mu}{}^{\beta} \partial_{\nu}\Omega 
+ 24 K_{\mu}{}^{\beta} \Omega^{-7} \partial^{\alpha}\Omega \partial_{\beta}\partial_{\alpha}\Omega \partial_{\nu}\Omega 
\nonumber\\
&-  \tfrac{2}{3} \Omega^{-5} \partial_{\beta}\partial_{\alpha}\partial_{\mu}K^{\alpha \beta} \partial_{\nu}\Omega - 3 \Omega^{-6} \partial^{\alpha}\Omega \partial_{\beta}\partial^{\beta}K_{\mu \alpha} \partial_{\nu}\Omega + 12 K_{\mu \alpha} \Omega^{-7} \partial^{\alpha}\Omega \partial_{\beta}\partial^{\beta}\Omega \partial_{\nu}\Omega 
\nonumber\\
&+ \Omega^{-5} \partial_{\beta}\partial^{\beta}\partial_{\alpha}K_{\mu}{}^{\alpha} \partial_{\nu}\Omega + 4 \Omega^{-6} \partial^{\alpha}\Omega \partial_{\beta}\partial_{\mu}K_{\alpha}{}^{\beta} \partial_{\nu}\Omega + 2 K^{\alpha \beta} \Omega^{-6} \partial_{\beta}\partial_{\mu}\partial_{\alpha}\Omega \partial_{\nu}\Omega 
\nonumber\\
&- 60 K_{\mu \beta} \Omega^{-8} \partial_{\alpha}\Omega \partial^{\alpha}\Omega \partial^{\beta}\Omega \partial_{\nu}\Omega + 24 \Omega^{-7} \partial^{\alpha}\Omega \partial_{\beta}K_{\mu \alpha} \partial^{\beta}\Omega \partial_{\nu}\Omega - 3 K_{\mu \beta} \Omega^{-6} \partial^{\beta}\partial_{\alpha}\partial^{\alpha}\Omega \partial_{\nu}\Omega 
\nonumber\\
&- 6 \Omega^{-6} \partial_{\alpha}K_{\mu \beta} \partial^{\beta}\partial^{\alpha}\Omega \partial_{\nu}\Omega - 8 \Omega^{-7} \partial^{\alpha}\Omega \partial^{\beta}\Omega \partial_{\mu}K_{\alpha \beta} \partial_{\nu}\Omega + 2 \Omega^{-6} \partial^{\beta}\partial^{\alpha}\Omega \partial_{\mu}K_{\alpha \beta} \partial_{\nu}\Omega 
\nonumber\\
&- 16 \Omega^{-7} \partial^{\alpha}\Omega \partial_{\beta}K_{\alpha}{}^{\beta} \partial_{\mu}\Omega \partial_{\nu}\Omega + 2 \Omega^{-6} \partial_{\beta}\partial_{\alpha}K^{\alpha \beta} \partial_{\mu}\Omega \partial_{\nu}\Omega - 8 K^{\alpha \beta} \Omega^{-7} \partial_{\beta}\partial_{\alpha}\Omega \partial_{\mu}\Omega \partial_{\nu}\Omega 
\nonumber\\
&+ 40 K_{\alpha \beta} \Omega^{-8} \partial^{\alpha}\Omega \partial^{\beta}\Omega \partial_{\mu}\Omega \partial_{\nu}\Omega - 16 K_{\alpha}{}^{\beta} \Omega^{-7} \partial^{\alpha}\Omega \partial_{\mu}\partial_{\beta}\Omega \partial_{\nu}\Omega - 6 \Omega^{-6} \partial^{\alpha}\Omega \partial_{\beta}K_{\mu}{}^{\beta} \partial_{\nu}\partial_{\alpha}\Omega 
\nonumber\\
&- 6 K_{\mu \beta} \Omega^{-6} \partial^{\beta}\partial^{\alpha}\Omega \partial_{\nu}\partial_{\alpha}\Omega - 3 K_{\mu}{}^{\beta} \Omega^{-6} \partial_{\alpha}\partial^{\alpha}\Omega \partial_{\nu}\partial_{\beta}\Omega + 12 K_{\mu}{}^{\beta} \Omega^{-7} \partial_{\alpha}\Omega \partial^{\alpha}\Omega \partial_{\nu}\partial_{\beta}\Omega 
\nonumber\\
&+ 24 K_{\mu \alpha} \Omega^{-7} \partial^{\alpha}\Omega \partial^{\beta}\Omega \partial_{\nu}\partial_{\beta}\Omega - 16 K_{\alpha}{}^{\beta} \Omega^{-7} \partial^{\alpha}\Omega \partial_{\mu}\Omega \partial_{\nu}\partial_{\beta}\Omega + 4 K^{\alpha \beta} \Omega^{-6} \partial_{\mu}\partial_{\alpha}\Omega \partial_{\nu}\partial_{\beta}\Omega 
\nonumber\\
&- 6 \Omega^{-6} \partial^{\alpha}\Omega \partial_{\beta}K_{\mu \alpha} \partial_{\nu}\partial^{\beta}\Omega + 2 \Omega^{-6} \partial^{\alpha}\Omega \partial^{\beta}\Omega \partial_{\nu}\partial_{\mu}K_{\alpha \beta} -  \tfrac{2}{3} \Omega^{-5} \partial^{\beta}\partial^{\alpha}\Omega \partial_{\nu}\partial_{\mu}K_{\alpha \beta} 
\nonumber\\
&+ 4 \Omega^{-6} \partial^{\alpha}\Omega \partial_{\beta}K_{\alpha}{}^{\beta} \partial_{\nu}\partial_{\mu}\Omega -  \tfrac{2}{3} \Omega^{-5} \partial_{\beta}\partial_{\alpha}K^{\alpha \beta} \partial_{\nu}\partial_{\mu}\Omega + 2 K^{\alpha \beta} \Omega^{-6} \partial_{\beta}\partial_{\alpha}\Omega \partial_{\nu}\partial_{\mu}\Omega
\nonumber\\
& - 8 K_{\alpha \beta} \Omega^{-7} \partial^{\alpha}\Omega \partial^{\beta}\Omega \partial_{\nu}\partial_{\mu}\Omega.
\label{F1}
\end{align}
Despite its 151 terms we can rewrite (\ref{F1}) identically as the compact
\begin{eqnarray}
\delta W_{\mu\nu}&=&\frac{1}{2}\Omega^{-2}\bigg{(}\partial_{\sigma}\partial^{\sigma}\partial_{\tau}\partial^{\tau}[\Omega^{-2}K_{\mu\nu}]
-\partial_{\sigma}\partial^{\sigma}\partial_{\mu}\partial^{\alpha}[\Omega^{-2}K_{\alpha\nu}]
-\partial_{\sigma}\partial^{\sigma}\partial_{\nu}\partial^{\alpha}[\Omega^{-2}K_{\alpha\mu}]
\nonumber\\
&+&\frac{2}{3}\partial_{\mu}\partial_{\nu}\partial^{\alpha}\partial^{\beta}[\Omega^{-2}K_{\alpha\beta}]+\frac{1}{3}\eta_{\mu\nu}\partial_{\sigma}\partial^{\sigma}\partial^{\alpha}\partial^{\beta}[\Omega^{-2}K_{\alpha\beta}]\bigg{)}.
\label{F2}
\end{eqnarray} 
Then, in analog to (\ref{E38}),  using the transverse-traceless projector we can write $\delta W_{\mu\nu}$ even more compactly as
\begin{eqnarray}
\delta W_{\mu\nu}=\frac{1}{2}\Omega^{-2}\eta^{\sigma\rho}\eta^{\alpha\beta}\partial_{\sigma}\partial_{\rho} \partial_{\alpha}\partial_{\beta}[\Omega^{-2}h_{\mu\nu}]^{T\theta},
\label{F3}
\end{eqnarray}
This one-term expression for $\delta W_{\mu\nu}$ involves no choice of gauge, and is exact without approximation for conformal gravity fluctuations around any geometry whatsoever that is conformal to flat.

\section{On the Unitarity of Quantum Conformal Gravity}
\label{SG}

\subsection{The Nature of the Problem}

While our interest in this paper is in classical aspects of conformal gravity fluctuations, one needs to be assured that these results survive quantization. Since conformal gravity is a fourth-order derivative theory we need to address two potential concerns that higher derivative theories are thought to have, an Ostrogradski instability concern that there might be states of negative energy, and a unitarity concern that there might be states of negative norm. Both of these issues have been resolved in  \cite{Bender2008a,Bender2008b,Mannheim2011a,Mannheim2018}.

To see what is involved, we note that in fluctuations around flat spacetime the conformal gravity gravitational fluctuation function $\delta W_{\mu\nu}$ reduces (cf. (\ref{AP25})) to $\delta W_{\mu\nu}=(1/2)\eta^{\sigma\rho}\eta^{\alpha\beta}\partial_{\sigma}\partial_{\rho} \partial_{\alpha}\partial_{\beta}K_{\mu \nu}$. With all the components of $K_{\mu\nu}$ being decoupled from each other in $\delta W_{\mu\nu}$, they thus propagate independently. For each of these components the  propagator takes the form $D(k_{\mu})=1/k^4$ in momentum space. On writing it as the limit 
\begin{eqnarray}
D(k_{\mu})=\frac{1}{k^4}=\lim_{M_1^2,M_2^2\rightarrow 0} \frac{1}{M_1^2-M_2^2}\left(\frac{1}{k^2-M_1^2}-\frac{1}{k^2-M_2^2}\right),
\label{G1}
\end{eqnarray}
we would obtain two standard second-order propagators with a relative minus sign between them, and thus obtain some poles in the complex $k_0$ plane that have negative residues. We would thus anticipate the presence of states with negative norm, and a unitarity-violating completeness relation for energy eigenstates of the form 
\begin{eqnarray}
\sum |n\rangle\langle n|-\sum |m\rangle\langle m|=I,
\label{G2}
\end{eqnarray}
since in the scalar field theory to be described below its insertion into $D(x)=i\langle \Omega|T(\phi(x)\phi(0))|\Omega\rangle$ would lead to (\ref{G1}). The good renormalizable, ultraviolet-convergent structure of a $1/k^4$ propagator would thus appear to be accompanied by an unacceptable negative norm structure. However, the presence of such negative residues is not actually indicative of the existence of ghost states with negative norm, since as  shown in \cite{Bender2008a,Bender2008b} and as described below, one can actually produce such negative residues in a Hilbert space in which all inner products are positive.

If we choose the standard Feynman contour $i\epsilon$ prescription, then on setting  $\omega_1=+({\bf k}^2 +M_1^2)^{1/2}$, $\omega_2=+({\bf k}^2 +M_2^2)^{1/2}$, the propagator would take the form  
\begin{eqnarray}
&&D(k_{\mu})=\frac{1}{(k_0^2-{\bf k}^2+i\epsilon)^2}=\lim_{M_1^2,M_2^2\rightarrow 0} \frac{1}{M_1^2-M_2^2}\left[\frac{1}{k_0^2-\omega_1^2+i\epsilon}-\frac{1}{k_0^2-\omega_2^2+i\epsilon}\right]
\nonumber\\
&&=\lim_{M_1^2,M_2^2\rightarrow 0} \frac{1}{M_1^2-M_2^2}\left[
\frac{1}{2\omega_1}\left(\frac{1}{k_0-\omega_1+i\epsilon}-\frac{1}{k_0+\omega_1-i\epsilon}\right)
-\frac{1}{2\omega_2}\left(\frac{1}{k_0-\omega_2+i\epsilon}-\frac{1}{k_0+\omega_2-i\epsilon}\right)
\right],
\label{G3}
\end{eqnarray}
with positive energy states propagating forward in time (poles below the real $k_0$ axis) and negative energy states  (poles above the real $k_0$ axis) propagating backward in time. With there being no forward in time propagation of negative energies there is no Ostrogradski instability associated with the standard Feynman contour. As noted in \cite{Bender2008b}, one can find a contour that would lead to forward in time propagation of negative energy states (poles below the real $k_0$ axis), viz.  
\begin{eqnarray}
&&D(k_{\mu})=\lim_{M_1^2,M_2^2\rightarrow 0} \frac{1}{M_1^2-M_2^2}\left[
\frac{1}{2\omega_1}\left(\frac{1}{k_0-\omega_1+i\epsilon}-\frac{1}{k_0+\omega_1-i\epsilon}\right)
-\frac{1}{2\omega_2}\left(\frac{1}{k_0-\omega_2-i\epsilon}-\frac{1}{k_0+\omega_2+i\epsilon}\right)
\right],~~~
\label{G4}
\end{eqnarray}
in a contour in which all poles below the real $k_0$ axis have positive residues. Thus we can trade negative residues for negative energies. However (\ref{G4}) is not the conventional Feynman contour and so we do not consider it further here. Thus we only need to address the negative residues given in (\ref{G3}).

\subsection{Lack of Normalizability of the Energy Eigenstates}

To explore the negative residue issue, we note that the propagator given in (\ref{G1}) can be associated with an equivalent flat spacetime scalar field theory with action
\begin{eqnarray}
I_S&=&\frac{1}{2}\int d^4x\bigg{[}\partial_{\mu}\partial_{\nu}\phi\partial^{\mu}
\partial^{\nu}\phi -(M_1^2+M_2^2)\partial_{\mu}\phi\partial^{\mu}\phi+M_1^2M_2^2\phi^2\bigg{]},
\label{G5}
\end{eqnarray}
where $\phi(x)$ is a neutral scalar field (and where now we use a metric with ${\rm diag}[\eta_{\mu\nu}]=(1,-1,-1,-1)$). For this action the equation of motion is given by
\begin{eqnarray}
&&(\partial_t^2-\nabla^2+M_1^2)(\partial_t^2-\nabla^2+M_2^2)\phi(x)=0,
\label{G6}
\end{eqnarray}
with (\ref{G1}) then following. For the theory the phase space Hamiltonian is given by $\int d^3x T_{00}$, where $T_{\mu\nu}$ and the canonical variables are constructed by the Ostrogradski method that is used for higher derivative theories, and are of the form 
\begin{eqnarray}
T_{\mu\nu}&=&\pi_{\mu}\phi_{,\nu}+\pi_{\mu}^{\phantom{\mu}\lambda}\phi_{,\nu,\lambda}-\eta_{\mu\nu}{\cal L},
\nonumber\\ 
 \pi^{\mu}&=&\frac{\partial{\cal L}}{\partial \phi_{,\mu}}-\partial_{\lambda
}\left(\frac{\partial {\cal L}}{\partial\phi_{,\mu,\lambda}}\right)=-\partial_{\lambda}\partial^{\mu}\partial^{\lambda}\phi- (M_1^2+M_2^2)\partial^{\mu}\phi,
\qquad \pi^{\mu\lambda}=\frac{\partial {\cal L}}{\partial \phi_{,\mu,\lambda}}=\partial^{\mu}\partial^{\lambda}\phi,
\nonumber\\
T_{00}&=&\pi_{0}\dot{\phi}+\frac{1}{2}\pi_{00}^2+\frac{1}{2}(M_1^2+M_2^2)\dot{
\phi}^2-\frac{1}{2}M_1^2M_2^2\phi^2
-\frac{1}{2}\pi_{ij}\pi^{ij}+\frac{1}{2}(M_1^2+M_2^2)\phi_{,i}\phi^{,i},
\label{G7}
\end{eqnarray}
\begin{eqnarray}
[\phi(\textbf{x},t),\pi_0(\textbf{y},t)]&=&i\delta^3(\textbf{x}-\textbf{y}),\qquad [\dot{\phi}(\textbf{x},t),\pi_{00}(\textbf{y},t)]=i\delta^3(\textbf{x}-\textbf{y}).
\label{G8}
\end{eqnarray}

Given the Hamiltonian, one can solve the Schrodinger equation, and when the theory is reexpressed in terms of two oscillators with frequencies $\omega_1$, $\omega_2$ (by freezing the linear momentum ${\bf k}$ to a fixed value) Bender and Mannheim found \cite{Bender2008a} that none of the energy eigenstates are normalizable. For the oscillators the Hamiltonian (to be labelled $K$) reduces to 
\begin{eqnarray}
K=p_zx+\frac{p_x^2}{2}+\frac{1}{2}\left(\omega_1^2+\omega_2^2 \right)x^2-\frac{1}{2}\omega_1^2\omega_2^2z^2,
\label{G9}
\end{eqnarray}
where we have set  $z=\phi$, $p_z=\pi_0$, $x=\dot{\phi}$, $p_x=\pi_{00}$,  with $[z,p_z]=i$, $[x,p_x]=i$.
On setting $p_z=-i\partial_z$, $p_x=-i\partial_x$,  the ground state wave function $\psi_0(x,z)$ with energy $E_0=(\omega_1+\omega_2)/2$ is found to take the form
\begin{eqnarray}
\psi_0(x,z)={\rm exp}\left[\frac{1}{2}(\omega_1+\omega_2)\omega_1\omega_2
z^2+i\omega_1\omega_2zx-\frac{1}{2}(\omega_1+\omega_2)x^2\right].
\label{G10}
\end{eqnarray}
The ground state energy eigenfunction thus behaves as a divergent ($\exp(+z^2)$) Gaussian rather than as a convergent one, to thus not be normalizable. One could not have inferred this merely by inspection of the propagator given in (\ref{G1}). Nonetheless, since it is the case, one cannot take the states in (\ref{G2}) to be normalized since  
\begin{eqnarray}
\langle\Omega|\Omega\rangle=\int dxdz\langle\Omega|x,z\rangle\langle x,z| \Omega\rangle
=\int dxdz\psi^*_0(x,z)\psi_0(x,z)=\infty,
\label{G11}
\end{eqnarray}
and thus the (\ref{G2}) completeness relation could not be valid. (Excited oscillator states are polynomials times the ground state wave function, and they are not normalizable either.) Thus in introducing (\ref{G2}) one is assuming that the states are normalizable without first having determined whether or not they are. 

\subsection{Lack of Hermiticity of the Hamiltonian}

With the states not being normalizable, one could not throw away surface terms in an integration by parts, and thus despite its appearance the Hamiltonian could not be Hermitian. (Whether or not surface terms can be ignored is a property of the states in which matrix elements are calculated and not a property of the operators that appear in those matrix elements.) However, all the poles in the propagator $D(k_{\mu})$ lie on the real axis in the complex $k_0$ plane, and thus all the energy eigenvalues are real. Now while Hermiticity of a Hamiltonian implies the reality of its energy eigenvalues there is no theorem that states that the eigenvalues of a non-Hermitian Hamiltonian must be complex. Hermiticity is thus sufficient for the reality of eigenvalues but not necessary.  In \cite{Mannheim2018b,Mostafazadeh2002,Solombrino2002}  a necessary condition has been given: the Hamiltonian must possess an antilinear symmetry. The fourth-order theory Hamiltonian thus falls into the class of non-Hermitian but $PT$-symmetric Hamiltonians ($P$ being the parity operator and $T$ being the antilinear time reversal operator) that have been found by Bender and collaborators to have all eigenvalues real (see e.g. the review of \cite{Bender2007}), with $H=p^2+ix^3$ being the canonical example \cite{Bender1998,Bender1999}. The surprise of the work of Bender and collaborators is that while not being Hermitian, the eigenvalues of $H=p^2+ix^3$ are all real. The surprise of the fourth-order theory is that the Hamiltonian $K$ is not Hermitian even though it appears to be (viz. no telltale factors of $i$), and then while not being Hermitian its eigenvalues are nonetheless real.

\subsection{The Resolution of the Problem -- Continuing into the Complex Plane}

To deal with the lack of normalizability of the fourth-order theory energy eigenstates it was pointed out  in \cite{Bender2008a,Bender2008b} that  if one continued the theory into the complex plane one could find a domain known as a Stokes wedge in which the wave functions are normalizable (an $\exp(z^2)$ Gaussian that is divergent on the real $z$ axis becomes convergent on the imaginary $z$ axis). And it is in such Stokes wedges that the theory is well-defined and one is able to throw surface terms away in an  integration by parts. It is thus in such Stokes wedges that the fourth-order theory has to be formulated, since there one can construct an inner product that is normalizable. That one can make a continuation into the complex plane at all is because such continuations preserve canonical commutation relations  and are thus legitimate. While one ordinarily represents a commutation relation such as $[z,p_z]=i$ by $[z, -i\partial/\partial z]=i$, one could just as legitimately represent it by $[e^{i\theta}z, -i\partial/\partial (e^{i\theta}z)]=i$. However, in order for the momentum operator to be representable as a differential operator at all it is necessary that it act on an appropriate normalizable test function. Thus one could set $[z, -i\partial/\partial z]\psi(z)=i\psi(z)$ or $[e^{i\theta}z, -i\partial/\partial (e^{i\theta}z)]\psi(e^{i\theta}z)=i\psi(e^{i\theta}z)$, and one must choose those Stokes wedge domains in the complex plane for which $\psi(e^{i\theta}z)$ is bounded at infinity, since otherwise one could not throw away surface terms at infinity and establish Hermiticity. In the appropriate complex $z$ plane Stokes wedge $\int dxdz\psi^*_0(x,z)\psi_0(x,z)$ is finite.

In the same way that the Hamiltonian is not Hermitian when acting on eigenstates such $\exp(z^2)$ when $z$ is real, the same is true of the position and momentum operators when they act on the selfsame states. Thus even though the position and momentum operators are Hermitian when acting on their own eigenstates, they are not Hermitian when acting on $\exp(z^2)$ type eigenstates of the Hamiltonian when $z$ is real, since for those states one cannot throw away surface terms in an integration by parts for a momentum operator that is represented by $-i\partial/\partial z$.  And in fact it is precisely such a mismatch between the action of an operator on its own eigenstates and on those of the Hamiltonian that is central to the $PT$ symmetry program, and one has to find an appropriate Stokes wedge for which one can throw surface terms away in integrations by parts for all the operators of interest in the theory. For the fourth-order theory the requisite Stokes wedge does not include the real $z$ axis. Instead it includes the imaginary $z$ axis, and as is shown below, it is because of this that one is able to obtain a unitary theory. 

\subsection{Implementation of the Complex Plane Continuation for Operators}

For operators one can implement the continuation into the complex plane by a commutation-preserving similarity transformation $T=\exp(\pi p_z z/2)$, to obtain
\begin{eqnarray}
TzT^{-1}=-iz,\qquad Tp_zT^{-1}=ip_z.
\label{G12}
\end{eqnarray}
On setting $y=-iz$, $q=ip_z$, the $y$ and $q$ position and momentum operators are now Hermitian in the Stokes wedge that contains the imaginary $z$ axis, with $y$ and $q$ obeying $[y,q]=i$. On setting $p_x=p$ for notational simplicity, so that $[x,p]=i$, we find that the Hamiltonian $K$ transforms into 
\begin{eqnarray}
TKT^{-1}=H=-iqx+\frac{p^2}{2}+\frac{1}{2}\left(\omega_1^2+\omega_2^2 \right)x^2+\frac{1}{2}\omega_1^2\omega_2^2y^2,
\label{G13}
\end{eqnarray}
and through the emergence of the factor of $i$, its lack of Hermiticity is now apparent. (We transform $z$ and $p_z$ but not $x$ or $p_x$, so $x$ and $p_x$ start off Hermitian and stay Hermitian since $\psi_0(x,z)$ is well-behaved at large real $x$.) Thus in taking care of the lack of normalizability of the eigenfunctions we obtain a Hamiltonian  that has the same structure as $p^2+ix^3$, a Hamiltonian that also is not Hermitian but has all eigenvalues real. With its real eigenvalues the non-Hermitian Hamiltonian $H$ given in (\ref{G13}) thus has an antilinear symmetry, and in \cite{Bender2008a,Bender2008b} it was identified as $PT$.

\subsection{Antilinear Symmetry}

The general idea behind antilinear symmetry is that if there exists an antilinear operator $A$ that commutes with a Hamiltonian and if one has an eigenket that obeys $H|n\rangle=E_n|n\rangle$, then as first noted by Wigner in his study of time reversal invariance, one obtains $AH|n\rangle=AE_n|n\rangle$, i.e. $HA|n\rangle=E_n^*A|n\rangle$. Thus for every state with eigenvalue $E_n$ there is a state with eigenvalue $E_n^*$, and thus operators with an antilinear symmetry can have all eigenvalues real. The study of the implications of antilinearity thus goes hand in hand with the identification of appropriate Stokes wedges, i.e. with the identification of domains in which one can impose convergent asymptotic boundary conditions, so that energy eigenfunctions can serve as good test functions.

\subsection{The Positive Definite $V$-norm}

In general it was noted in  \cite{Mostafazadeh2002,Solombrino2002,Mannheim2018b} that if a Hamiltonian (taken here to be time independent) has an antilinear symmetry there will always exist a time independent operator $V$ that obeys the so-called pseudo-Hermiticity condition $VH=H^{\dagger}V$. If $V$ is invertible (this being the case for the fourth-order theory \cite{Bender2008a,Bender2008b}), then $H$ and $H^{\dagger}$ are isospectrally related according to $H^{\dagger}=VHV^{-1}$, to thus have the same set of eigenvalues,  and thereby permit all energy eigenvalues to be real. If $H$ is not Hermitian but the $E_n$ are  real, then from $i\partial_t|n\rangle=H|n\rangle=E_n|n\rangle$ we obtain 
\begin{eqnarray}
-i\partial_t\langle n|=\langle n|H^{\dagger}=\langle n|VHV^{-1}=\langle n|E_n,\qquad -i\partial_t\langle n|V=\langle n|VH=\langle n|VE_n,
\label{G14}
\end{eqnarray}
with it being the state $\langle n|V$ that is the left eigenstate of $H$ and not the bra $\langle n|$ itself. Consequently in the non-Hermitian case the standard Dirac norm $\langle n(t)|n(t)\rangle=\langle n(0)|e^{iH^{\dagger}t}e^{-iHt}|n(0)\rangle$ is not time independent (i.e. not equal to $\langle n(0)|n(0)\rangle$), and one cannot use it as an inner product. However, the $V$ norm is time independent since 
\begin{eqnarray}
i\partial_t\langle n(t)|V|n(t)\rangle=\langle n(t)|(VH-H^{\dagger} V)|n(t)\rangle=0,\qquad 
\langle n(t)|V|n(t)\rangle=\langle n(0)|Ve^{iHt}e^{-iHt}|n(0)\rangle=\langle n(0)|V|n(0)\rangle.~
\label{G15}
\end{eqnarray}
It is thus the $V$-norm that is needed in order to implement conservation of probability (unitarity of time development), with  the completeness relation being given not by (\ref{G2}) but by 
\begin{eqnarray}
\sum |n\rangle\langle n|V=I
\label{G16}
\end{eqnarray}
instead. As shown in \cite{Mannheim2018c}, when charge conjugation ($C$) is separately conserved,  the $V$-norm is the same as the overlap of state with its $PT$ conjugate. As we discuss below and in \cite{Mannheim2018b}, more generally the $V$-norm is the same as the overlap of state with its $CPT$ conjugate.

In the appropriate Stokes wedge the fourth-order theory $V$ norm is finite. With all the energy eigenvalues of the Hamiltonian $H$ given in (\ref{G13}) being real and with harmonic oscillator wave functions being complete, as shown in \cite{Bender2008a,Bender2008b}, while not Hermitian itself, $H$ must be similarity equivalent to a Hamiltonian $H^{\prime}=SHS^{-1}$ that is Hermitian. In \cite{Bender2008a,Bender2008b} $S$ was explicitly constructed according to
\begin{eqnarray}
S=\exp(-Q/2),\qquad Q=\alpha pq+\beta xy,\qquad \alpha=\frac{1}{\omega_1\omega_2}{\rm log}\left(\frac{\omega_1+\omega_2}{\omega_1-\omega_2}\right),\qquad \beta=\alpha\omega_1^2\omega_2^2,
\label{G17}
\end{eqnarray}
as $S$ implements
\begin{eqnarray}
SHS^{-1}=H^{\prime}=\frac{p^2}{2}+\frac{q^2}{2\omega_1^2}+
\frac{1}{2}\omega_1^2x^2+\frac{1}{2}\omega_1^2\omega_2^2y^2,
\label{G18}
\end{eqnarray}
to yield a Hamiltonian $H^{\prime}$ that is a manifestly Hermitian, conventional two-oscillator Hamiltonian, one for which all eigenstates have a  standard positive norm. Consequently,  the fourth-order theory is unitary.

With $H^{\prime\dagger}=H^{\prime}$, it follows that $S^{\dagger}SH(S^{\dagger}S)^{-1}=H^{\dagger}$, with the eigenstates of $H^{\prime}$ being related to those of $H$ by $|n^{\prime}\rangle =S|n\rangle$, $\langle n^{\prime}|=\langle n|S^{\dagger}$. We can thus identify $V$ with $S^{\dagger}S$, with the $V$-norm not just being finite but even being positive definite, with the states obeying 
\begin{eqnarray}
\langle n^{\prime}|m^{\prime} \rangle=\langle n|S^{\dagger}S|m \rangle=\langle n|V|m \rangle=\delta_{m,n}.
\label{G19}
\end{eqnarray}
Analogously, the propagator is given not by $D(x)=i\langle \Omega|T(\phi(x)\phi(0))|\Omega\rangle$ but by
\begin{eqnarray}
D(x)=i\langle \Omega|VT(\phi(x)\phi(0))|\Omega\rangle
\label{G20}
\end{eqnarray}
instead, with the insertion of (\ref{G16}) into (\ref{G20}) leading to (\ref{G1}) \cite{Bender2008b}. Consequently, the relative minus sign in (\ref{G1}) is generated by the presence of the $V$ operator and not by any possible minus signs associated with the states themselves. 

\subsection{Extension to include Loop Diagrams}

When an $I_{int}=-\int d^4x \lambda \phi^4$ interaction is added on to the action $I_S$ given in (\ref{G5}), as shown in \cite{Mannheim2018} positivity is not lost in loop diagrams  and unitarity is preserved. This is of course to be anticipated since once the Hilbert space associated with the action $I_S$ given in (\ref{G5}) is free of negative norm states, the theory must  remain free of negative norm states  when interactions are included since one cannot change the signature of a Hilbert space in perturbation theory. We refer the reader to \cite{Mannheim2018} where one can find a detailed analysis of how unitarity is specifically maintained in loop diagrams. Quantum conformal gravity is thus a unitary theory in lowest perturbative order (analog of $I_S$) and remains so under radiative gravitational corrections (analog of $I_{int}$).

\subsection{Fundamentality of $CPT$ Symmetry}

As regards the issue of whether there might be some privileged antilinear symmetry that might actually always be required in quantum field theory, in \cite{Mannheim2016b,Mannheim2018b} it was shown quite generally that if one requires only that inner products be time independent and that the theory be invariant under the complex Lorentz group, the symmetry of the Hamiltonian is then uniquely prescribed to be the antilinear $CPT$. These requirements are quite  minimal, and hold regardless of whether the Hamiltonian may or may not be Hermitian. The $CPT$ theorem is thus extended to the non-Hermitian case, and the needed relevant time independent inner products are those between states and their $CPT$ conjugates, and not those between states and their Hermitian conjugates. For theories that are separately charge conjugation invariant such as the above fourth-order scalar field theory (the scalar field being neutral), or analogously the conformal gravity theory itself (the metric being neutral), $CPT$ symmetry then reduces to $PT$ symmetry. Conformal gravity thus falls into the class of non-Hermitian but $PT$-symmetric theories studied by Bender and collaborators. Since the $V$ norm is equivalent to the $CPT$ norm and thus to the $PT$ norm when $C$ is separately conserved, conformal gravity is thus unitary to all quantum perturbative orders and none of its states has negative norm. 

\subsection{Conclusion}

To conclude, we note that the propagator given in (\ref{G1}) is itself purely a  c-number. It is not a q-number operator. From a knowledge of the c-numbers that follow solely from the structure of differential equations of motion  one cannot infer the structure of the underlying quantum Hilbert space theory that would generate them, and one could not have inferred that the quantum-mechanical energy eigenfunctions would not have not been normalizable merely by inspection of the c-number propagator given in (\ref{G1}). Thus a priori one cannot identify the propagator in (\ref{G1}) with a matrix element such as $D(x)=i\langle \Omega|T(\phi(x)\phi(0))|\Omega\rangle$. One has to first construct the quantum Hilbert space and then determine the c-number matrix elements and not the other way round. And it does not follow that because one is used to working with theories where the propagator is given by  $D(x)=i\langle \Omega|T(\phi(x)\phi(0))|\Omega\rangle$ that this will always be the case. And indeed, when Bender and Mannheim did construct the fourth-order theory quantum Hilbert space they found that the propagator was not given by $D(x)=i\langle \Omega|T(\phi(x)\phi(0))|\Omega\rangle$ at all but by $D(x)=i\langle \Omega|VT(\phi(x)\phi(0))|\Omega\rangle$ instead, with all unitarity concerns then being completely resolved.

\end{document}